\documentclass[structabstract]{aa} 
\usepackage[utf8]{inputenc}
\usepackage{txfonts} 
\usepackage{graphicx} 
\usepackage{natbib} 
\usepackage{url} 
\usepackage{float}
\usepackage{graphicx}	
\usepackage{amsmath}	
\usepackage{amssymb}	
\usepackage[dvipsnames]{xcolor}

\usepackage{comment}

\usepackage[colorlinks = true,
            linkcolor = blue,
            urlcolor  = blue,
            citecolor = blue,
            anchorcolor = blue]{hyperref}

\usepackage{ulem}

\usepackage{longtable}

\usepackage{threeparttable}

\usepackage{colortbl}

\usepackage{subcaption}
\usepackage{mwe}

\usepackage{enumitem}

\usepackage{pifont}
\usepackage{dirtytalk}
\usepackage[switch]{lineno}

\newcommand{\be}{\begin{equation}}
\newcommand{\ee}{\end{equation}}

\newcommand{\rstar}{R_{\star}}

\newcommand{\mdot}{\dot{M}}

\def\msun{\, \mathrm{M}_{\hbox{$\odot$}}}
\def\rsun{\, \mathrm{R}_{\hbox{$\odot$}}}
\def\Zsun{\, \mathrm{Z}_{\hbox{$\odot$}}}

\newcommand{\rbb}{R_{\rm BB}}

\newcommand{\orcid}[1]{\href{https://orcid.org/#1}
{\includegraphics[width=10pt]{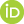}}}

\DeclareRobustCommand{\pjp}{\hyperlink{cite.Pessi2023}{P23 }}
\DeclareRobustCommand{\pjpnospc}{\hyperlink{cite.Pessi2023}{P23}}

\begin{document}

   \title{Luminous, rapidly declining supernovae as stripped transitional objects in low metallicity environments: the case of SN~2022lxg}

    \titlerunning{SN~2022lxg}

    \author{P. Charalampopoulos\inst{1}\fnmsep\thanks{Contact e-mail: \href{mailto:pachar@utu.fi}{pachar@utu.fi}}\orcid{0000-0002-0326-6715}\
    \and
    R. Kotak\inst{1}\
    \orcid{0000-0001-5455-3653}
    \and
    J. Sollerman\inst{2}
    \orcid{0000-0003-1546-6615}
    \and
    C.~P. Guti\'errez\inst{3,4}
    \orcid{0000-0003-2375-2064}
    \and
    M. Pursiainen\inst{5}
    \orcid{0000-0003-4663-4300}
    \and
    T. L. Killestein\inst{1}
    \orcid{0000-0002-0440-9597}
    \and
    S. Schulze\inst{6}
    \orcid{0000-0001-6797-1889}
    \and
    P. J. Pessi\inst{2}
    \orcid{0000-0002-8041-8559}
    \and
    K. Maeda\inst{7}
    \orcid{0000-0003-2611-7269}
    \and
    T. Kangas\inst{8,1}
    \orcid{0000-0002-5477-0217}
    \and
    Y.-Z. Cai\inst{9,10,11}
    \orcid{0000-0002-7714-493X}
    \and
    C. Fremling\inst{12,13}
    \orcid{0000-0002-4223-103X}
    \and
    K. R. Hinds\inst{14}
    \orcid{0000-0002-0129-806X}
    \and
    T. Jegou du Laz\inst{13}
    \orcid{0009-0003-6181-4526}
    \and
    E. Kankare\inst{1}
    \orcid{0000-0001-8257-3512}
    \and
    M. M. Kasliwal\inst{13}
    \orcid{0000-0002-5619-4938}
    \and
    H. Kuncarayakti\inst{1}
    \orcid{0000-0002-1132-1366}
    \and
    P. Lundqvist\inst{2}
    \orcid{0000-0002-3664-8082}
    \and
    F. J. Masci\inst{15}
    \orcid{0000-0002-8532-9395}
    \and
    S. Mattila\inst{1,16}
    \orcid{0000-0002-8532-9395}
    \and
    D. A. Perley\inst{14} 
    \orcid{0000-0001-8472-1996}
    \and
    A. Reguitti\inst{17,18}
    \orcid{0000-0003-4254-2724}
    \and 
    T. M. Reynolds\inst{1,19,20}
    \orcid{0000-0002-1022-6463}
    \and
    M. Stritzinger\inst{21}
    \orcid{0000-0002-5571-1833}
    \and
    L. Tartaglia\inst{22}
    \orcid{0000-0003-3433-1492}
    \and
    J. Van Roestel\inst{23}
    \orcid{0000-0002-2626-2872}
    \and
    A. Wold\inst{15}
    \orcid{0000-0002-9998-6732}
     \
    }

    \institute{Department of Physics and Astronomy, University of Turku, FI-20014 Turku, Finland
    \and
    The Oskar Klein Centre, Department of Astronomy, Stockholm University, AlbaNova, 10691, Stockholm, Sweden
    \and
    Institut d’Estudis Espacials de Catalunya (IEEC), E-08034  
    Barcelona, Spain
    \and
    Institute of Space Sciences (ICE, CSIC), Campus UAB, Carrer de  
    Can Magrans, s/n, E-08193 Barcelona, Spain
    \and
    Department of Physics, University of Warwick, Gibbet Hill Road, Coventry, CV4 7AL, UK
    \and
    Center for Interdisciplinary Exploration and Research in Astrophysics (CIERA), Northwestern University, 1800 Sherman Ave, Evanston, IL 60201, USA
    \and
    Department of Astronomy, Kyoto University, Kitashirakawa-Oiwake-cho, Sakyo-ku, Kyoto 606-8502, Japan
    \and
    Finnish Centre for Astronomy with ESO (FINCA), FI-20014 University of Turku, Finland
    \and
    Yunnan Observatories, Chinese Academy of Sciences, Kunming 650216, P.R. China
    \and
    International Centre of Supernovae, Yunnan Key Laboratory, Kunming 650216, P.R. China
    \and
    Key Laboratory for the Structure and Evolution of Celestial Objects, Chinese Academy of Sciences, Kunming 650216, P.R. China
    \and
    Caltech Optical Observatories, California Institute of Technology, Pasadena, CA 91125, USA
    \and
    Division of Physics, Mathematics and Astronomy, California Institute of Technology, Pasadena, CA 91125, USA
    \and
    Astrophysics Research Institute, Liverpool John Moores University, 146 Brownlow Hill, Liverpool L3 5RF, UK
    \and
    IPAC, California Institute of Technology, 1200 E. California
             Blvd, Pasadena, CA 91125, USA
    \and
    School of Sciences, European University Cyprus, Diogenes street, Engomi, 1516 Nicosia, Cyprus
    \and
    INAF –- Osservatorio Astronomico di Brera, Via E. Bianchi 46, I23807 Merate (LC), Italy
    \and
    INAF –- Osservatorio Astronomico di Padova, Vicolo dell’Osservatorio 5, I-35122 Padova, Italy
    \and
    Cosmic Dawn Center (DAWN), Niels Bohr Institute, University of Copenhagen, 2200, Denmark
    \and
    Niels Bohr Institute, University of Copenhagen, Jagtvej 128, 2200 København N, Denmark
    \and
    Department of Physics and Astronomy, Aarhus University, Ny Munkegade 120, DK-8000 Aarhus C, Denmark
    \and
    INAF -- Osservatorio Astronomico d’Abruzzo, via Mentore Maggini snc I-64100 Teramo, Italy
    \and
    Anton Pannekoek Institute for Astronomy, University of Amsterdam, 1090 GE Amsterdam, The Netherlands
    \
             }

   \date{Received - ; accepted -}

 

  \abstract
   {We present an analysis of the optical and near-infrared properties of SN~2022lxg, a bright ($\rm M_{g\, \mathrm{peak}}=-19.41$\,mag) and rapidly evolving SN. It was discovered within a day of explosion, and rose to peak brightness in $\sim$10\,d. Two distinct phases of circumstellar interaction are evident in the data. The first is marked by a steep blue continuum (T $>15,000$\,K) with flash-ionisation features due to hydrogen and \ion{He}{II}. The second, weaker phase is marked by a change in the colour evolution accompanied by changes in the shapes and velocities of the spectral line profiles. Narrow P-Cygni profiles ($\sim150\,$km\,s$^{-1}$) of \ion{He}{I} further indicate the presence of slow-moving unshocked material and suggesting partial stripping of the progenitor. The fast decline of the light curve from peak (3.48$\pm$ 0.26\,mag\,$\rm (50\,d)^{-1}$ in $g$-band) implies that the ejecta mass must be low. Spectroscopically, until $+35$\,d there are similarities to some Type IIb SNe but then there is a transition to spectra that are more reminiscent of an interacting SN II. However, metal lines are largely absent in the spectra, even at epochs of $\sim$80\,d. Its remote location ($\sim$4.6\,kpc projected offset) from the presumed host galaxy, a dwarf with $\rm M_B \sim -14.4$ mag, is consistent with our metallicity estimate -- close to the Small Magellanic Cloud value -- obtained from scaling relations. Furthermore, several lines of evidence (including intrinsic polarisation of $p\sim(0.5-1.0) \%$) point to deviations from spherical symmetry. We suggest that a plausible way of uniting the observational clues is to consider a binary system that underwent case C mass transfer. This failed to remove the entire H-envelope of the progenitor before it underwent core-collapse. In this scenario, the progenitor itself would be more compact and perhaps straddle the boundary between blue and yellow supergiants, tying in with the early spectroscopic similarity to Type IIb SNe.
   }

   \keywords{supernovae: general --- supernovae: individual: SN~2022lxg -- stars:mass loss --- stars: circumstellar matter }

   \maketitle
%

\clearpage
\section{Introduction} \label{sec:intro}

It is well accepted that the evolution of massive ($\gtrsim$8\,$\msun$) stars is driven primarily by mass loss, be it via line or continuum driven winds, eta Carina type giant outbursts, other variability, or even due to a binary companion \citep[e.g.][]{Meynet1994,Langer1998,Woosley2002,Langer2012}. Thus, the immediate environment into which a massive star explodes is modified by these processes. Regardless of how the mass loss happens, the distribution and extent of this circumstellar material (CSM), as well as its composition, velocity, and amount can be indelibly imprinted onto observations of the supernova (SN).

A clear signature of interaction between SN ejecta and the surrounding medium is the presence of narrow (tens to hundreds of km\,s$^{-1}$)
emission lines \citep{Schlegel1990}. At the earliest epochs after explosion, radiation from the shock breakout ionises this material resulting in prominent narrow emission features associated with species such as \ion{He}{II}, \ion{C}{IV}, and \ion{N}{iii/iv}
(often dubbed \say{flash features}). As the shock propagates through the CSM, a fraction of the kinetic energy of the ejecta is converted to energetic photons; these contribute to maintaining the high ionisation levels, compensating for the relatively short recombination timescales of the ionised species. As the CSM gas is swept up by the fast-moving ejecta, the temperature drops, and the narrow emission lines become weaker, ultimately giving way to a nearly featureless continuum, followed by the emergence of broad SN features \citep[e.g.,][]{Chugai2001, Fransson2005, Gal-Yam2014, Shivvers2015, Khazov2016, Yaron2017, Dessart2017, Bruch2021, Bruch2023, Jacobson-Galan2024}. The duration of this phase is strongly dependent on the spatial extent and density of the surrounding material, but typically lasts for approximately a week in most cases \citep{Bruch2021}. This implies that the material is spatially confined, and originates from presumably enhanced mass-loss shortly (months to a few years) prior to core-collapse. 

In the single star framework and roughly solar metallicities, we expect an inverse correlation between progenitor mass and the amount of hydrogen remaining in the envelope at the time of explosion. This translates into a sequence of SN subtypes with the Type II-plateau (IIP) progenitors having the thickest H envelopes, while the Type Ic progenitors, at the other extreme, have lost both H and He layers. Between these, we find the II-linear (IIL), IIb, and Ib subtypes that reflect a transition from a H-dominated envelope to a He-dominated one. Although the above framework is appealing and borne out by observations, we both expect and find a significant contribution from binary systems. Indeed, several studies have argued that a close binary companion is necessary to efficiently remove the H and He layers \citep[e.g.,][]{Nomoto1993,Nomoto1995,Claeys2011,Yoon2017,Ercolino2024}. 
Within either single or binary progenitor frameworks, we might also expect to find a continuum of observed properties (e.g. peak luminosity, duration). These will depend primarily on the specifics of the mass and mass loss history, and the conditions in the core at the time of collapse, for each case.

Within this multi-dimensional parameter space, SNe that display extreme properties usually in terms of peak absolute brightness or decline rate from peak, often stand out \citep[e.g.][]{Barbary2009, Miller2009, Gezari2009}. While several systematic studies of regular SNe II have been conducted, incorporating an increasing number of events over the years, most do not include objects with rest-frame light curve peaks exceeding $\sim-18.5$\,mag in the $V$-band, that is luminous SNe (LSNe II; e.g., \citealt{Anderson2014,Valenti2016}). However, a growing number of such events have been identified. These luminous objects were already noted by \citet{Patat1994}, who analysed a sample of 51 SNe II and highlighted a gap between regular SNe II and brighter events ($-18.5$\,mag in the $B$-band). More recently, \citet{Pessi2023} (\pjp hereafter) considered a sample of six SNe II with peak $V$ band magnitudes brighter than $-18.5$, persistent blue colours and fast decline rates. Spectroscopically, the Balmer lines showed broad, multi-component emission profiles over the time span of the observations ($\sim10-100$\,d for most of the sample), and metal lines were weak if at all present. Type II SNe that are brighter than $-18.5$\,mag at peak in the optical region are unlikely to be missed by transient discovery surveys, and must therefore be rare. It is important to understand whether they arise from some unique combination of parameters, or whether they simply represent the tail of the Type II parameter distribution with some preferred viewing angle.

\begin{figure}
  \centering
  \includegraphics[width=0.5 \textwidth]{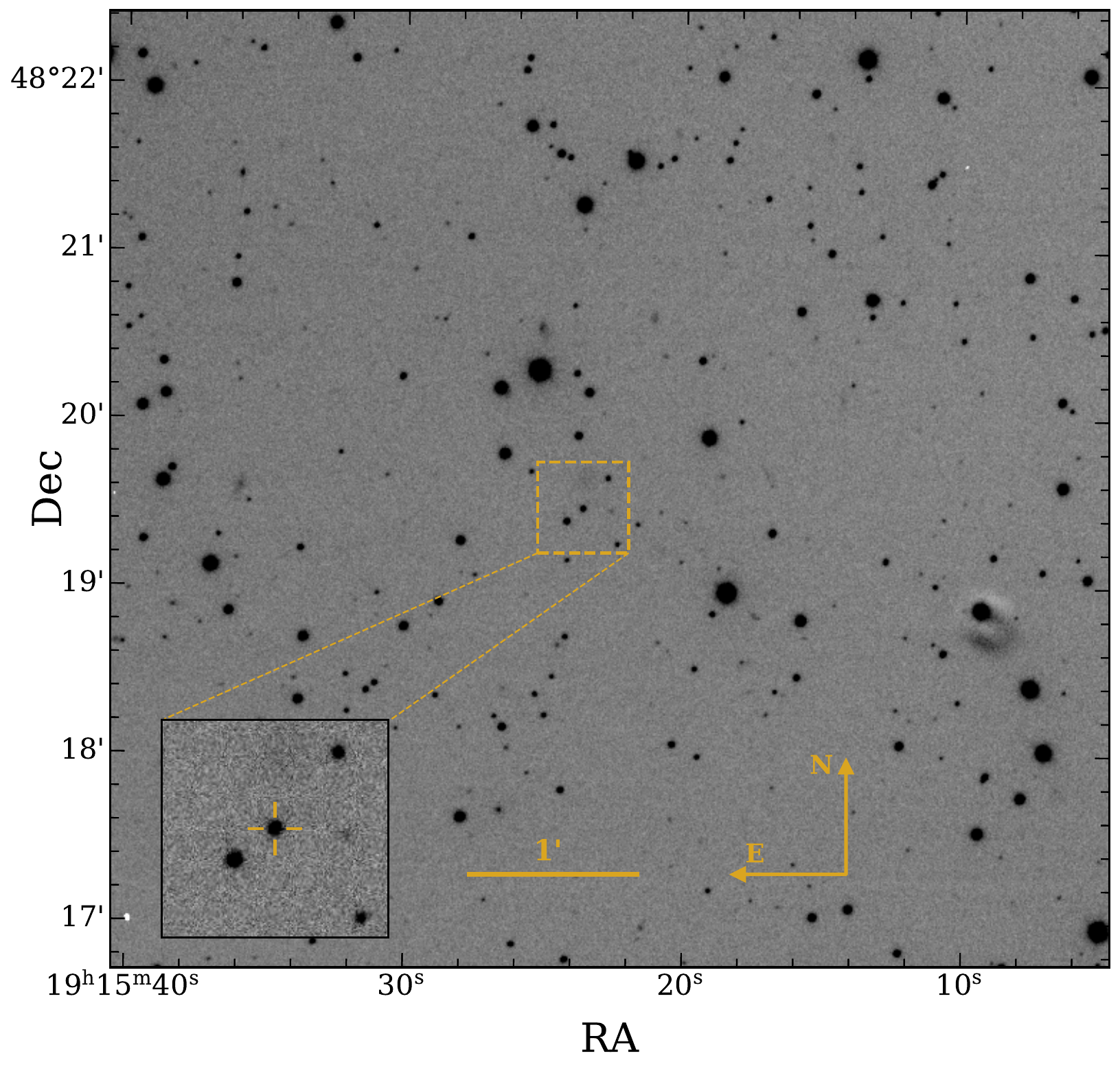}
  \caption{$r$-band image showing SN~2022lxg ($\alpha = 19^{\rm h}15^{\rm m}23.630^{\rm s}$, $\delta = +48^{\circ}19'27.70''$, J2000), taken with NOT+ALFOSC on 2022 Jul 25
($+54.2$\,d). The inset shows the region around the SN; no obvious host galaxy is apparent.}
  \label{fig:position}
\end{figure}

On 2022 Jun 04, the All-Sky Automated Survey for Supernovae (ASAS-SN; \citealt{Shappee2014}) collaboration reported the discovery of a rising transient (ASASSN-22hp, IAU name: AT~2022lxg) in an uncatalogued host galaxy \citep{Stanek2022} to the Transient Name Server. Four days later, on 2022 Jun 08, the transient was classified as a SN II (SN~2022lxg) based on a blue and featureless spectrum \citep{Ashall2022}. Given the excellent explosion epoch constraints, the rapid rise to a peak absolute brightness of $-$19.3 in the $r$ band, and the presence of narrow emission lines in the spectrum at +2\,d, we embarked on an observational campaign to monitor its evolution. We show the field of the SN in Fig. \ref{fig:position}. As no redshift information was available for the presumed host galaxy, we measured the redshift of SN~2022lxg to be ${z = 0.0214 \pm 0.0006}$, from the centroid of H$\alpha$ and H$\beta$ lines in the late time spectra ($+65$ and $+80$\,d). Correcting the spectra with this value results in the flash-ionisation lines in the early spectra (H$\alpha$, H$\beta$ and \ion{He}{II} $\lambda4686$) being at their respective rest wavelengths (Sect. \ref{subsec:spec_analysis}).

In what follows, we present the follow-up and analysis of SN~2022lxg. 
We assume a Planck Collaboration $\Lambda$CDM cosmology with \mbox{H$_{0}$ = 67.4\,km\,s$^{-1}$ Mpc$^{-1}$}, ${\Omega_{\rm m}}$ = 0.315 and ${\Omega_{\rm \Lambda}}$ = 0.685 \citep{Aghanim2020}. The redshift, as inferred above, corresponds to a distance of 96.6\,Mpc and distance modulus of $\mu=34.925$. 
All phases are given relative to the estimated explosion epoch (MJD=59\,731.37), in the transient rest-frame ( Sect. \ref{subsubsec:bb_lc}). Magnitudes are in the AB system \citep{Oke1983} unless noted otherwise, and the reported uncertainties correspond to 68\% ($1\sigma$) and upper-limits to $3\sigma$.

\section{Observations and data reduction} \label{sec:observations}

We acquired well-sampled imaging and spectroscopy, and three epochs of imaging polarimetry from a range of telescopes (\S \ref{sec:observations}). 
There is no evidence for host reddening in the spectra given the blue slope at early times and lack of absorption features due to the \ion{Na}{I} D lines; hence we consider the host reddening to be negligible. Throughout this work, we assume a \citet{Cardelli1989} extinction law with $R_{V}$ = 3.1 and a foreground Galactic extinction of $A_{V}$ = 0.1815\,mag \citep{Schlafly2010}, to deredden our photometry and spectra.  

\begin{figure*}
\centering
\includegraphics[width=1 \textwidth]{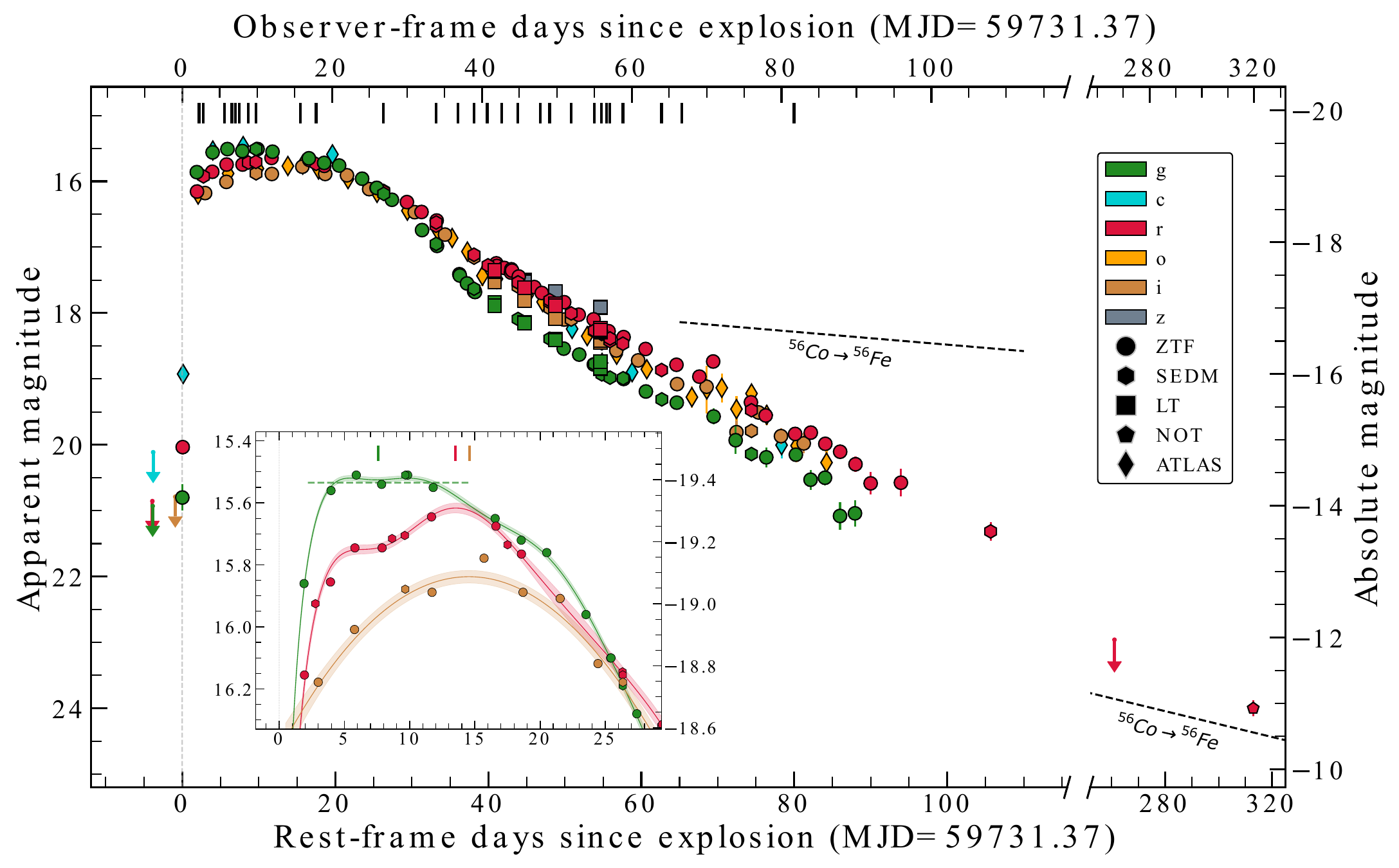}
\caption{Light curves of SN~2022lxg (corrected for MW extinction; $\rm E(B-V)_{MW}=0.059$\,mag). The explosion epoch is constrained to MJD 59\,731.37. The short black vertical dashes denote the epochs of spectroscopy (see Table \ref{tab:spec_log}). Non-detections are shown as small downward-facing arrows. The $^{56}$Co decay rate is shown as a black dashed line. Note the break in the x-axis to accommodate the latest epochs. The inset focuses on the phases around the peak light (for the $gri$ bands). The $g$-band seems to plateau for $\sim$ 8 days at its peak (horizontal dashed line is a linear fit to guide the eye), while the $r$ band shows a bump around the time of the start of the $g$-band plateau, but then continues to rise to its main peak. We show the Gaussian process interpolations used to infer the peak epochs in the $gri$ bands (with 1$\sigma$ uncertainties as shaded regions), which are marked with short vertical dashes (see also Table \ref{tab:basic_props}).}
\label{fig:photometry}
\end{figure*}

\subsection{Ground-based imaging} \label{subsec:gbi}

We obtained $gri$ imaging with a roughly $2-3$\,d cadence via the Zwicky Transient Facility (ZTF; \citealt{Graham2019,Bellm2019a,Dekany2020}), the Palomar Schmidt 48-inch (P48) Samuel Oschin and the Spectral Energy Distribution Machine Rainbow Camera on the Palomar 60-inch telescope (SEDM; \citealt{Blagorodnova2018a,Kim2022}). These data were processed using the ZTF forced photometry service\footnote{\url{https://ztfweb.ipac.caltech.edu/cgi-bin/requestForcedPhotometry.cgi}} \citep{Masci2019} and \texttt{FPipe} \citep{Fremling2016}, respectively. Further imaging obtained at the Liverpool Telescope (LT; \citealt{Steele2004}) with the IO:O imager in the $griz$ filters was reduced using custom pipelines, while light curves using imaging from the Asteroid Terrestrial-impact Last Alert System (ATLAS; \citealt{Tonry2018,Smith2020}) survey in the $o$ and $c$ bands were generated using the ATLAS Forced Photometry\footnote{\url{https://fallingstar-data.com/forcedphot/}} service \citep{Shingles2021}. Two epochs of late-time ($>250$\,d) imaging were obtained with the Alhambra Faint Object Spectrograph and Camera (ALFOSC) mounted on the 2.56 m Nordic Optical Telescope (NOT) on La Palma, Spain. The complete, dereddened optical light curves are shown in Fig. \ref{fig:photometry} and tabulated (non-dereddened) in Table \ref{tab:phot}.

\subsection{Optical spectroscopy} \label{subsec:opt_spec}

\begin{figure*}
\centering
\includegraphics[width=1 \textwidth]{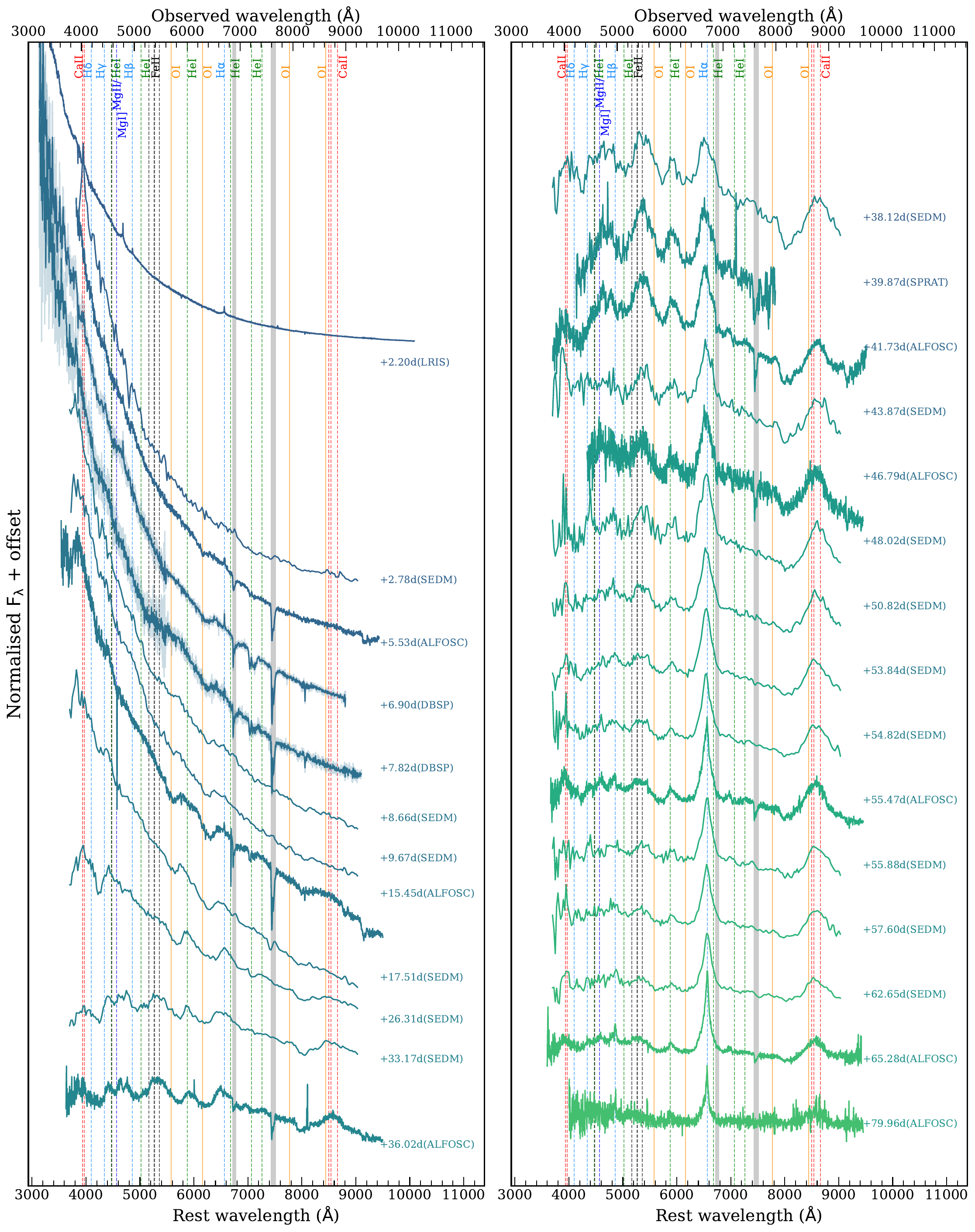}
\caption{Spectral series of SN~2022lxg (scaled with the photometry and corrected for MW extinction). Emission lines are marked with vertical dashed lines; throughout the evolution of SN~2022lxg we detect Balmer lines, \ion{He}{I} $\lambda5876$, a broad blend of \ion{Fe}{II} ($\sim$5300 \AA) and the \ion{Ca}{II} NIR triplet. Other marked species are to guide the eye. Telluric features with grey shaded vertical lines.}
\label{fig:spectra}
\end{figure*}

We were able to collect 28 spectra of SN~2022lxg spanning $\sim2-90$\,d. 
The earliest spectrum was obtained using the Low-Resolution Imaging Spectrometer (LRIS; \citealt{Oke1995}) on the Keck I 10-m telescope and reduced using \texttt{lpipe} \citep{Perley2019}. Spectra obtained using SEDM were reduced as 
described in \citet{Rigault2019}; those collected with the Double Beam Spectrograph (DBSP) on the Palomar 200-in telescope were reduced using custom pipeline \citep{Mandigo-Stoba2022} based on \texttt{PypeIt} \citep{Prochaska2020}. An LT spectrum was obtained with SPRAT instrument \citep{Piascik2014} and reduced using the automated LT pipeline \citep{Barnsley2012}. All other spectra were obtained using the ALFOSC instrument on the NOT as part of the ZTF and NUTS2 (NOT Un-biased Transient Survey 2) collaborations. Reductions were performed using \texttt{Foscgui}\footnote{Foscgui is a graphic user interface aimed at extracting SN spectroscopy and photometry obtained with FOSC-like instruments. It was developed by E. Cappellaro. A package description can be found at \url{https://sngroup.oapd.inaf.it/foscgui.html}}. The spectra were scaled with the available $gri$ photometry. The spectral series, scaled with the photometry and dereddened for Milky Way (MW) extinction, are presented in Fig. \ref{fig:spectra} and a spectroscopic log is provided in Table \ref{tab:spec_log}.

\subsection{Imaging polarimetry} \label{subsec:gbi}

We were also able to obtain three epochs of imaging polarimetry (2022 Jun 18, 2022 Jul 07, 2022 Jul 18) with ALFOSC at the NOT in the $V$ and $R$ filters. All observations were obtained at four half-wave retarder plate (HWP) angles (0$^\circ$, 22.5$^\circ$, 45$^\circ$, 67.5$^\circ$). The data were reduced with a custom pipeline that uses \texttt{photutils} \citep{Bradley2024} for the photometry. The optimal aperture size was chosen to be 2 times the full-width at half-maximum (FWHM) in order to enclose the majority of the flux in the aperture and avoid inducing spurious polarisation due to the different Point Spread Function (PSF) elongation of the sources in the ordinary and extraordinary beams respectively, that is a known effect in the imaging polarimetry mode of ALFOSC (\citealt{Leloudas2017} and discussions therein). The third epoch was of insufficient signal-to-noise ratio (S/N) to provide a useable measurement and thus was not included in the analysis as observations with a S/N $\lesssim120$ are unreliable \citep{Pursiainen2023}. A log of the polarimetric observations is provided in Table \ref{tab:pol_log}.

\section{Analysis} \label{sec:analysis}

\subsection{Host galaxy} \label{subsec:host}

\begin{figure}
  \centering
  \includegraphics[width=0.5 \textwidth]{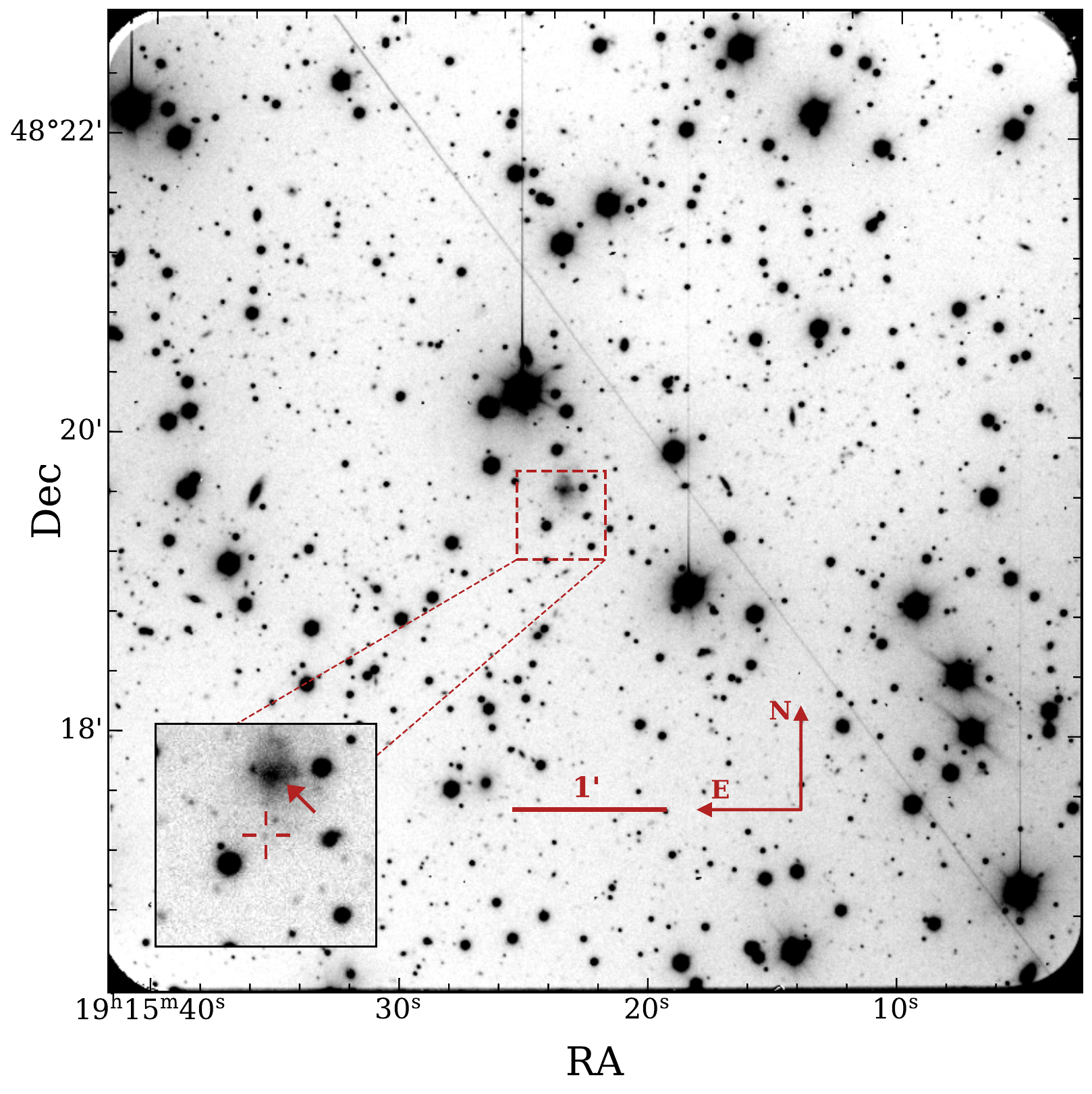}
  \caption{Deep (1 hr) $r$-band image taken with NOT+ALFOSC on 2023 Apr 16
($+313.1$\,d; Table \ref{tab:phot}). A faint source is visible at the location of SN~2022lxg. Interestingly, a known galaxy (WISEA J191523.71+481938.5), 9\farcs78\,NW of the SN is now apparent. At our adopted luminosity distance of the SN, this separation corresponds to a projected distance of $\sim$4.6\,kpc. The inset zooms in on the position of the SN (indicated by a red cross), and the host galaxy (indicated by a red arrow).}
  \label{fig:position_deep}
\end{figure}

There is no obvious bright host galaxy near SN~2022lxg (Fig. \ref{fig:position}). A faint and diffuse source is located at 9\farcs78\,North-West (NW) from the SN. In the NASA/IPAC Extragalactic Database (NED\footnote{\url{https://ned.ipac.caltech.edu/}}), this object is WISEA J191523.71+481938.5. The late, deep $r$-band image taken with the NOT when the SN has faded away is shown in Fig. \ref{fig:position_deep}, where this galaxy becomes more evident. We retrieved archival photometry of this galaxy in stacked Kron magnitudes from the Panoramic Survey Telescope and Rapid Response System (PanSTARRS) catalogue \citep{Huber2015} in the $g,\,r,\,i,\,z$ filters. The magnitudes are: $m_{g}=20.09\pm0.08$\,mag, $m_{r}=19.70\pm0.01$\,mag, $m_{i}=19.50\pm0.04$\,mag, and $m_{z}=19.17\pm0.12$\,mag. Unfortunately, there is no photometric or spectroscopic redshift available for this galaxy. At the luminosity distance of the SN, this separation corresponds to a projected distance of 4.58 kpc. In the absence of other candidate hosts, we assume that this galaxy is the host galaxy of SN~2022lxg.

\subsection{Photometric analysis} \label{subsec:phot_analysis}

\subsubsection{Broadband light curve evolution} \label{subsubsec:bb_lc}

In the following subsection, we present the features of the broadband light curves of SN~2022lxg, such as the explosion epoch estimate, the peak epochs and magnitudes in the various bands, the rise and decline timescales and the colour evolution.

The last non-detection (in the $i$-band) was at MJD 59\,730.45 while the first detection was at MJD~59\,731.39 in the $g$-band. Moreover, the first three detections (in ZTF $g$, $r$ and ATLAS $c$ bands) are within $\sim$ one day from the last non-detection (at 59\,731.44 in $r$-band and at 59\,731.50 in $c$-band). However, due to the very fast early rise and the very blue colours of SNe at epochs so close to the explosion, the $i$-band last non-detection is not particularly constraining. In order to determine the explosion epoch, we applied the following steps: The first estimates from the blackbody fits some days after these very early detections ($>2$\,d), return temperatures $\sim 20\,000$\,K (see Sect. \ref{subsubsec:Bol}). However, right after the explosion ($\lesssim1$\,d), SNe temperatures can decline very rapidly from several 10\,000\,K \citep{Yaron2017}. Hence we assume a temperature of 40000\,K for these early epochs and using the flux densities in the various bands, we obtained an estimate of the radius (3.9$\times10^{13}$\,cm, 7.0$\times10^{13}$\,cm, 9.9$\times10^{13}$\,cm respectively). Applying the Stefan-Boltzmann law yields the blackbody luminosities of these points and we propagated the uncertainties of the luminosities in the standard way. Given the inherent assumptions and systematic uncertainties in this process (especially since we do not have UV photometry to tightly constrain the early temperatures), we allowed for generous errors on the temperature ($\sigma_{\rm T}=25\,000$\,K) and radius ($\sigma_{\rm R}=10^{14}$\,cm). 
We fit a linear model to the obtained luminosities using a custom Monte Carlo routine with 10\,000 iterations (and assuming uniform distributions for the priors), thereby retrieving a posterior distribution on when the fits cross the zero luminosity level (i.e. explosion epoch); we use the median of this distribution as the explosion epoch estimate, and the 16th and 84th percentiles as the explosion epoch uncertainties. The median fit is within 1$\sigma$ from all the next, rising points (i.e., those not included in the fit), hence a linear model suffices for the purpose of estimating the explosion epoch. We find MJD$_{\rm expl}=59731.37^{+0.02}_{-0.07}$ and adopt this as our reference epoch throughout the manuscript. Therefore, the first $g$-band detection was made within $\sim$ 1 hour from the explosion.

In order to determine the epochs of maximum light in each of the optical bands, we performed numerical interpolation of the light curves using a Gaussian process regression algorithm \citep{Rasmussen2004}. We used the \texttt{Python} package \texttt{GPy}\footnote{\url{https://gpy.readthedocs.io/en/deploy/}}, employing a radial basis function (RBF) kernel. The uncertainty in the peak epoch was estimated as the time range when the GP light curve is brighter than the 1$\sigma$ lower bound on the peak brightness. For the peak epochs of the three different bands ($gri$), we find MJD = 59\,739.14$^{+2.23}_{-2.23}$, 59\,745.17$^{+1.47}_{-1.47}$ and 59\,746.26$^{+2.81}_{-2.89}$, respectively. We note a lag between the peak epochs with the bluer bands peaking earlier; 6.0\,d between $g$ and $r$, and 1.1\,d between $r$ and $i$. The peak epochs are denoted as short vertical dashes in the inset of Fig. \ref{fig:photometry} where we show the shape of the light curves around this time and the GP interpolations that were employed to measure them. In terms of peak absolute magnitudes, we obtain $\rm M_{g}=-19.41$ $\pm$ 0.01\,mag, $M_{\rm r}=-19.31$ $\pm$ 0.02\,mag and $\rm M_{i}=-19.09$ $\pm$ 0.02\,mag respectively. This brightness makes SN~2022lxg a LSN, in between the SNe II and the superluminous SNe II (SLSNe-II). The photometric and spectroscopic properties of SN~2022lxg show remarkable similarities with those of the sample of LSNe Type II studied by \pjpnospc. These properties will be highlighted further as we present and study the properties of SN~2022lxg.

\begin{figure*}
  \includegraphics[width=0.5 \textwidth]{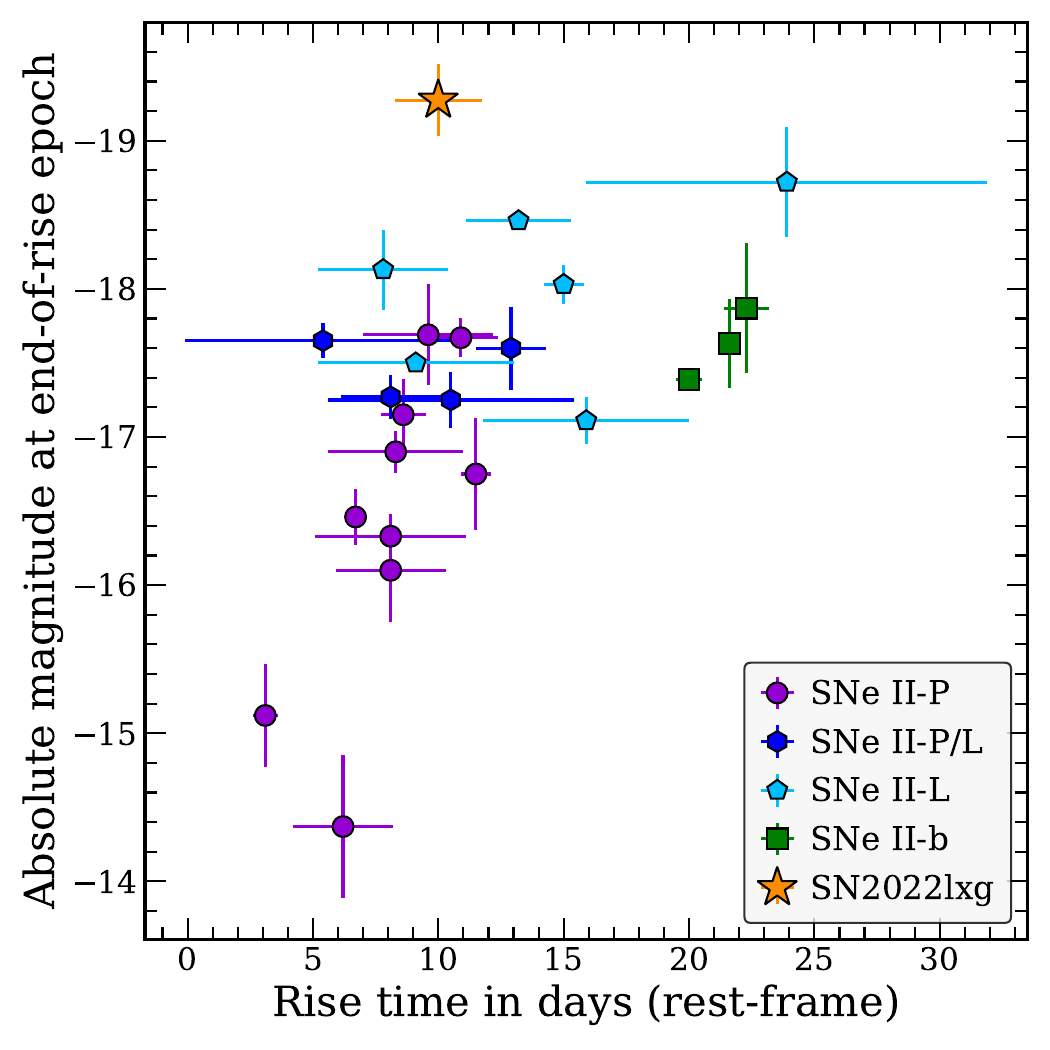}
  \includegraphics[width=0.5 \textwidth]{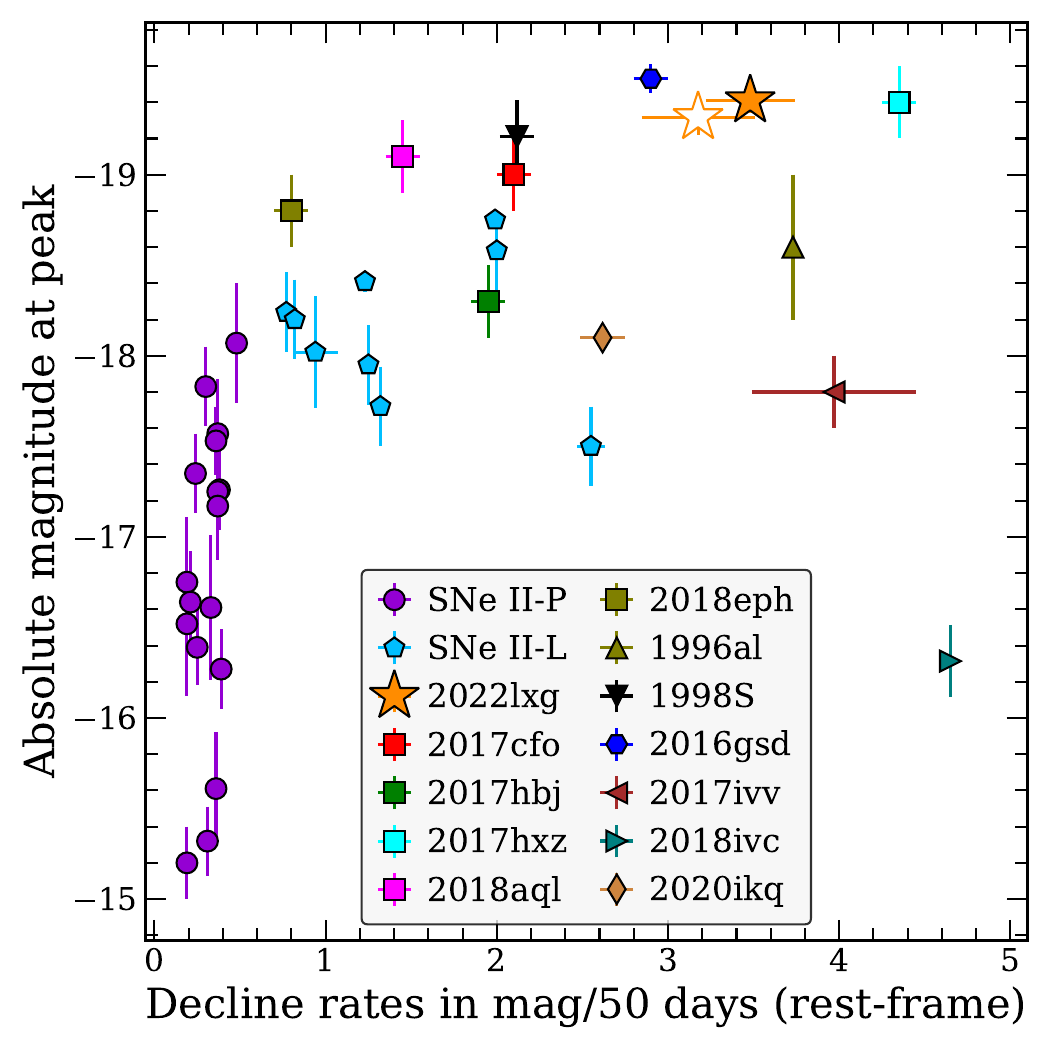}
  \caption{light curve rise and decline timescales against absolute magnitude. \textbf{Left}: Rise time against the \citet{Gall2015} sample measured from the $r$-band. The \pjp sample does not have explosion constraints, and is therefore not included here. \textbf{Right}: Decline rate of SN~2022lxg (filled marker in $g$-band, open marker in $V$-band), compared to other luminous Type II SNe from the \pjp sample ($V$-band), the Type IIP/IIL sample of \citet{Faran2014} ($V$-band), and several unusual transitional and/or interacting Type II SNe; SN~1996al \citep{Benetti2016} ($B$-band), SN~1998S \citep{Leonard1999,Fassia2000} ($V$-band), SN~2016gsd \citep{Reynolds2020} ($B$-band), SN~2017ivv \citep{Gutierrez2020} ($V$-band), SN~2018ivc \citep{Bostroem2020,Maeda2023a,Reguitti2024} ($B$-band), and the fast and faint Type IIb SN SN~2020ikq \citep{Ho2023} ($g$-band).}
  \label{fig:rise_times}
\end{figure*}

The rest-frame rise time from explosion to peak in the $g$-band is $7.6\pm2.2$ days (and 13.5 and 14.6 days in the $r$ and $i$ bands respectively), consistent with the median of Type II SNe ($7.5\pm0.3$\,d; \citealt{Gonzalez-Gaitan2015}). During this time frame, it rose by 5.4 mags with a rate of 0.56\,mag\,$\rm day^{-1}$. However, the rate considerably slows down close to the peak, and especially in the $g$-band, there is a plateau of $\sim$ 8 days at $\sim$ $-19.38$ mags that we mark with a horizontal dashed line in the inset of Fig. \ref{fig:photometry} (a linear fit to the plateau). That plateau is not seen in the $r$ and $i$ bands, the former however shows a \say{bump} at those epochs, while the latter shows a smoother evolution around peak (see also discussion in Sect. \ref{subsubsec:IIb_SNe}). Hence, the fast rise can be better appreciated by measuring the same quantities from the first to the second $g$-band detection, where within $\sim$ 1.9 days the $g$-band rose by 4.9 mags, a rate of 2.6\,mag\,$\rm day^{-1}$. 
In order to account for this change in the slope of the rise and fairly compare to other SNe with different light curve morphologies and tight explosion constraints, we follow the approach of \citet{Gall2015} where they define an epoch termed \say{end-of-rise}, as the epoch at which the $r$-band magnitude rises by less than 0.01\,mag\,$\rm d^{-1}$. This is estimated by fitting a low-order polynomial to the data, with an iteratively chosen step-size in time. For the $r$- band, we measure the \say{end-of-rise} at MJD 59741.59 $\pm$ 1.73 (i.e. 10.00 d post explosion in rest-frame) with an absolute magnitude of $-19.27$ $\pm$ 0.10. In the left panel of Fig. \ref{fig:rise_times}, we plot the measured \say{end-of-rise} time of SN~2022lxg in the $r$-band versus the respective absolute magnitude at this epoch, and we compare those to the sample of \citet{Gall2015}, a compilation of 23 Type II SNe of various subtypes. SN~2022lxg is the brightest of them with an intermediate \say{end-of-rise} time.

By 50\,d from peak, SN~2022lxg has declined by $3.48\pm0.26$\,mag\,$\rm (50\,d)^{-1}$ in the $g$-band; it continued to decline at this rate until $\sim+100$\,d (i.e. $\sim$ 6.96\,mag\,$\rm (100\,d)^{-1}$), when it became too faint for further observations. In order to compare with $V$-band literature measurements, we employed the following process: We obtained synthetic photometry in the $g$ (ZTF) and $V$ (Bessel) filters (using the filter curves hosted at the SVO Filter Profile Service (\citealt{Rodrigo2012,Rodrigo2020,Rodrigo2024}) from our dense spectral series. In order to retrieve the $V$-band magnitudes at the $g$-band light curve epochs, we linearly interpolated the derived $g-V$ synthetic colour curve. In that way we created a \say{transformed} $V$-band light curve and measured the decline rate to be $3.18\pm0.33$\,mag\,$\rm (50\,d)^{-1}$.
The decline is significantly faster than the fastest declining ($\sim2.5$\,mag\,$\rm (50\,d)^{-1}$.) Type IIL SNe reported in \citet{Faran2014}. Interestingly, the decline rate of SN~2022lxg is reminiscent of Type IIb SNe ($5-9$\,mag\,$\rm (100\,d)^{-1}$; e.g. \citealt{Gutierrez2020} and references therein). Up to an epoch of $\sim$100\,d, the pseudo-bolometric luminosity of SN~2022lxg did not settle onto the expected decline rate for $^{56}$Co decay of 0.98\,mag\,$\rm (100\, day)^{-1}$ (assuming complete gamma-ray trapping; \citealt{Woosley1989}). Our attempt to obtain a late-time constraint at $+261.3$\,d and subsequently at $+313.1$\,d, resulted in a non-detection and an upper-limit of 22.96\,mag, and in a detection of $24.00 \pm 0.12$\,mag ($M_{\rm r}=-10.91$\,mag) respectively (Fig. \ref{fig:photometry}). We searched for archival images that could be used as templates for difference imaging. However, the deepest image available is from the PanSTARRS1 survey \citep{Kaiser2002}; we measure a limiting magnitude of $r\sim+22.8$\ at the SN location. As this is significantly shallower than our deepest science image, we do not perform template subtraction. The late-time detection implies that the initial fast decline rate reported above must have slowed down, but it is difficult to firmly attribute the cause of this based on a single data point.

\begin{figure}
  \centering
  \includegraphics[width=0.5 \textwidth]{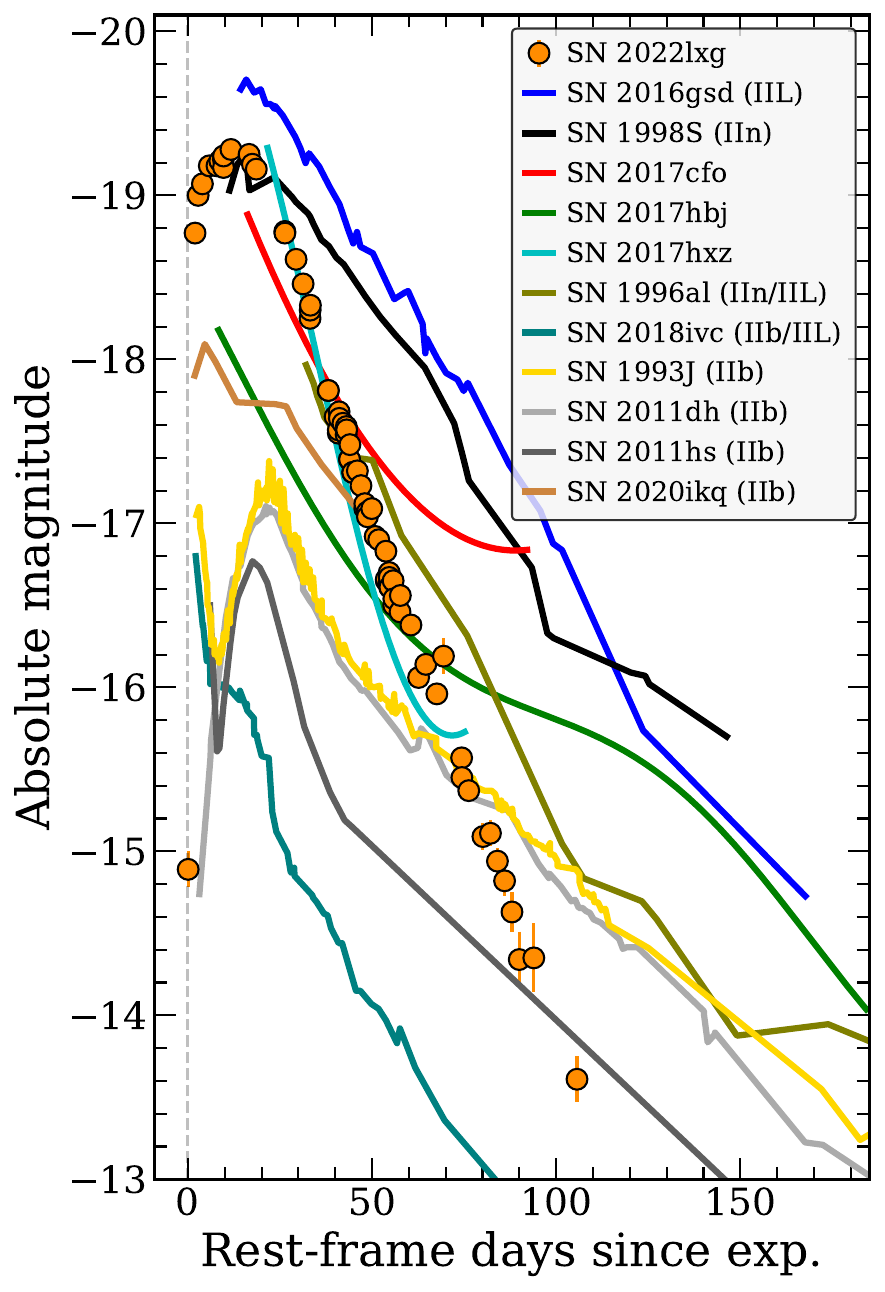}
  \caption{Comparison of the absolute magnitude $r$-band light curve of SN~2022lxg to three LSNe from the \pjp sample in the $V$-band (SN~2017cfo, SN~2017hbj, SN~2017hxz) and to other fast declining transients of various subtypes: SN~1998S, SN~2016gsd, SN~2018ivc in $V$-band, and SN~1996al, SN~2020ikq and two more Type IIb SNe (SN~2011dh; \citealt{Arcavi2011,Bersten2012}, SN~2011hs; \citealt{Bufano2014}) in the $r/R$ band. Visual extinctions (for both MW and host galaxies) and distance moduli are retrieved from the referenced works (see also Fig. \ref{fig:rise_times} for the references). There is a remarkable similarity in the decline rate (and luminosity) with SN~2017hxz of the \pjp sample.}
  \label{fig:mag_LII_comp}
\end{figure}

In the right panel of Fig. \ref{fig:rise_times}, we show the magnitude decline rate per 50 days versus the peak absolute magnitude. We compare with two samples: the Type IIP/IIL sample of \citet{Faran2014} and the one of \pjp (both measure the decline rates from the $V$-band). The latter are all characterised as fast declining. Here, similar to \pjpnospc, we consider as fast, those SNe with decline rates $\gtrsim$1.4\,mag\,$\rm (100\,d)^{-1}$, an arbitrary limit that has previously been used to separate slow and fast declining SNeII (\citealt{Davis2019} and references therein). We also include several unusual transitional and/or interacting Type II SNe. The only SN from the \pjp sample that declines faster than SN~2022lxg is SN~2017hxz. We show the morphology of the $r/R$-band light curves (and few in the $V-$band) in Fig. \ref{fig:mag_LII_comp} compared to brighter, slower declining Type II SNe, and fainter, rapidly declining ones. Although there is significant heterogeneity in the comparison objects, as we discuss below, SN~2022lxg has photometric and spectroscopic properties in common with both fainter and brighter Type II SNe.

\begin{figure}
\centering
\includegraphics[width=0.5 \textwidth]{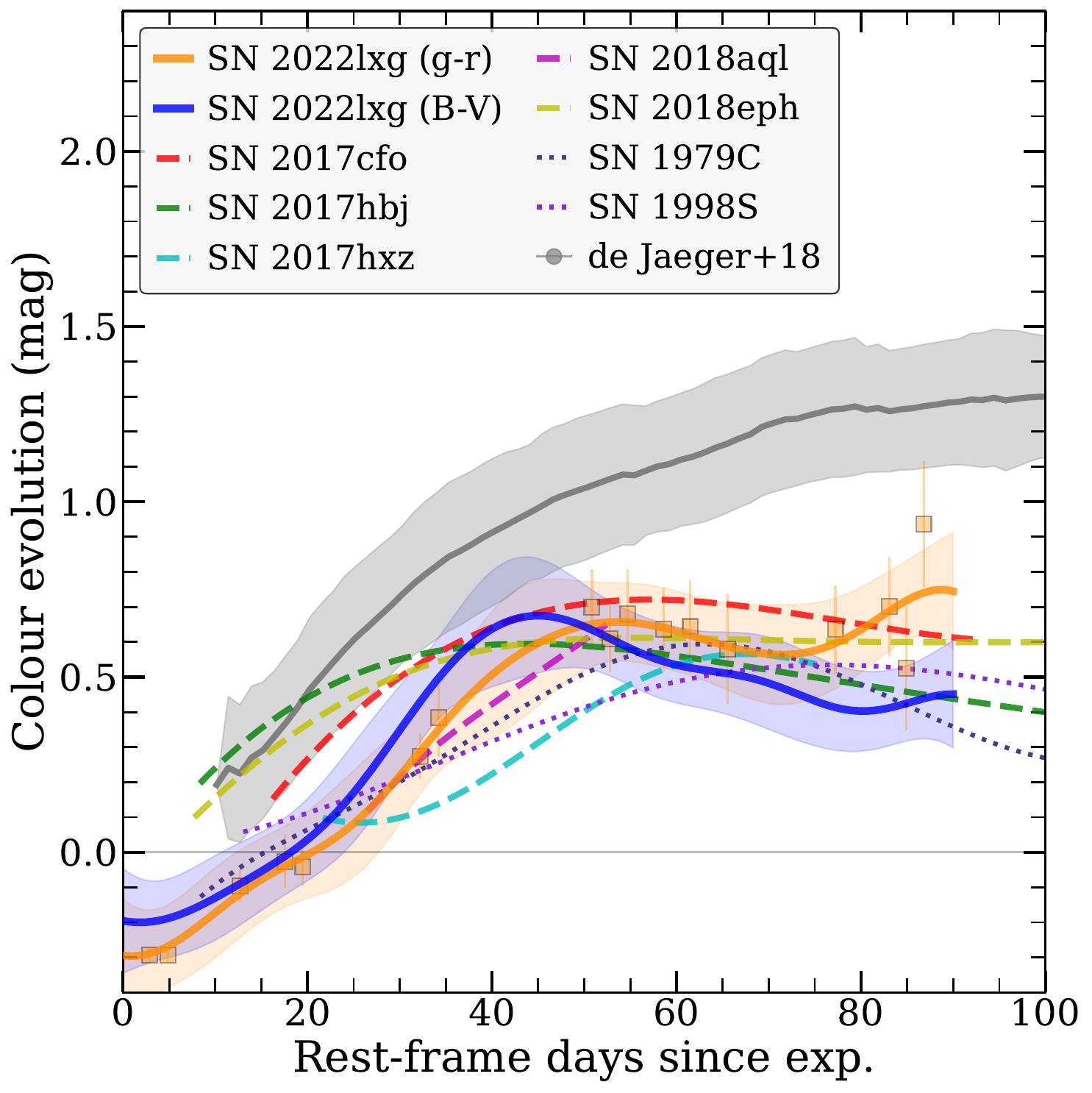}
\caption{Gaussian process interpolations of the colours ($g-r$ in orange and $B-V$ in blue) of SN~2022lxg compared to those of other luminous Type II SNe from the \pjp sample. We overplot the actual measured $g-r$ colours of SN~2022lxg with faint squares and the 1$\sigma$ uncertainties of the interpolations as shaded lines. Mean values of $B-V$ colours of the sample of SNe II studied by \citet{DeJaeger2018} are presented in grey for comparison (with the 1$\sigma$ standard deviation plotted as a shaded region).}
\label{fig:colors}
\end{figure}

Regarding the colour evolution, SN~2022lxg has rather blue colours ($g-r$ $\lesssim -0.2$\,mag) for the first $\sim$ 10 days after the explosion, then until +40 days, the colours redden rapidly and reach a maximum value of $g-r \sim$ 0.6\,mag. Then we see the cooling stop abruptly and the colour evolution plateaus, and even colours become slowly bluer until $+65$\,d, followed by a more gradual cooling until the SN becomes too faint to observe. 
An almost identical colour evolution is seen in the \pjp sample. We showcase this in Fig. \ref{fig:colors} by reproducing their Figure 9 and including SN~2022lxg in the comparison as well as a few well-studied luminous SNe with similar colours, e.g., SNe~1998S and 1979C \citep{Branch1981,DeVaucouleurs1981}. In order to compare with $B-V$ colours from the literature, we transformed these measurements to $g-r$ using a procedure analogous to that described above for transforming to the $V$-band. We plot the Gaussian process interpolations of the $g-r$ and $B-V$ light curves (performed using the \texttt{Python} package \texttt{GPy}) and compare with the sample of \pjpnospc, and also with the $B-V$ colours of the sample of SNe II studied by \citet{DeJaeger2018}. It is clear that SN~2022lxg follows the same pattern as the LSNe II of the \pjp sample, and differs from the gradual cooling shown by the majority of Type II SNe. In Table \ref{tab:basic_props}, we tabulate various photometric properties of SN~2022lxg that were presented in this section.

\begin{table}
    \def\arraystretch{1.1}%
    \setlength\tabcolsep{3pt}
    \centering
    \begin{tabular}{ll}
    \hline
        Property & Value \\
        \hline
         $z$ &  0.0214 $\pm$  0.0006\\
         $\rm E(B-V)_{MW}$ & 0.059\,mag\\
         MJD of last non detection ($i$-band) & 59730.45\\
         MJD of first detection ($g$-band) & 59731.39\\
         MJD of explosion & 59731.37$^{+0.04}_{-0.06}$ \\
         Peak $g$ MJD (phase)& 59739.14$^{+2.23}_{-2.23}$ ($+7.6$\,d) \\
         Peak $g$ magnitude & $-19.41$ $\pm$ 0.01\,mag \\
         Peak $r$ MJD (phase)& 59745.17$^{+1.47}_{-1.47}$ ($+13.5$\,d)\\
         Peak $r$ magnitude & $-19.31$ $\pm$ 0.02\,mag \\
         Peak $i$ MJD (phase)& 59746.26$^{+2.81}_{-2.89}$ ($+14.6$\,d)\\
         Peak $i$ magnitude & $-19.09$ $\pm$ 0.02\,mag \\
         End-of-rise $r$ MJD & 59741.59 $\pm$ 1.73 \\
         End-of-rise $r$ magnitude  & $-19.27$ $\pm$ 0.10\,mag \\
         Decline ($g$-band) & 3.48 $\pm$ 0.26\,mag\,$\rm (50\,d)^{-1}$ \\
         Half-flux duration ($g$-band) & 25.70 $\pm$ 0.26 days\\
         $\rm M(^{56}Ni)$ (from peak) & $<$0.729 $\pm$ 0.163\,$\msun$ \\
         $\rm M(^{56}Ni)$ (from tail) & $\geq$0.003 $\pm$ 0.002\,$\msun$ \\
         
         \hline
    \end{tabular}
    \caption{Basic properties of SN~2022lxg. MJDs are in observer frame, durations are in rest-frame. Unless indicated otherwise, we use the $g$-band as the reference. The peak epochs and magnitudes are inferred as described in Sect. \ref{subsubsec:bb_lc}. We derive the $^{56}$Ni mass estimates from the luminosities calculated with two different methods, either from the blackbody fits and the Stefan-Boltzmann law, or from the pseudo-bolometric luminosity estimates (see Sect. \ref{subsubsec:Bol} for details). We provide the median values of the different tail ($313.1$\,d; $0.001-0.009\,\msun$) and peak estimates ($0.247-1.210\,\msun$).}
    \label{tab:basic_props}
\end{table}

\subsubsection{Bolometric light curve} \label{subsubsec:Bol}

\begin{figure}
\centering
\includegraphics[width=0.5 \textwidth]{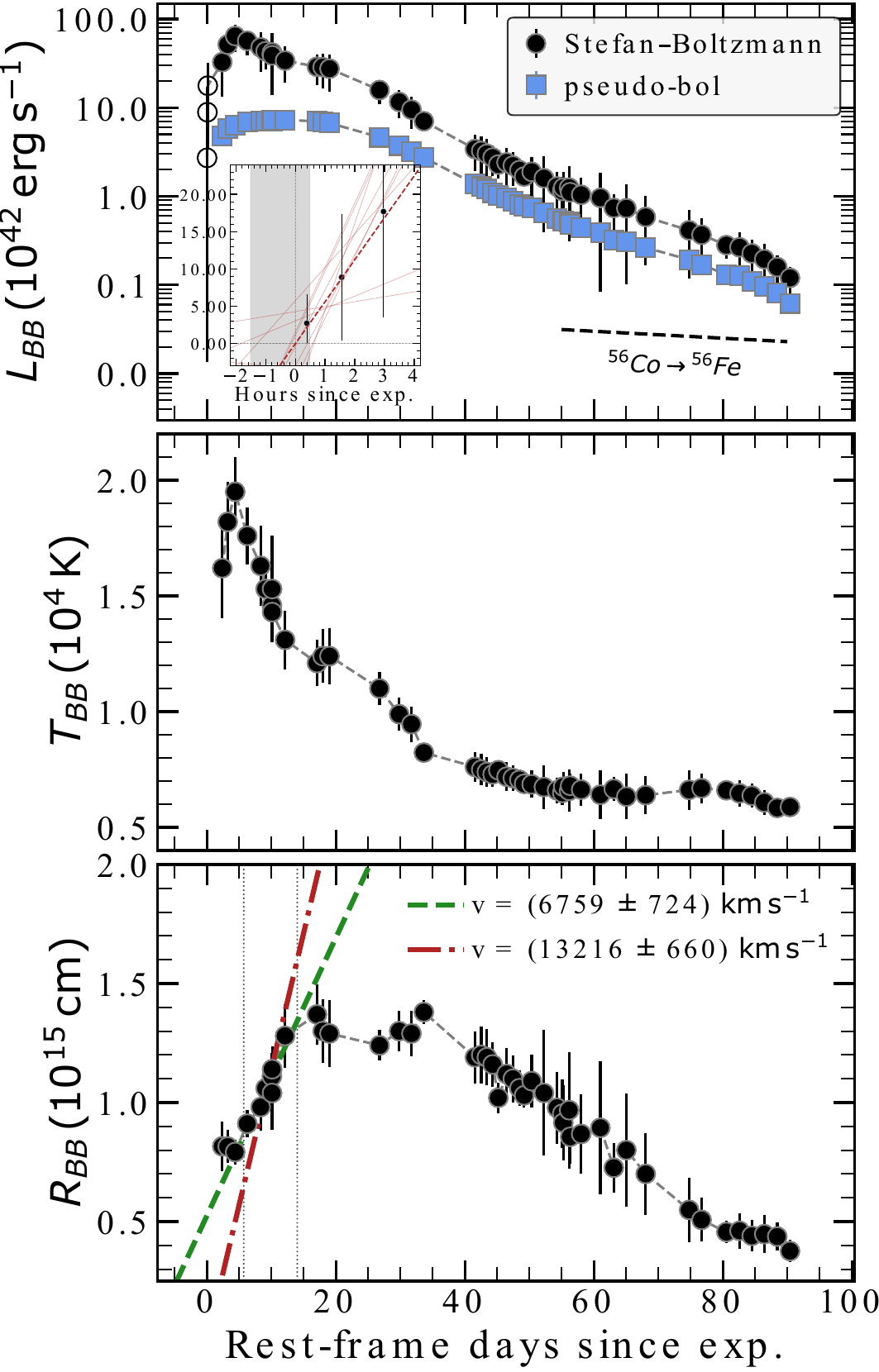}
\caption{Pseudo-bolometric $gcroiz$ light curve of SN~2022lxg (top panel) and blackbody temperature and radius evolution (middle and bottom panels) derived from blackbody fits to the SEDs. The bolometric luminosity derived with the Stefan-Boltzmann law is also shown in the top panel with the open markers showing the luminosity estimates of the early first three detections (within 3 hours from the explosion), converted from temperature to luminosity using the Stefan-Boltzmann law. The inset zooms-in on those points where we plot with red 10 random samples from the posterior distribution of linear Monte Carlo fits. The grey vertical line denotes the explosion epoch estimate (median of the posterior distribution) while the grey shaded region denotes the uncertainty on the estimate (16th and 84th percentiles). In the bottom panel, we show linear fits to the rising part of the expanding photosphere (the exact fitted region is within the vertical dotted lines), the red dash-dotted fit has $\rm R(t=0)$ fixed at zero, while the green dashed one does not.}
\label{fig:BB}
\end{figure}

We constructed a pseudo-bolometric light curve of SN~2022lxg using \texttt{SUPERBOL} \citep{Nicholl2018a}. The flux in the available bandpasses was estimated at the epochs of our $r$-band photometry. We interpolate the light curves using polynomials of third to fifth order and integrate under the spectral energy distribution (SED) of each epoch to get the luminosity. We fit a blackbody function to the SED at each reference epoch in order to estimate the temperature and the radius, and also to calculate the missing energy outside of the observed wavelength range. The pseudo-bolometric light curve, as well as the blackbody temperature, radius and luminosity evolution are shown in Fig. \ref{fig:BB}. We note that the results should be interpreted with caution as we have photometric coverage for the full evolution of the SN in only five optical bands ($gcroi$) and only 4 points between $+40$ to $+60$\,d in the $z$-band; thus, the temperature estimates at early times are almost certainly underestimated. In the inset of the top panel of Fig. \ref{fig:BB} we visualise how we estimated the explosion epoch (Sect. \ref{subsubsec:bb_lc}).

The bolometric light curve analysis results in a blackbody temperature that peaks at $\sim 20\,000$\,K at 5 days post explosion. Then the temperature drops fast for the next 10 days, followed by a $\sim10$ day break in the cooling and then the cooling rate becomes fast again until $+35$\,d. From Between roughly $+35$ to $+75$\,d the cooling rate significantly drops and even plateaus, only to drop again until the SN becomes too faint to observe. The blackbody photosphere expands linearly until $\sim$ $+20$\,d, and a linear fit to this expansion results in an photospheric velocity of $v_{\rm ph} = 6759 \pm 724$\,km\,s$^{-1}$. Fixing $\rm R(t=0)$ at zero, returns a higher velocity of $v_{\rm ph} = 13216 \pm 660$\,km\,s$^{-1}$ however the fit is not as good. Between $+15$ to $+35$ days, the radius almost plateaus, that is it slightly drops during the cooling rate break, and then peaks again at $\rm\rbb\sim1.4\times10^{15}$\,cm, at the end of the second fast cooling phase. After that, the photosphere contracts until the SN becomes too faint to observe. The pseudo-bolometric luminosity slowly peaks at $\sim7.2\times10^{42}\rm\,erg\,s^{-1}$ at 12 days post explosion and then smoothly declines. The luminosity derived from the Stefan–Boltzmann (SB) law follows a similar evolution, with a sharper peak of $\sim6.4\times10^{43}\rm\,erg\,s^{-1}$ at 5 days post explosion.

Using the bolometric luminosity, we make some $^{56}$Ni mass estimates produced in SN~2022lxg. For H-rich SNe, it is very difficult to estimate how much of the power comes from nickel during the peak times when the hydrogen recombines, leading to inaccurate estimates. This also applies to SN~2022lxg; by plugging the above peak estimates into the Arnett rule \citep{Arnett1982}, we measure 0.25\,$\msun$ and 1.21\,$\msun$ for the $^{56}$Ni mass, using the pseudo-bolometric and the SB estimate respectively. These values should be seen as rough upper-limits. Based on the colours of the SN before it faded, we construct a pseudo-SED for our late $r$-band detection at $+313.1$\,d. From that, we measure a pseudo-bolometric luminosity of $(1.7\pm 0.7) \times10^{39} \rm\,erg\,s^{-1}$, and from a blackbody fit and the Stefan–Boltzmann law, we get a luminosity of $(6.2\pm 2.7) \times10^{39}\rm\,erg\,s^{-1}$. We use these to estimate the $^{56}$Ni mass from the tail of the luminosity using various prescriptions (Tail, \citealt{Hamuy2003}, SN~1987A ratio) and we always measure values $<0.009\msun$. Those values should be seen as a rough lower-limits since we do not have a good estimate of the bolometric luminosity at these epochs. The $^{56}$Ni masses estimated from the tail are very low, further highlighting that the estimates derived from the peak are not trustworthy and that the peak is not dominated by $^{56}$Ni-heating. We tabulate the median values of the above $^{56}$Ni mass estimates in Table \ref{tab:basic_props}, and all the individual values in Table \ref{tab:nickel}.

\subsubsection{Light curve fits} \label{subsubsec:mosfit}

\begin{figure}
  \centering
  \includegraphics[width=0.5 \textwidth]{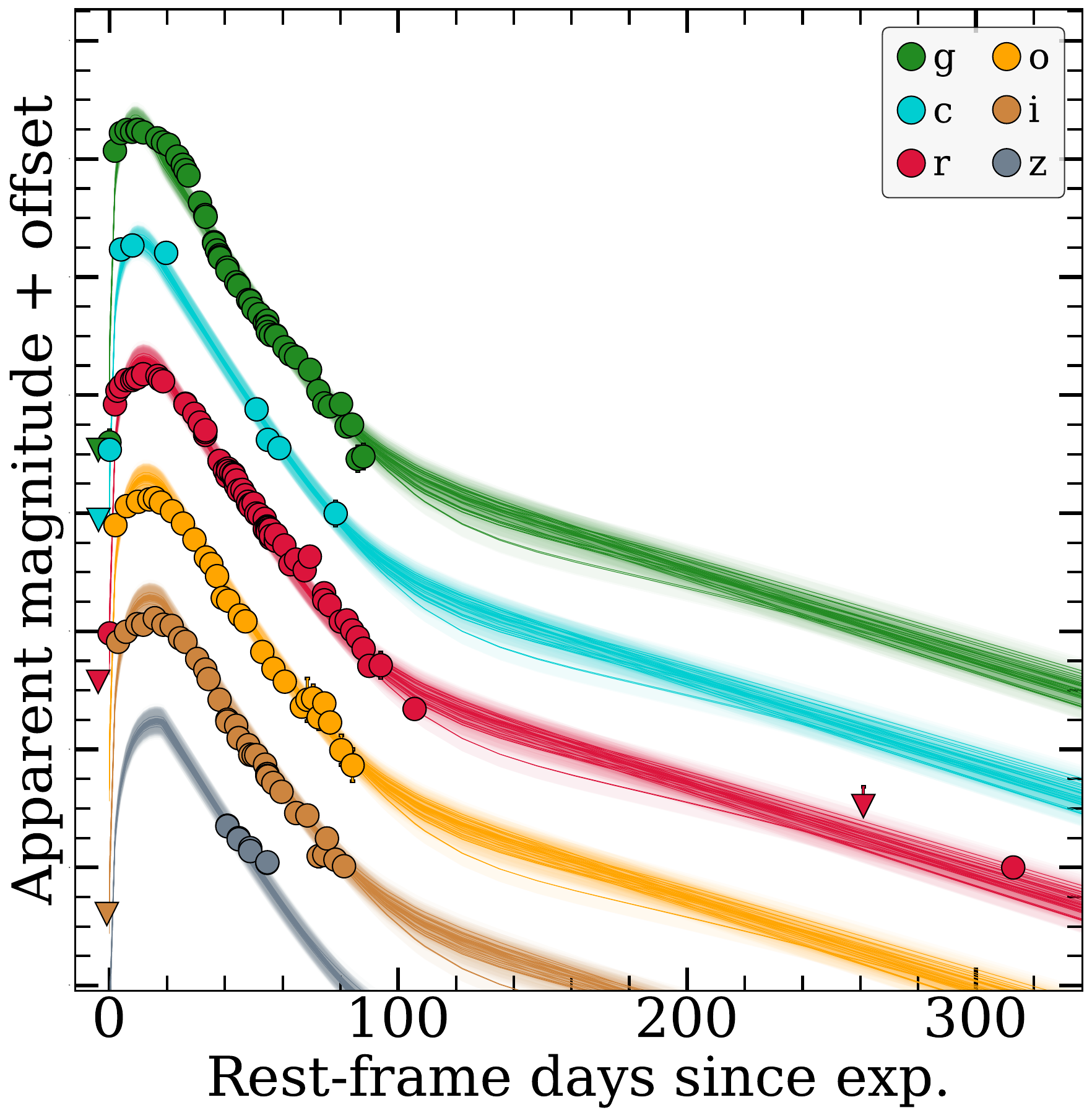}
  \caption{Fits to the multi-band light curve using the \texttt{csmni} model in \texttt{MOSFiT} \citep{Guillochon2018}. The relevant parameters are listed in Table \ref{tab:mosfit}.}
  \label{fig:mosfit}
\end{figure}

\begin{table}
  \centering
  \begin{tabular}{cccc}
  \hline
  Parameter & Prior & Posterior & Units\\
  \hline
$\log{(f_{\mathrm{Ni}})}$ & $[-4, 0]$ & $-1.87^{+0.20}_{-0.15} $  \\
$\log{(\kappa_{\rm \gamma})}$ & $[-2, 1]$ & $0.53^{+0.21}_{-0.18}$ & cm$^2$\,g$^{-1}$ \\
$v_{\rm ej}$ & $[1.7, 2.3]$ & $2.00^{+0.04}_{-0.05}$ & $10^{4}$\,km\,s$^{-1}$ \\
$\log{(M_{\rm CSM})}$ & $[-2, 1.5]$ & $-1.79^{+0.12}_{-0.12}$ & M$_\odot$ \\
$\log{(M_{\rm ej})}$ & $[-1, 1.0]$ & $0.01^{+0.13}_{-0.18}$ & M$_\odot$ \\
$\log{(n_{\rm H,\mathrm{host}})}$ & $[16,21]$ & $16.93^{+0.76}_{-0.59}$ & cm$^{-2}$ \\
$\log{(R_0)}$ & $[-1, 2.4]$ & $0.14^{+0.14}_{-0.20}$ & AU \\
$\log{(\rho_{0})}$ & $[-15, -7]$ & $-8.61^{+0.17}_{-0.16}$ & g\,cm$^{-3}$ \\
$\log{(T_{\rm min})}$ & $[3.7, 3.9]$ & $ 3.82^{+0.01}_{-0.01}$ & K  \\
$ t_{\rm exp}$ & $[-1, 0]$ & $-0.21^{+0.03}_{-0.03}$ & days  \\
$\log{(\sigma)} $ & $[-4, 2] $ & $ -0.65^{+0.02}_{-0.02}$  &   \\
$\kappa$ & 0.34 & - & cm$^2$\,g$^{-1}$ \\
$n$ & 12 & - & \\
$\delta$ & 0 & - & \\
$s$ & 2 & - & \\

  \hline
\end{tabular}
  \caption{Priors and marginalised posteriors for the \texttt{MOSFiT} \texttt{csmni} model. The posterior results are the median of each distribution, and the uncertainties are the 16th and 84th percentiles (which are purely statistical). A dash in the posterior value and a single prior value, denote that the parameter was fixed to that value.}
  \label{tab:mosfit}
\end{table}

We used the publicly available Modular Open Source Fitter for Transients \citep[\texttt{MOSFiT}\footnote{\url{https://mosfit.readthedocs.io/en/latest/index.html}};][]{Guillochon2018} to fit the multi-band light curves. \texttt{MOSFiT} takes as input the multi-band photometry and priors on the parameters of the model that is being fit to the data. We used the built-in model \texttt{csmni} that combines the luminosity from the decay of radioactive $^{56}$Ni and additional luminosity from CSM interaction, wherein a fraction of the kinetic energy of the SN ejecta is converted to radiative energy through collision with the CSM. The $^{56}$Ni decay model is from \citet{Nadyozhin1994}, while the CSM interaction model is based on the semi-analytic treatment of \citet{Chatzopoulos2013}. The model is set up such that the contribution from CSM interaction starts at time $t_{\rm int}=R_{\rm 0}/v_{\rm ej}$, where $R_{\rm 0}$ is the inner radius of the CSM shell and $v_{\rm ej}$ is the bulk velocity of SN ejecta. Assuming $v_{\rm ej}$ to be the average photospheric velocity of the SN ejecta, the kinetic energy $E_{\rm k}$ of the ejecta is inferred from the free parameters $M_{\rm ej}$ (the ejecta mass) and $v_{\rm ej}$, assuming a constant density \citep{Arnett1982}, using $E_{\rm k}\approx\frac{3}{10}M_{\rm ej}{v_{\rm ej}}^{2}$. The model has 11 free parameters, namely $^{56}$Ni mass fraction ($f_{\rm Ni} \equiv M_{\rm Ni}/M_{\rm ej}$), $\gamma$-ray opacity ($\kappa_\gamma$), bulk velocity of SN ejecta ($v_{\rm ej}$), mass of the CSM shell ($M_{\rm CSM}$), total ejecta mass ($M_{\rm ej}$), host galaxy hydrogen column density ($n_{\rm H,\mathrm{host}}$), inner radius of the CSM shell ($R_{\rm 0}$), CSM density at the initial radius $R_{\rm 0}$ ($\rho_{0}$), minimum temperature ($T_{\rm min}$) that the expanding and cooling photosphere settles down to, time of explosion relative to first epoch of observation ($t_{\rm exp}$) and a white-noise variance term ($\sigma$) representing the additional uncertainty (in mag) that would make the reduced $\chi^{2}=1$. A power-law density profile for the CSM shell is adopted with $\rho(r)=qr^{-s}$, where the scaling factor $q = \rho_{0}{R_{0}}^{s}$ \citep{Chatzopoulos2012}. The power-law index was fixed to s = 2 corresponding to a steady-wind CSM model \citep{Chevalier2011}. Furthermore, there are three more parameters that we fix; the Thomson scattering opacity ($\kappa$) at 0.34\,cm$^2$\,g$^{-1}$, a typical value for hydrogen-rich ejecta (close to the result of \citealt{Nagy2018}), and the density power-law parameters in the inner ($\rho_{ej} \propto r^{-\delta}$) and outer ($\rho_{ej} \propto r^{-n}$) ejecta, $\delta=0$ and $n=12$, respectively \citep[typical values in H-rich ejecta;][]{Chatzopoulos2013}.

We set simple uninformative uniform or log-uniform priors for each free parameter of the model. We set a well-constrained uniform explosion time prior ($t_{\rm exp}>-1$) since we have put tight constraints on the explosion time and we also use the last non-detections for the fits. We also set a uniform velocity prior around our estimate of 20\,000\,km\,s$^{-1}$ (from the minima of absorption lines; see Sect. \ref{subsec:spec_analysis}): between 17\,000 and 23,000\,km\,s$^{-1}$. Based on the lack of narrow \ion{Na}{I} D absorption lines in the spectra and the very faint (potential) host, we also set an upper limit for the host galaxy extinction, $A_{V,\mathrm{host}} \leq 0.5$~mag, converted from the column density of neutral hydrogen as $n_{H,\mathrm{host}} \leq 10^{21}$\,cm$^{-2}$ based on \citet{Guver2009}. We also have a good constraint on the $T_{\rm min}$ from the blackbody fits, and we set a prior between 5\,000 and 8\,000\,K. Finally, we assume a hydrogen-rich progenitor, but not necessarily an extended envelope such as that of a red supergiant (RSG). Thus the minimum inner radius of the CSM, $R_{\rm 0}$, is set at 0.1 AU ($\sim20\,R_\odot$), roughly half the radius of the blue supergiant progenitor of SN~1987A \citep{Podsiadlowski1992} but larger than a Wolf-Rayet progenitor of a stripped-envelope (SE) SN.

We ran \texttt{MOSFiT} using dynamic nested sampling with \texttt{DYNESTY}\footnote{\url{https://dynesty.readthedocs.io/en/latest/}} \citep{Speagle2020} in order to evaluate the posterior distributions of the model. We list the free parameters of the model, their priors and their posterior probability distributions in Table \ref{tab:mosfit}, and we present the model light curves in Fig. \ref{fig:mosfit}, with two-dimensional posteriors shown in Fig. \ref{fig:corner} of the Appendix. The fit has fully converged with well-constrained parameters and the logarithm of the Bayesian evidence $Z$ (which quantifies the quality of the fit) is equal to $\log{(Z)}=191$. The model is successful in reproducing the multi-band light curves. The only deviation is that the model light curves return a sharper peak (in all bands) than the smoother peaks of the data. Finally, the $r$-band model light curves successfully fit the late-time, deep $r$-band epoch ($+313.1$\,d). 

\begin{figure*}[t]
\includegraphics[width=0.60\textwidth]{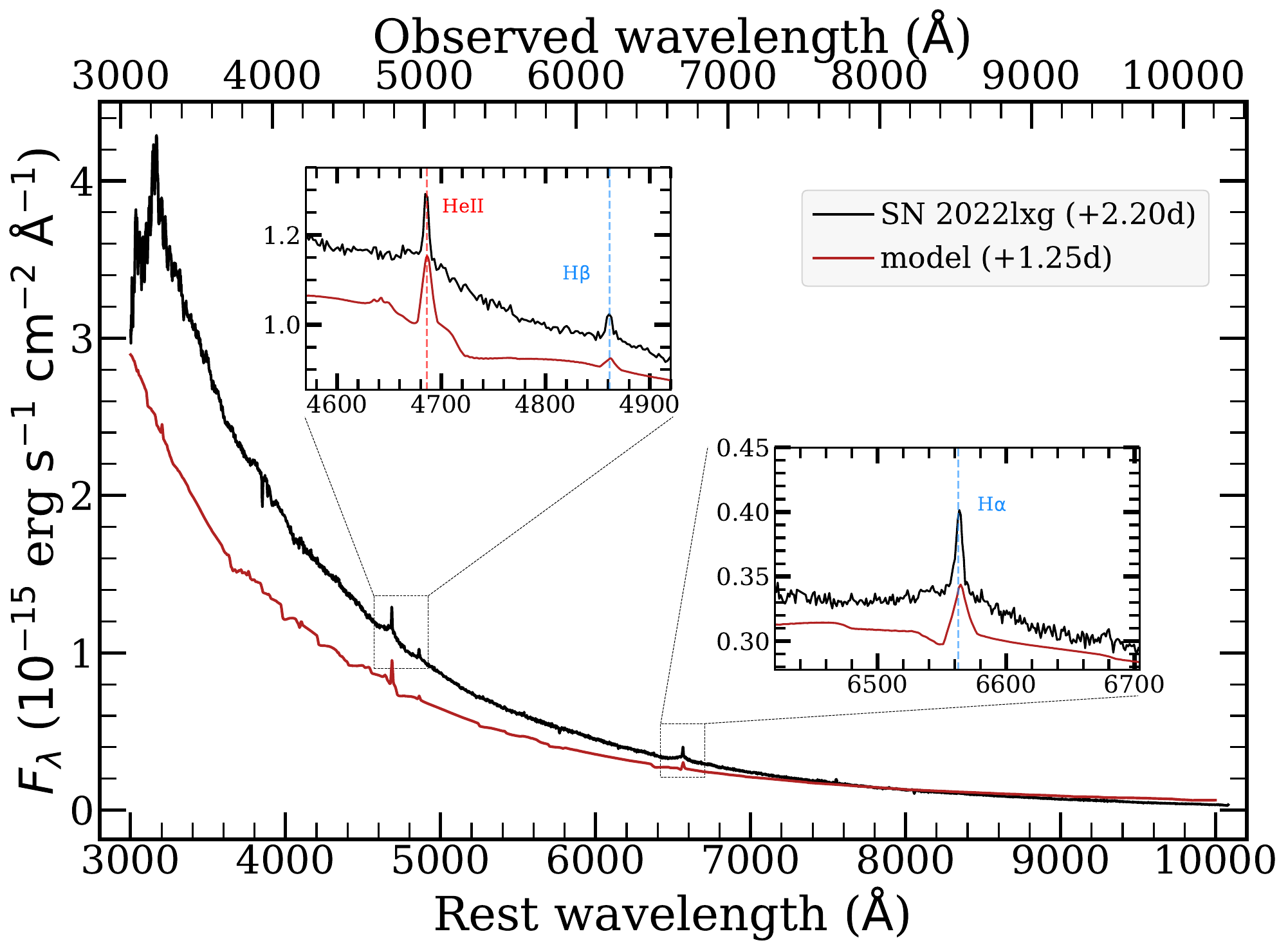}
\includegraphics[width=0.40 \textwidth]{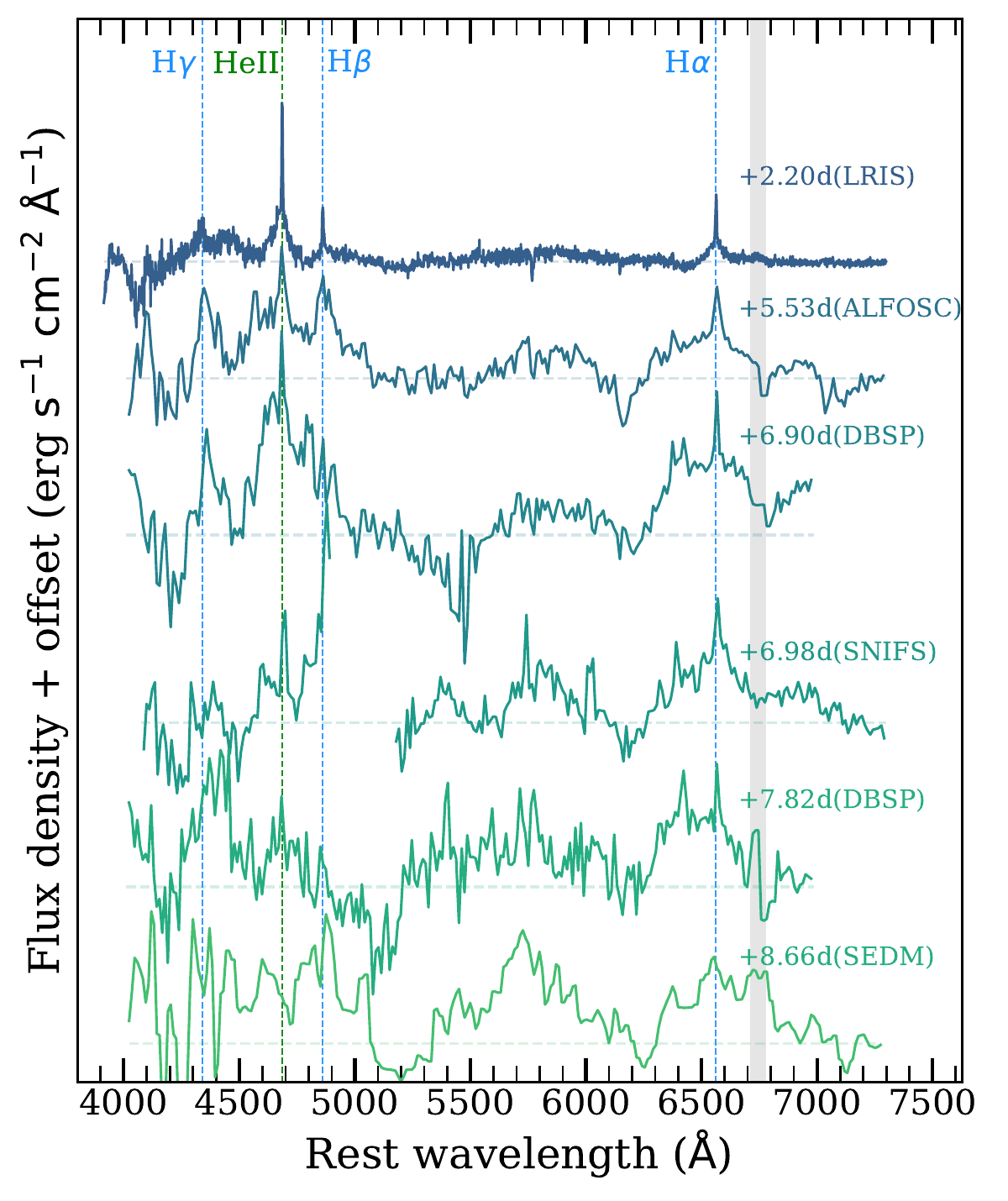}
\caption{Flash-ionisation features in SN~2022lxg. Left panel: Keck+LRIS spectrum, $+2.20$\,d after the explosion. We see flash-ionisation lines of H$\alpha$, H$\beta$, H$\gamma$ and \ion{He}{II} $\lambda4686$, on top of a blue continuum. The red spectrum below is a model from \citet{Dessart2023} (`\textit{mdot1em3early\_nb5}') with a steady pre-explosion wind of ${\rm 1 \times 10^{-3}\,\msun\, \rm yr^{-1}}$, that successfully reproduces the early continuum shape and the flash-ionisation lines. The insets focus on the flash-ionisation lines of H$\beta$ and \ion{He}{II} (left inset) and of H$\alpha$ (right inset), with the dashed vertical lines denoting the rest wavelengths of those lines. Right panel: Early, continuum-subtracted spectra of SN~2022lxg. Flash-ionisation lines are marked with dashed vertical lines. From left to right, we detect H$\gamma$, \ion{He}{II}, H$\beta$ and H$\alpha$. The spectra (apart from the LRIS one) are binned to 12 \AA\, for visual purposes. The horizontal dashed lines denote the zero flux level of each spectrum after the approximate continuum removal. The telluric features are marked with grey shaded vertical lines.}
\label{fig:Keck}
\end{figure*}

Some key explosion parameters are: $M_{\rm ej}=1.02^{+0.36}_{-0.35}\,M_\odot$ a very low fraction of which ($\sim1\%$) is $^{56}$Ni (${M_{\rm Ni}=0.013^{+0.009}_{-0.006}\,M_\odot}$; within 1$\sigma$ from our SB tail estimates), and $v_{\rm ej}=2.00^{+0.04}_{-0.05}\times10^{4}$\,km\,s$^{-1}$, which combined with the ejecta mass leads to $E_{\rm k}=2.44^{+0.92}_{-0.84}\times10^{51}$\,erg. The very low $^{56}$Ni mass and the high ejecta velocity are fully consistent with the results derived from other observables. However, the ejecta velocity of the model (that fully agrees with what we derive from the spectroscopic lines; see Sect. \ref{subsec:spec_analysis}) is much higher compared to the rather low photospheric expansion velocity derived from the blackbody fits ($\sim 7\,000$\,km\,s$^{-1}$). This discrepancy is further discussed in Sect. \ref{subsec:prog}. The very low ejecta mass could be somewhat under-estimated; however, a broadly low ejecta mass is consistent with the fast nature of SN~2022lxg and with the lack of typical metal lines in the spectra (see Sect. \ref{subsec:spec_analysis}). Some key CSM parameters are $M_{\rm CSM}=0.016^{+0.005}_{-0.004}\,M_\odot$, with an inner CSM radius and density of $R_{\rm 0}=2.07^{+0.79}_{-0.76}\times10^{13}$\,cm ($\sim1.38$\,AU) and ${\rho_{\rm 0}=2.45^{+1.18}_{-0.76}\times10^{-9}}$\,g\,cm$^{-3}$ respectively. In order to reproduce the fast evolution and luminous peak of SN~2022lxg, the model favours a low-mass, dense CSM close to the progenitor, blasted by the low-mass, fast-moving ejecta. The inner CSM radius $R_{\rm 0}$ can put an upper-limit on the radius of the progenitor (${\rstar\lesssim297\,\rsun}$). The high density and low mass of the CSM implies that it occupies a small volume (i.e. not extended). The ejecta interact with the CSM immediately after explosion ($t_{\rm int}\sim2.9$\,hrs after explosion), and due to the small volume of the CSM, the fast ejecta sweep it up quickly. If indeed the CSM is that dense, that could explain why the evolution of the light curves slows down around the peak epochs (even showing a small plateau around peak in $g$-band; see Sect. \ref{subsubsec:bb_lc}), but then the light curves decline rapidly. Finally, the model predicts a negligible host extinction and sets the explosion epoch at MJD$_{\rm expl}=59731.18^{+0.03}_{-0.03}$ within 3$\sigma$ from our estimate.

We note here several caveats of the model fits above. Concerning the CSM configuration, the \citet{Chatzopoulos2013} model assumes optically thick interaction that would not be appropriate for low CSM masses. Additionally, regardless of what the value of $s$, the power-law index (e.g. $s=2$ for a wind-like CSM and $s=0$ for a shell of constant density), the CSM is assumed spherically symmetric c.f. Sect. \ref{subsec:pola_analysis}. 
Another caveat is that the $\gamma$-ray opacity ($\kappa_\gamma$) is a free parameter with a higher prior up to 10. However, this value is usually assumed fixed at 0.027\,cm$^2$\,g$^{-1}$ (e.g. \citealt{Cappellaro1997}). Our best model fit returns a value of 3.4\,cm$^2$\,g$^{-1}$. If we fix this value to 0.027, the model under-fits the late-time $r$-band detection by $\sim$ an order of magnitude and suggests that the interaction with the CSM starts $\sim$ 2\,d after explosion, which is at odds with the observations. In order to assess the influence of $\kappa_\gamma$ and our last detection ($+313.1$\,d) of the SN, we recomputed the fits with $\kappa_\gamma$ fixed to 0.027 \,cm$^2$\,g$^{-1}$ and excluding the $+313.1$\,d data point. We find mostly similar values for the ejecta and CSM parameters, but with an order of magnitude higher $^{56}$Ni mass. Overall, the fit fails to match the rise or peak magnitude in any band, and for some bands the decline as well. It is possible that weak ongoing CSM interaction that is not accounted for by \texttt{MOSFiT} is manifested as an increased value of $\kappa_\gamma$. Furthermore, we tried several configurations for the $s$, $d$, and $n$ parameters but the fit that we present in this section is the best (in terms of $\log{(Z)}$ but also in capturing the photometric evolution both at early and late times). We also tried relaxing the ejecta velocity priors (e.g. setting the minimum to 5,000\,km\,s$^{-1}$) consistently returns velocities between 17\,000 and 23,000\,km\,s$^{-1}$. Finally, we also tried individually the \texttt{default} model (luminosity powered only by the decay of radioactive $^{56}$Ni) and the \texttt{csm} model (luminosity powered only by interaction) and both models returned bad fits, both in terms of $\log{(Z)}$ and fitting the data points, but also by returning unphysically high $^{56}$Ni masses (the former) or very low ejecta masses ($\sim 0.5 \msun$) the latter. All the above considered, the model fit results should be interpreted with caution and the struggle of different configuration to explain the data might be a hint that the CSM is indeed non-spherical.

\begin{figure*}
        \centering
        \begin{subfigure}[b]{0.94\textwidth}
            \centering
        \includegraphics[width=0.33 \textwidth]{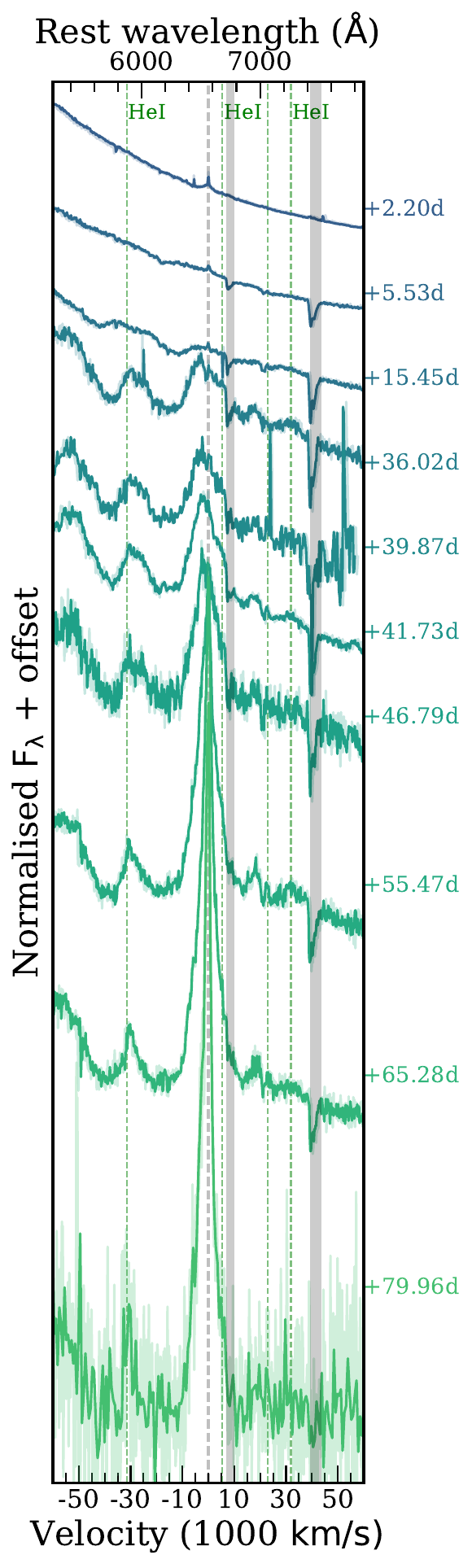}
        \includegraphics[width=0.33 \textwidth]{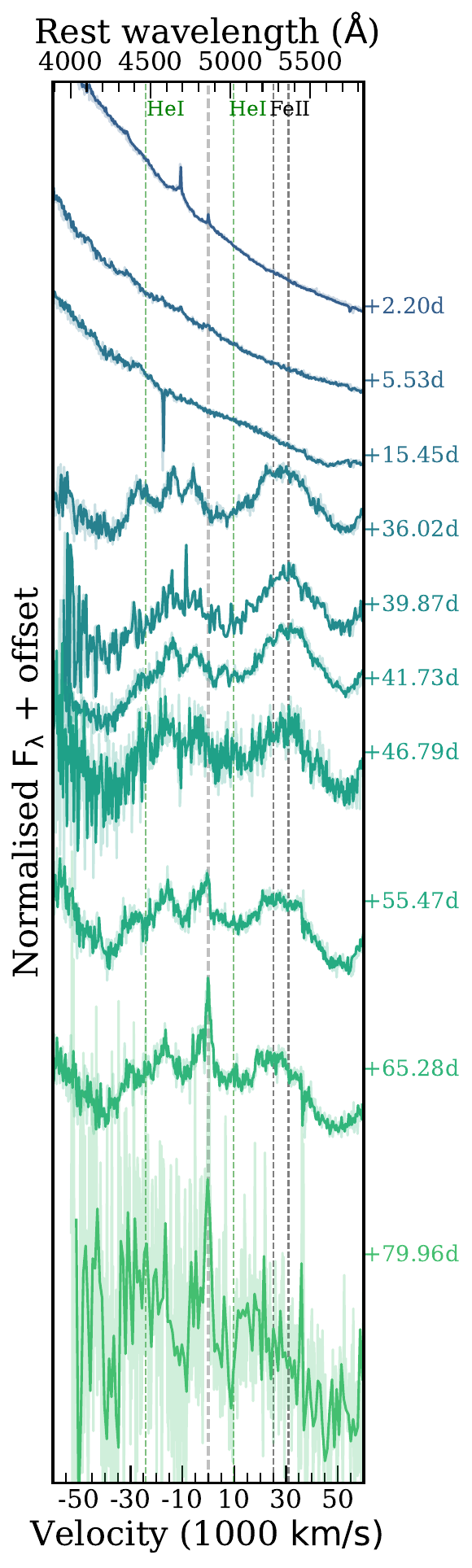}
        \includegraphics[width=0.33 \textwidth]{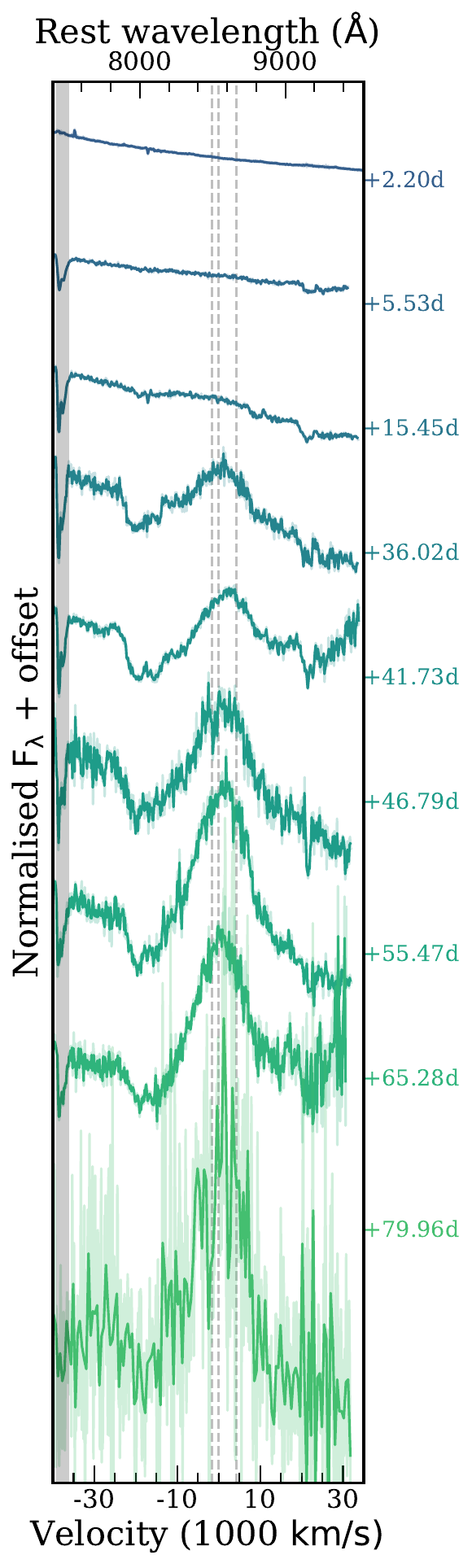}
        \end{subfigure}
        \caption{Evolution of optical spectroscopic lines in velocity space. \textbf{Left}: Region around H$\alpha$. Note that for the epochs $\gtrsim +45$\,d, we seem to see a break around 5700\AA, with the continuum becoming suddenly stronger blue-wards of that wavelength. \textbf{Middle}: Region around H$\beta$. \textbf{Right}: Region around \ion{Ca}{II} NIR triplet (centred in the middle line but all three lines are shown). The spectra are binned to 5 \AA\, for visual purposes and the original spectra are plotted with lighter colours in the background. The grey dashed vertical line shows the central wavelength. The colours of the vertical lines denote the different elements, green is for helium and black for iron. Although only \ion{He}{I} $\lambda5876$ is robustly identified, we mark other helium lines to guide the eye. The telluric features are marked with grey shaded vertical lines.}
        \label{fig:lines}
    \end{figure*}

\begin{figure*}
\centering
\includegraphics[width=0.95 \textwidth]{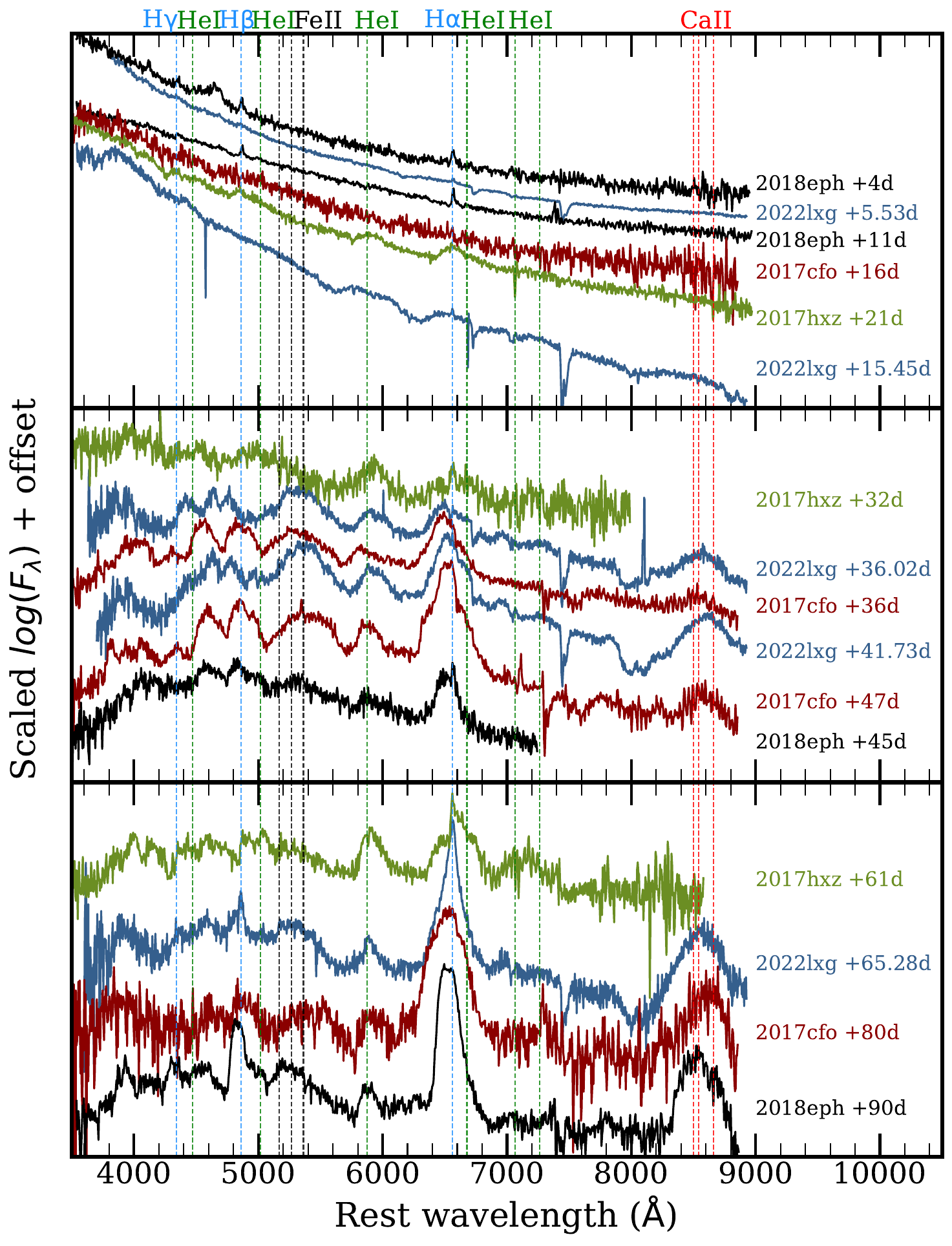}
\caption{Spectral comparison of SN~2022lxg to other luminous Type II SNe from  \pjpnospc, at different epochs. Emission lines are marked with vertical dashed lines. Although only \ion{He}{I} $\lambda5876$ is robustly identified, we mark other helium lines to guide the eye. There is a clear resemblance between SN~2022lxg and the comparison SNe, throughout different phases. SN~2017cfo in particular appears to be the best match.}
\label{fig:spec_comp_LII}
\end{figure*}

\subsection{Spectroscopic analysis} \label{subsec:spec_analysis}

\subsubsection{Early spectra and flash-ionisation features} \label{subsubsec:ff}

Our earliest spectrum ($+2.20$\,d, left panel of Fig. \ref{fig:Keck}) shows flash features on top of a $\sim$18000\,K continuum that we attribute to H$\alpha$, H$\beta$, H$\gamma$ and \ion{He}{II} $\lambda4686$. We simultaneously fit Lorentzian and Gaussian profiles to account for the narrow and underlying broad components, respectively. For H$\alpha$ we find a FWHM of $241\pm19\,\rm km\,s^{-1}$ and $2\,814\pm228\,\rm km\,s^{-1}$. For the \ion{He}{II} features, we find $226\pm37\,\rm km\,s^{-1}$ and $3\,717\pm356\,\rm km\,s^{-1}$, while for H$\beta$ we find $389\pm65\,\rm km\,s^{-1}$ and $3\,829\pm1014\,\rm km\,s^{-1}$.
A narrow component around H$\alpha$ appears in some of the later spectra, but at the resolution of our spectra, we cannot determine whether this is due to underlying emission from an \ion{H}{II} region. As shown in the continuum subtracted spectra in right hand panel of Fig. \ref{fig:Keck}, the flash features persist out to $\sim$8\,d. The two subsequent spectra ($+8.66$ and $+9.67$\,d) are of too low a spectral resolution to detect weak narrow features, but by the epoch of our next spectrum ($15.45$\,d), there is no sign of \ion{He}{II}. We therefore conclude that the flash-ionisation lines must have disappeared by $\sim+(8-9)$\,d.
Following \citet{Bruch2023}, we calculate the flash timescale as the time from the estimated explosion date to the mid-point between the last spectrum that shows the \ion{He}{II} $\lambda$4686 line and the first one that does. Choosing our $+8.66$\,d spectrum as the latter, leads to a timescale of $8.24\pm0.42$ days, while being more conservative and choosing our $+15.45$\,d spectrum, gives $11.63\pm3.82$ days. Both values are in agreement with \citet{Bruch2023} who argue that the duration of the flash-ionisation features is partially correlated with the rise time and the absolute peak magnitude. We show the location of SN~2022lxg with their sample in Fig. \ref{fig:FF_comp}.

\subsubsection{Spectral evolution and line identification} \label{subsubsec:spec_evol}

After $\sim$10\,d broad, shallow line profiles develop and become increasingly prominent as the continuum cools. The Balmer lines and the \ion{Ca}{II} NIR triplet ($\lambda\lambda$ 8498.02, 8542.09, 8662.14) can be easily identified (Fig. \ref{fig:spectra}). A feature close to \ion{He}{I} $\lambda5876$ and the \ion{Na}{I} D doublet is also present. If it were solely due to \ion{He}{I} $\lambda5876$, then it appears redshifted by $\sim 3\,000\rm\,km\,s^{-1}$. We would expect  \ion{Na}{I} D in emission to typically emerge at later epochs; attempting to de-blend the feature with two Gaussian profiles results in any contribution due to \ion{Na}{I}D being redshifted by $\sim 5\,000\rm\,km\,s^{-1}$. Hence, we conclude that it most likely is \ion{He}{I} $\lambda5876$ emission. There is also a broad feature around 5400 \AA\, that we tentatively identify as \ion{Fe}{II} ($\lambda\lambda$ 5169, 5267, 5363) and a blend of lines around and blue-wards the rest wavelength of H$\beta$. Both the \ion{He}{I} and \ion{Fe}{II} features appear to be strongest in the $+41.73$\,d spectrum before gradually fading. We used the tools \texttt{SNID} \citep{Blondin2007} and \texttt{GELATO} \citep{Harutyunyan2008} to search for objects with spectra similar to those of SN~2022lxg during these early and intermediate phases ($\lesssim +45$\,d). The returned matches are of a range of SN subtypes and epochs, and therefore inconclusive. Nevertheless, a reasonable match is with the type IIb SN~2011fu \citep{Morales-Garoffolo2015}. Interestingly, these tools also return matches with some Type Ic SNe with broad spectral features (SNe Ic-BL). Some examples are SN~2013dx \citep{DElia2015}, SN~2007ru \citep{Sahu2008} and SN~1998bw (e.g. \citealt{Kulkarni1998,Woosley1999,Patat2001}).

\begin{figure}
  \centering
  \includegraphics[width=0.5 \textwidth]{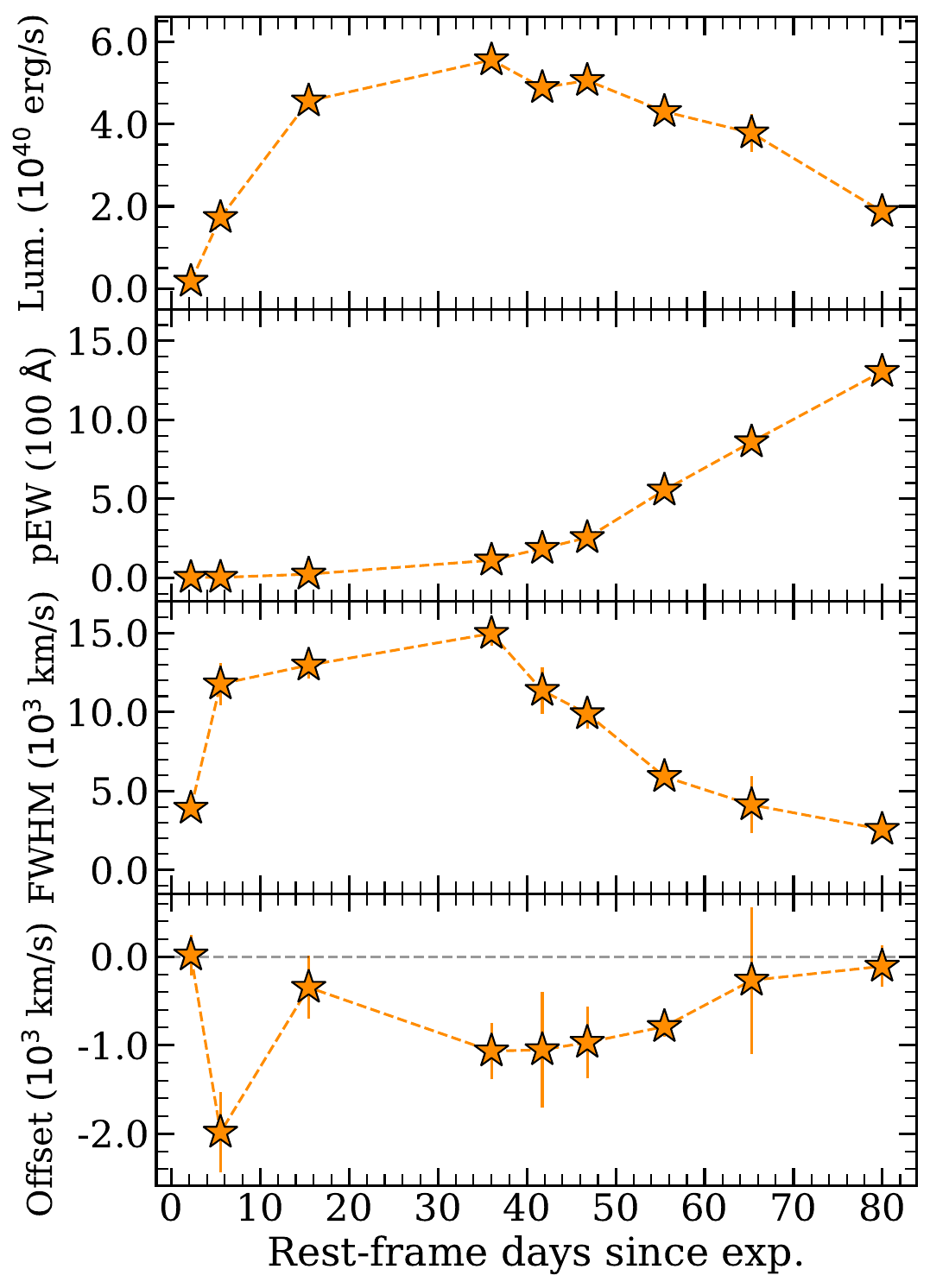}
  \caption{Luminosity, pEW, FWHM and velocity offset evolution of the H$\alpha$ line emission.}
  \label{fig:ha_prop}
\end{figure}

From about $\gtrsim +45$\,d, most features begin to fade with the exception of  H$\alpha$ and the \ion{Ca}{II} NIR triplet. The \ion{He}{I} $\lambda5876$ feature, though weak, is still present in our last spectrum. We defer a discussion of the evolution of the H$\alpha$ line to the next section (\ref{subsubsec:Ha_evol}), but note that it displays a persistent blueshift until $\sim +55$\,d after which it remains close to its rest wavelength, and gradually becomes narrower. In Fig. \ref{fig:lines} we show the evolution of H$\alpha$, H$\beta$, and the \ion{Ca}{II} NIR triplet in velocity space. The latter shows a P-Cygni profile, with the absorption minimum found at $\sim20\,000\rm\,km\,s^{-1}$, extending up to $\sim24\,000\rm\,km\,s^{-1}$. We note that for the epochs $\gtrsim +45$\,d, we start to see a break in the continuum around 5700 \AA, with the continuum becoming suddenly stronger blue-wards of that wavelength. We again tried to find spectral matches to these epochs of SN~2022lxg using the aforementioned tools and get matches with SNe Type IIn. For epochs between $+45$ to $+60$\,d where there is still \ion{He}{I} $\lambda5876$ in the spectra, we get good matches with SN~1996al \citep{Benetti2016}. After that, when H$\alpha$ dominates the spectrum, we get good matches with SN~2005ip \citep{Smith2009,Stritzinger2012}. Indeed, the properties of SN~2022lxg share similarities with transitional complex SNe like SN~1996al (Type IIn/IIL; \citealt{Benetti2016}) and SN~2018ivc (Type IIb/IIL; \citealt{Bostroem2020,Maeda2023a,Maeda2023,Reguitti2024}. They are fast rising and fast linearly declining (at least during their early $<+150$\,d evolution, see Fig. \ref{fig:mag_LII_comp}), with strong H$\alpha$ emission as well as a prominent feature likely due to \ion{He}{I} $\lambda5876$.

Allowing a free range in choice of epoch, we can also find superficial matches with bright, slowly-declining Type II SNe. For instance the $+103$\,d and $+200$\,d spectra of SNe~1998S and 2021irp \citep[][respectively]{Fassia2000,Reynolds2025} bear some similarity to the $+65$\,d spectrum of SN~2022lxg possibly indicative of similar ejecta or CSM conditions at those epochs. From the previous discussion on the light curve and spectral evolution, it is clear that SN 2022lxg shares some properties in common with all Type II SN subgroups. However, it consistently shares most properties in common mainly with the sample of objects in \pjpnospc. Therefore, in what follows, we restrict our comparisons primarily to this group, but invoke analogous behaviour for a small handful of other SNe for illustrative purposes.

\begin{figure}
  \centering
  \includegraphics[width=0.5 \textwidth]{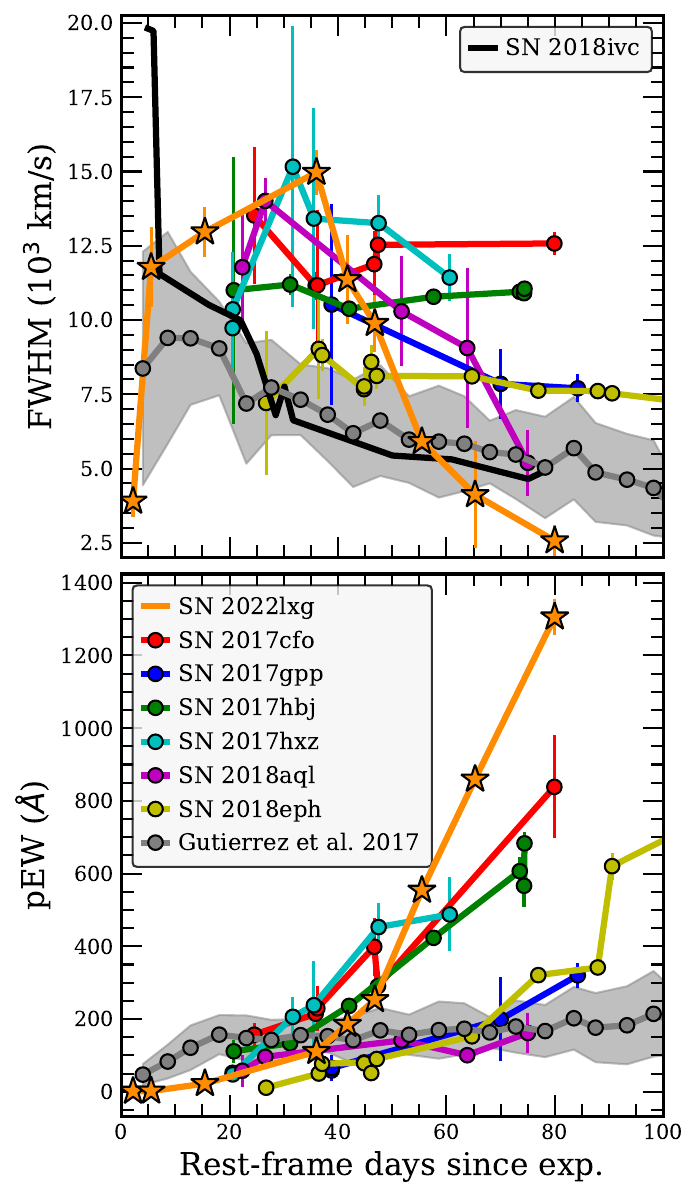}
  \caption{H$\alpha$ line emission FWHM (top) and pseudo-equivalent width (bottom) compared to other luminous Type II SNe from the \pjp sample (adapted from \pjpnospc). Mean values for regular SNe II from the sample of \citet{Gutierrez2017} are presented in grey (with the 1$\sigma$ standard deviation plotted as a shaded region). In the top panel, we also plot the H$\alpha$ FWHM of SN~2018ivc (from \citealt{Bostroem2020}).}
  \label{fig:ha_prop_LII}
\end{figure}

The spectral evolution of SN~2022lxg is also noteworthy in that typical metal lines seen in Type II SNe are conspicuous by their absence. The broad feature that we attributed to \ion{Fe}{II} ($\lambda\lambda$ 5169, 5267, 5363) is an exception. It is likely that our last spectrum ($\sim$80\,d) does not probe the nebular phase. Nevertheless, this lack of metal lines in the photospheric phase is another feature in common with \pjpnospc. Another is a persistent emission line at $\sim 5800$\,\AA, identified as \ion{He}{I} $\lambda5876$. The similarities between the H$\alpha$ profiles will be discussed in detail in Sect. \ref{subsubsec:Ha_evol}. In Fig. \ref{fig:spec_comp_LII}, we plot the spectral evolution of SN~2022lxg compared to LSNe II of the \pjp sample, in order to highlight the spectral similarities throughout the different phases of the evolution of those SNe.

\begin{figure}
\centering
\includegraphics[width=0.5 \textwidth]{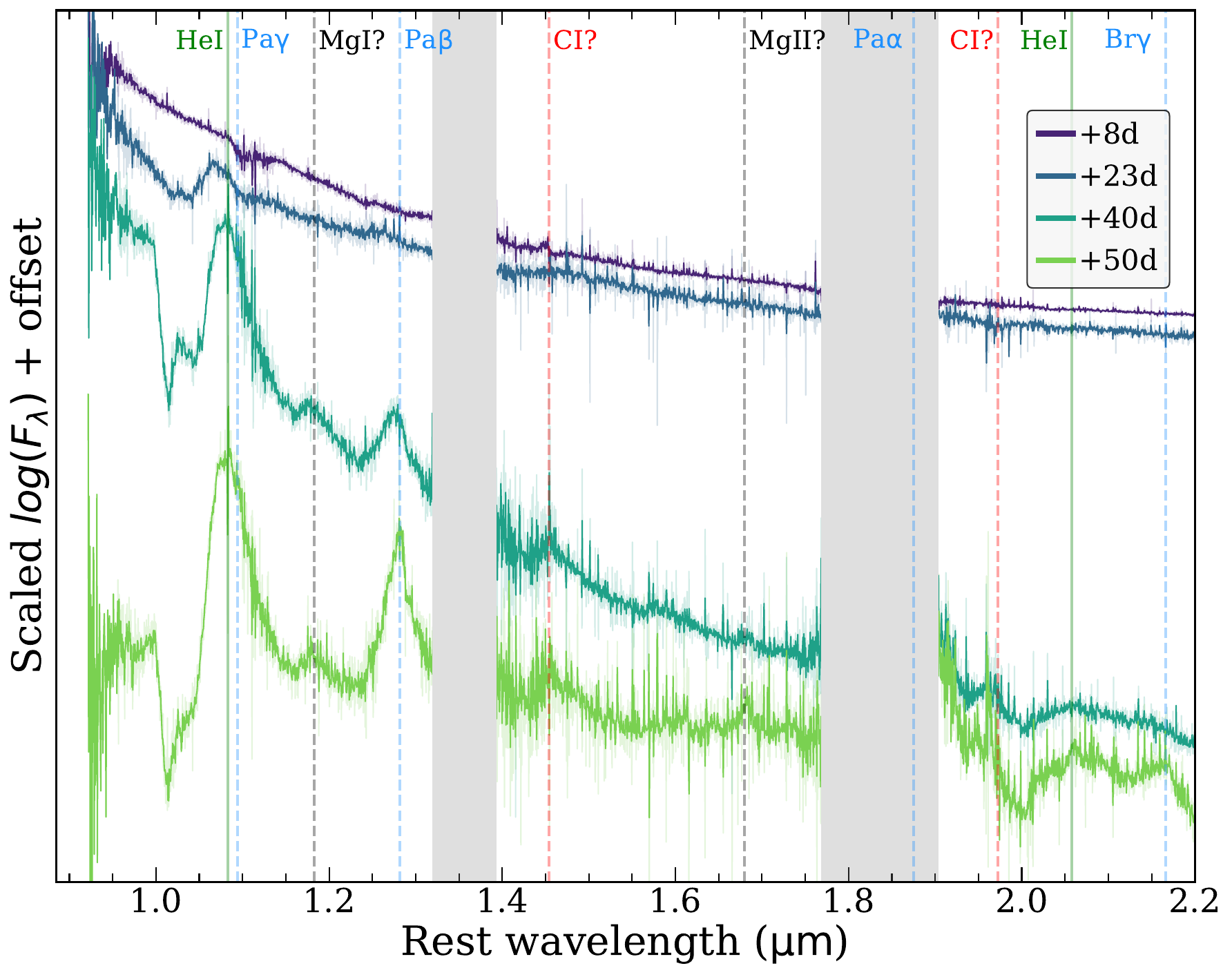}
\caption{NIR spectra of SN~2022lxg taken from \citet{Tinyanont2024}. The spectra are binned to 5 \AA\, for visual purposes and the original spectra are plotted with lighter colours in the background. The colours of the vertical lines denote the different elements, green is for helium and cyan for hydrogen. The tentatively identified magnesium and carbon are marked with black and red respectively (with a question mark next to the element name). The telluric features are marked with grey shaded regions.}
\label{fig:NIR}
\end{figure}

\subsubsection{H$\alpha$ properties} \label{subsubsec:Ha_evol}

Even during the flash-ionisation phase ($\lesssim+8$\,d), there is a broad shallow H$\alpha$ component underneath the narrow feature. The profile broadens quickly and gradually becomes stronger. The LSNe II in the \pjp sample show a weak or non-existent absorption component in their H$\alpha$ profile (i.e. a P-Cygni profile). Similarly, the H$\alpha$ profile of SN~2022lxg only shows a weak absorption component, more evident at the $+15.45$\,d spectrum, that appears even weaker as the SN evolves, probably due to contamination from the strong \ion{He}{I} $\lambda5876$ emission. In studies of Type IIP SNe, a weak or absent absorption component associated with the H$\alpha$ profile correlates brighter peak magnitudes, fast decline rates, and high velocities \citep{Gutierrez2014}, similar to what we observe for SN~2022lxg. As it evolves, the H$\alpha$ emission profile shows a blue excess, and a bump on its red side, coincident with the wavelength of \ion{He}{I} $\lambda6678$. Blueshifted emission peaks are indeed expected in Type II SNe \citep{Anderson2014a}. At $\gtrsim+55$\,d the profile narrows down and becomes centred to the rest wavelength, while the red bump completely disappears. We performed a spectroscopic line study in order to quantify the properties of the H$\alpha$ line using customized \texttt{Python} scripts. For this study, we use all our eight NOT+ALFOSC spectra that ensure consistency between the measurements and a good sampling of the evolution of the event from explosion to fading. Additionally, we include our first Keck+LRIS spectrum ($+2.20$\,d). For all the following measurements, we remove the (linear) continuum locally. We quantify the line luminosity and the pseudo-equivalent width (pEW) by integrating under the line profile. In order to measure the velocity width (FWHM) we used a custom script in \texttt{Python} which first smooths the spectrum, then locates the data points on the left and right of the maximum that have flux values closest to the half of the maximum and then calculates the distance between them on the x-axis. The observed central wavelength of the line is defined as the middle point of this distance, and its distance from the central wavelength of H$\alpha$ is defined as the line offset. We use a custom Monte Carlo method (10\,000 iterations of re-sampling the data assuming Gaussian error distribution) in order to calculate uncertainties for the flux (luminosity), line width and offset from the central wavelength.

\begin{figure*}
\centering
\includegraphics[width=0.45 \textwidth]{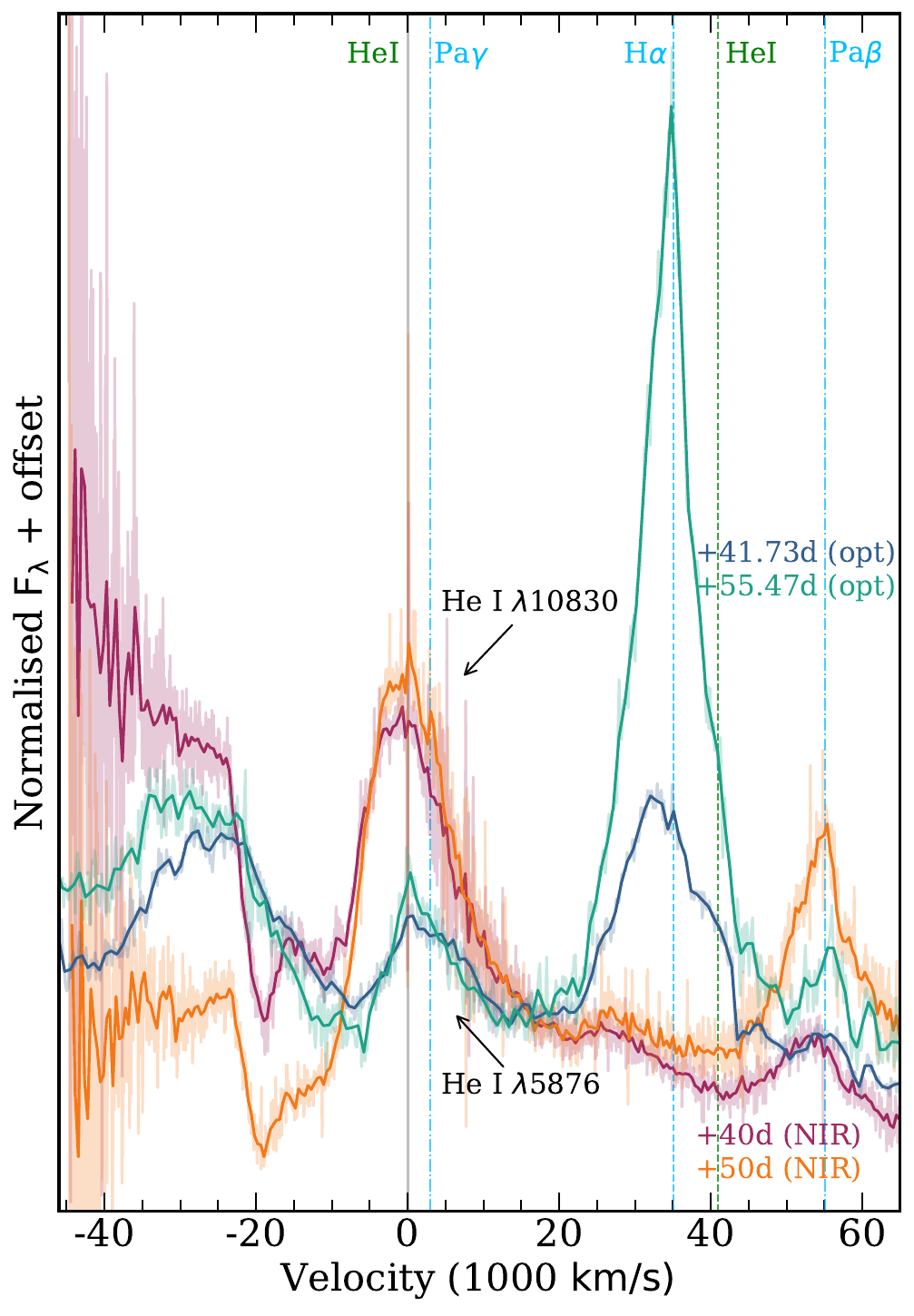}
\includegraphics[width=0.45 \textwidth]{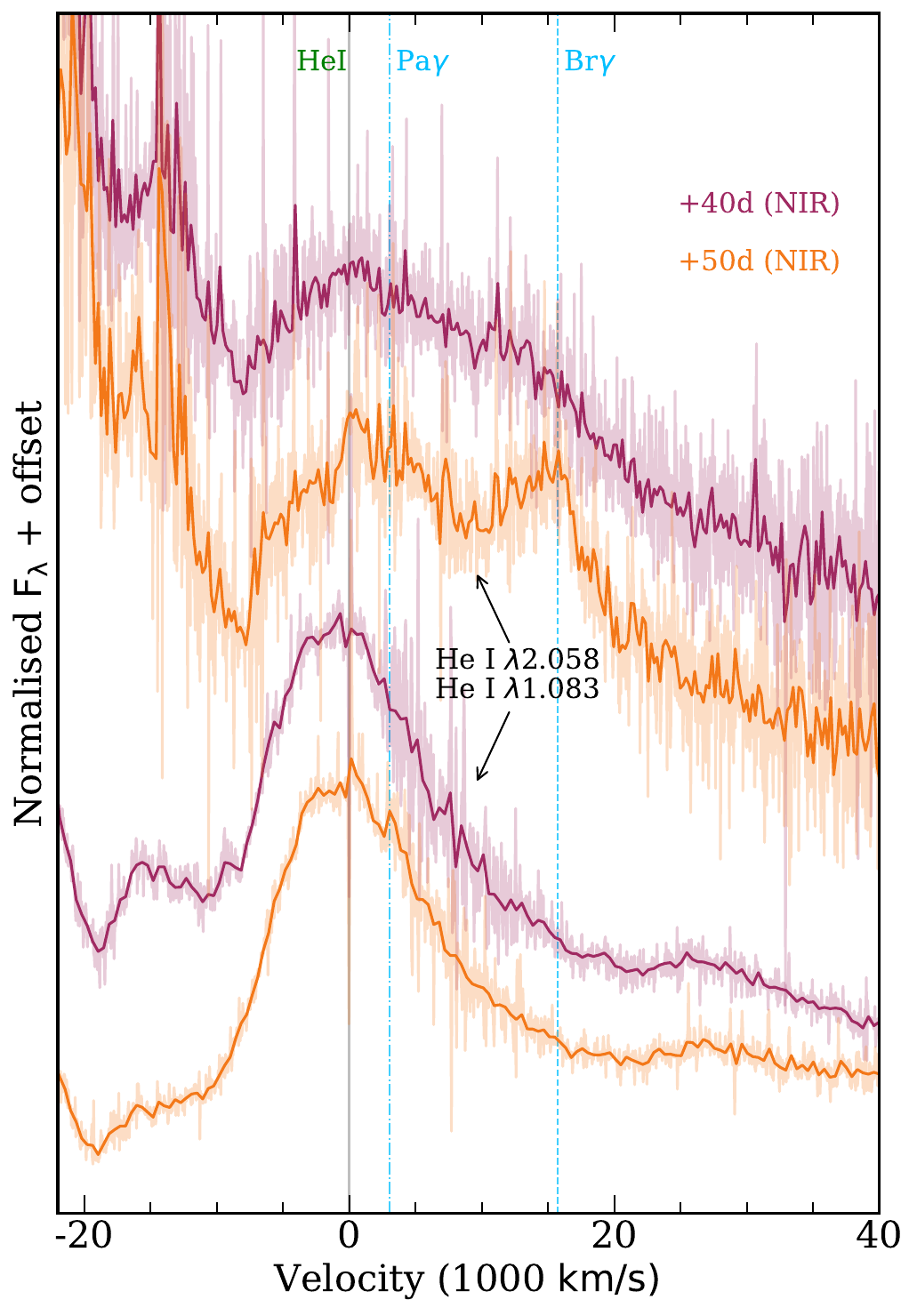}
\caption{Evolution of NIR spectroscopic lines in velocity space. The spectra are binned to 5 \AA\, for visual purposes and the original spectra are plotted with lighter colours in the background. The grey vertical line shows the central wavelength. The colours of the vertical lines denote the different elements, green is for helium and cyan for hydrogen. \textbf{Left:} optical (blue/green colourmap) and NIR (red/orange colourmap) spectra, centred at \ion{He}{I} $\lambda5876$ and $\lambda10830$ respectively. The spectra are normalised at $\sim 16\,000\,\rm km\,\rm s^{-1}$ red-wards of the central wavelengths. Dashed vertical lines mark elements in the optical while dashed-dotted ones mark elements in the NIR. \textbf{Right:} evolution of \ion{He}{I} $\lambda10830$ (bottom) and \ion{He}{I} $\lambda20581$ (top) region of the spectra in velocity space. Dashed vertical lines mark elements in the top spectra while dash-dotted ones mark elements in the bottom ones. Both \ion{He}{I} lines show multi-component profiles with absorption components at $\sim -10000$ and $-20000$\,km\,s$^{-1}$. }
\label{fig:NIR_vel}
\end{figure*}

In the sub-panels of Fig. \ref{fig:ha_prop}, we present the H$\alpha$ luminosity (top panel), pEW (second panel), velocity width (third panel) and offset from the rest wavelength (bottom panel), of SN~2022lxg. We find that the luminosity of the line is quickly rising until it peaks at $\sim+36$\,d with a luminosity of $\sim5.5 \times10^{40} \rm\,erg\,s^{-1}$. After that, it declines with a slower pace, measuring a luminosity of $\sim2.0 \times10^{40} \rm\,erg\,s^{-1}$ in our last spectrum ($+79.96$\,d). The pEW very slowly rises until the intermediate epochs ($\sim+36$\,d) and then it sharply rises. Since the pEW provides an indication of how strong the line is with respect to the underlying continuum, what we observe in SN~2022lxg is that, even though the luminosity of H$\alpha$ drops after $\sim+36$\,d, the luminosity of the continuum drops much faster. This is not surprising since we have already highlighted how fast the light curves of SN~2022lxg decline. The velocity width (FWHM) and offset of the line, indeed evolves like the LSNe II in the \pjp sample. The FWHM quickly ($\sim +10$\,d) rises to values around 10\,000\,$\rm km\,s^{-1}$ and then more slowly reaches its peak at $\sim$ 15\,000\,$\rm km\,s^{-1}$ in the same epoch that the luminosity peaks. Similarly, the velocity offset reaches a blueshift of $\sim -1\,000\, \rm km\,s^{-1}$ during the epochs that the line luminosity and FWHM peaks, before it gradually gets centred to the rest wavelength of H$\alpha$. There is one epoch (first NOT+ALFOSC spectrum at $+5.53$\,d) where the broad shallow profile seems to broaden too fast (a jump of $\sim 8\,000\, \rm km\,s^{-1}$ within $\sim 3.5$ days) and the offset reaches $\sim -2\,000\, \rm km\,s^{-1}$. The offset drops down to $\sim 350\, \rm km\,s^{-1}$ almost ten days after, and then smoothly reaches the local blueshift minimum at at $\sim+36$\,d. This same profile is seen in the $+6.98$\,d UH88+SNIFS spectrum, giving further confidence that this is not instrumental/artefact, but a real feature of the SN. The rapid light curve rise of SN~2022lxg, can potentially explain the fast evolution of the line profiles as well.

In Fig. \ref{fig:ha_prop_LII}, we present a direct comparison of the aforementioned H$\alpha$ properties of SN~2022lxg with those measured in the spectra of the LSNe in \pjpnospc. In the top panel we present the velocity width (FWHM) evolution (including available measurements of SN~2018ivc) while in the bottom panel we present the pEW evolution. For comparison, we also plot mean values for regular SNe II from the sample of \citet{Gutierrez2017}. The similarity between the evolution of both the FWHM and pEW of SN~2022lxg, with those of the \pjp sample is striking. Combined with the fact that the velocity offset in their sample shows the same behaviour as the one of SN~2022lxg (most likely without the \say{spike} at $+6$\,d though), and with the fact that they show a tentatively identified \ion{He}{I} $\lambda5876$ as well as a clear lack of metal lines, leads to the conclusion that the events are spectroscopically similar.

\begin{figure}
\centering
\includegraphics[width=0.47 \textwidth]{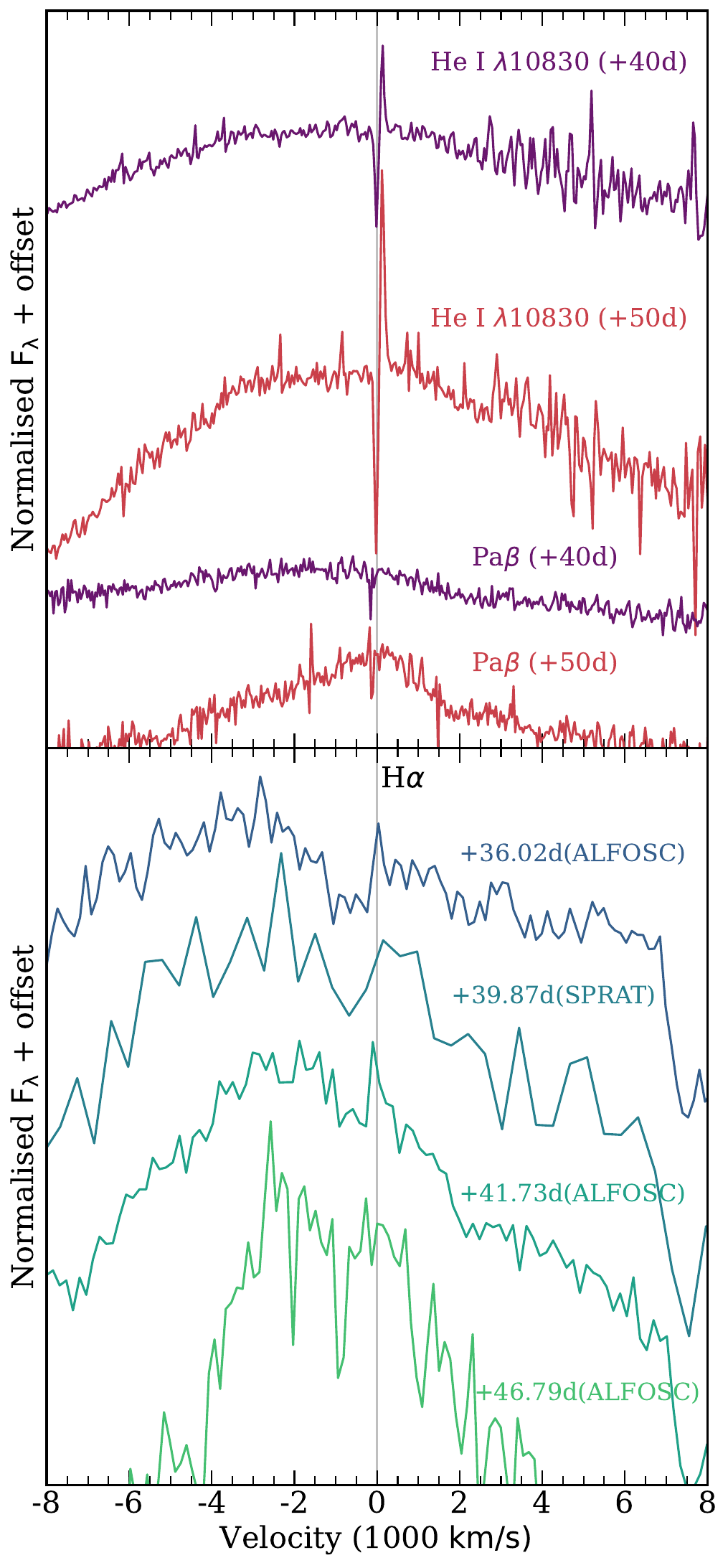}
\caption{Narrow P-Cygni profiles in velocity space. Top panel (LRIS NIR spectra): The two upper spectra clearly show narrow P-Cygni profiles of \ion{He}{I} 1.0830 $\mu$m with peak to peak separations of $\sim150\,\rm km\,s^{-1}$, pointing to unshocked CSM. In the two lower (same epoch) spectra, a narrow profile in the Pa$\beta$ line is tentatively detected. Bottom panel: four low-resolution optical spectra centred on H$\alpha$ spanning $\sim10$\, days.}
\label{fig:Pcyg}
\end{figure}

\subsubsection{Near-infrared spectroscopy} \label{subsubsec:NIR}

We present four publicly available near-infrared (NIR) spectra of SN~2022lxg in Fig. \ref{fig:NIR}. The spectra were presented as part of the Keck Infrared Transient Survey \citep{Tinyanont2024} and we collected them from the Weizmann Interactive Supernova data REPository \citep[WISeREP\footnote{\url{https://www.wiserep.org/}};][]{Yaron2012}. The spectra are taken at phases $+8$\,d, $+23$\,d, $+40$\,d, and $+50$\,d. The first spectrum is practically featureless and blue, in accordance with the optical spectra at that time. In the second spectrum, a broad P-Cygni profile of \ion{He}{I} 1.0830 $\mu$m line is formed, again in accordance with the optical when the \ion{He}{I} $\lambda5876$ line is emerging. In the last two epochs of $+40$\,d, and $+50$\,d, the main difference that we see in the NIR spectra is that hydrogen lines have emerged, again in accordance with the optical. We clearly detect strong Pa$\beta$ but unfortunately Pa$\alpha$ is blended with a strong telluric line. The \ion{He}{I} 1.0830 $\mu$m line has become much stronger and is blended with Pa$\gamma$, and in the last spectrum we also detect \ion{He}{I} 2.0581 $\mu$m which is also blended with Br$\gamma$. In Fig. \ref{fig:NIR_vel}, we visualise in velocity space, the comparison of these last two spectra with the optical spectra at similar epochs (left panel) and between the two helium NIR lines (right panel). The absorption minimum of the \ion{He}{I} 1.0830 $\mu$m P-Cygni profile lies at $\sim20\,000\rm\,km\,s^{-1}$ and extends up to $\sim23\,000\rm\,km\,s^{-1}$. That is fully consistent for what we find from the P-Cygni profiles of the \ion{Ca}{II} NIR triplet and it also agrees with the FWHM of the H$\alpha$ emission around the same epochs. All the above considered, we assume an ejecta velocity of $\sim$20\,000\,km\,s$^{-1}$ for SN~2022lxg.

Finally, in the last two spectra, there is an emission line at $\sim$ 1.82 $\mu$m that we identify as \ion{Mg}{I} 1.183 $\mu$m, and a line at $\sim$ 1.65 $\mu$m that we tentatively identify as \ion{Mg}{II} 1.680 $\mu$m. However, there is no sign of \ion{Mg}{I} 1.504 $\mu$m in the spectra. There also seem to be two lines at $\sim$ 1.455 $\mu$m and 1.965 $\mu$m that are harder to identify. Using the National Institute of Standards and Technology (NIST) database\footnote{\url{https://physics.nist.gov/PhysRefData/ASD/lines_form.html}}, we find two strong \ion{C}{I} transitions close to these wavelengths that we mark with a question mark in Fig. \ref{fig:NIR}. We do the same with the magnesium lines in order to highlight that all these lines are tentatively identified.

Another feature revealed by the NIR spectra, is the presence of a narrow P-Cygni profile at the peak of the emission of the \ion{He}{I} 1.0830 $\mu$m line, at $+40$ and $+50$ days. Such profiles are often detected in spectra of interacting supernovae provided that they are of sufficient resolution. This points to unshocked CSM along our line of sight (e.g. \citealt{Smith2002a,Kotak2004,Trundle2008,Andrews2025}). We fit two Gaussian profiles to the absorption and emission components and measure a peak-to-peak offset between the two of $143.4\pm3.8$ and $150.9\pm4.2\rm\,km\,s^{-1}$ respectively, which is larger than the instrumental resolution of Keck+NIRES ($\approx100\rm\,km\,s^{-1}$). The mixed H and He CSM suggests partial stripping of the progenitor. We searched for similar profiles in other lines (Fig. \ref{fig:Pcyg}) and tentatively identify an absorption component in the Pa$\beta$ line during the same epochs as for the \ion{He}{I} 1.0830 $\mu$m line. Although our optical spectra are of too low a spectral resolution to draw robust conclusions, there may be hints of an absorption trough (Fig. \ref{fig:Pcyg}) if we are guided by the near-IR spectra.

\subsection{Polarimetric analysis} \label{subsec:pola_analysis}

The intensity-normalised Stokes parameters ($q = Q/I$ and $u = U/I$,
where $Q$ and $U$ are the differences in flux with electric field oscillating in two perpendicular directions, and $I$ is the total flux) were used to calculate the polarisation degree (${p = \sqrt{q^{2} + u^{2}}}$), and the polarisation angle (${\chi = 0.5\arctan(u/q)}$). All values of p presented in this paper have been corrected for polarisation bias \citep[e.g.][]{Simmons1985,Wang1997} following \citet{Plaszczynski2014}. 

\begin{figure}
  \centering
      \includegraphics[width=0.5 \textwidth]{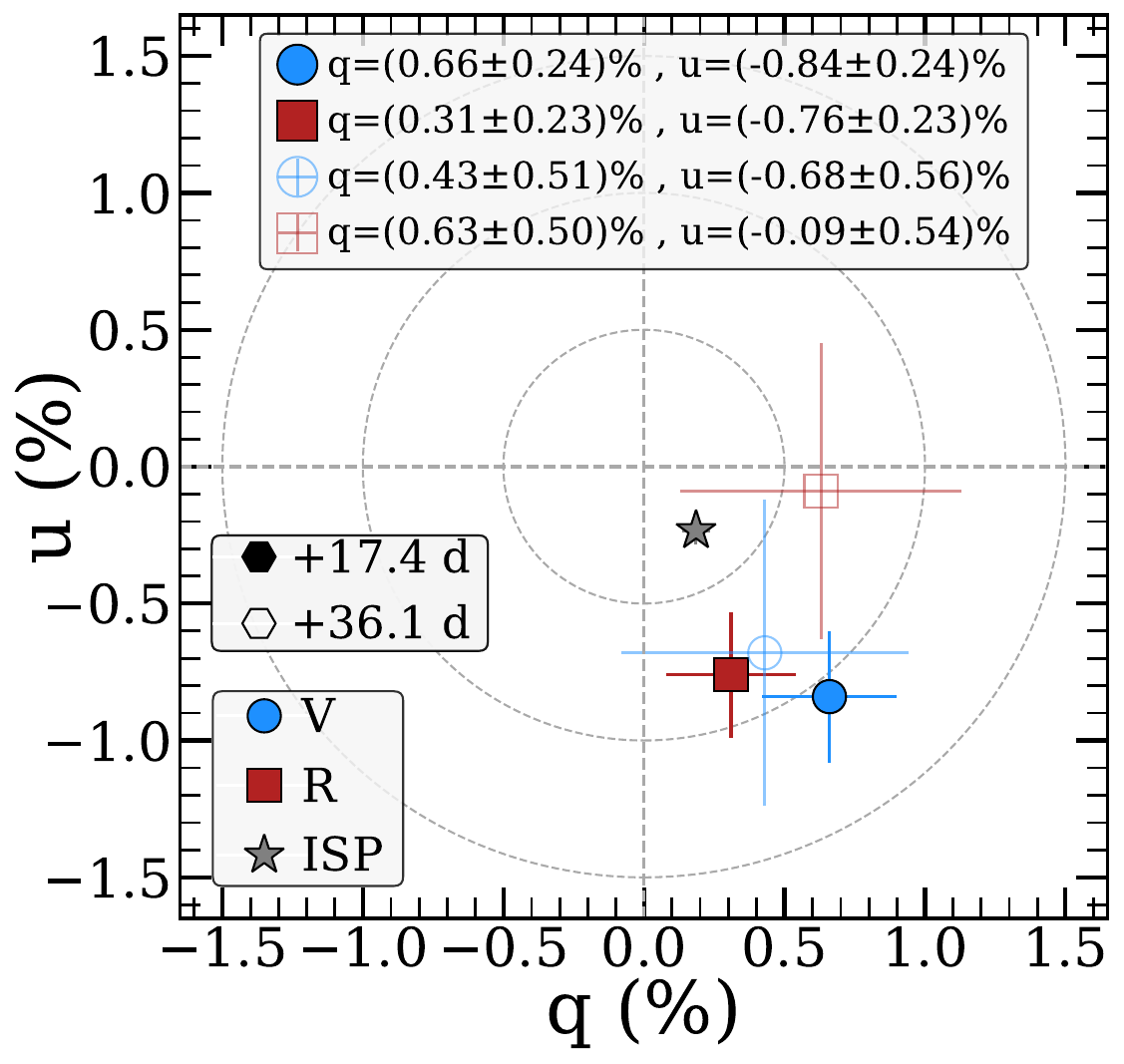}
  \caption{Intensity-normalised Stokes $q$ and $u$ parameters, from ALFOSC imaging polarimetry, in the $V$ (blue circles) and $R$ (red squares) bands, at phases $+17.4$\,d (filled markers) and $+36.1$\,d (open markers). The grey star marks our best ISP estimate and the dashed circles mark the 0.5\%, 1.0\% and 1.5\% polarisation values ($p$). The first epoch (with a robust S/N $\sim$ 300) shows that SN~2022lxg is intrinsically polarised.}
  \label{fig:impol}
\end{figure}

We have measurements in the $V$ and $R$ bands for two epochs, $+17.4$\,d and $+36.1$\,d. We present the Stokes $q$ – $u$ planes for the imaging polarimetry results in Fig. \ref{fig:impol} where we also plot the ISP estimate as a grey star. The measured values can be found in Table \ref{tab:pol_log}. For the first epoch, we measure $p =(1.04\pm0.24)\,\%$ and $p =(0.79\pm0.23)\,\%$ for the $V$ and $R$ band respectively, and a polarisation angle of $\chi =(-25.92\pm6.44)\,^{\circ}$ and $\chi =(-33.90\pm8.03)\,^{\circ}$. For the second epoch, we measure $p =(0.65\pm0.55)\,\%$ and $p =(0.46\pm0.50)\,\%$ for the $V$ and $R$ band respectively, and a polarisation angle of $\chi =(-28.85\pm19.45)\,^{\circ}$ and $\chi =(-4.01\pm22.55)\,^{\circ}$. The accuracy of the second epoch measurements are low ($p/\sigma_{p} \sim 1$), however the first epoch measurements are statistically significant ($p/\sigma_{p} \sim 3-4$) and we measure polarisation of $p \sim (0.8-1)\%$. 

In order to estimate the intrinsic polarisation of a source, the effect of the interstellar polarisation (ISP), introduced by dust grains in the line of sight, has to be estimated. Since the (potential) host galaxy is very faint and distant from the SN, we consider its contamination in the polarisation negligible. Hence we only treat contamination from the MW. SN~2022lxg lies in a crowded field with many stars, so we can put tight constraints on the MW ISP. For each filter, we make a weighted average estimate between the field stars within each epoch and then double confirm with the ISP estimate of the other epoch. We measure ${q_{ISP} = (0.19\pm0.05)\,\%}$ and ${u_{ISP} = (-0.23\pm0.05)\,\%}$ leading to $p_{ISP} =(0.29\pm0.05)\,\%$ and $\chi_{ISP} =(-25.75\pm4.92)\,^{\circ}$. Another way to roughly estimate an upper-limit for the Galactic ISP is with the empirical law $9 \times E(B-V) \,\%$ \citep{Serkowski1975}, and for SN~2022lxg that would be $\sim 0.53\,\%$, consistent with our estimate. We also checked for polarisation standard stars published in \citet{Heiles2000} that are close to the location of SN~2022lxg. We find one star within 3.5 degrees from the location of the SN with a polarisation value of ${p = (0.0\pm0.2)\%}$, again consistent with the above estimate. 

We perform vector subtraction in the $q-u$ plane in order to remove the ISP contribution and estimate the intrinsic polarisation of SN~2022lxg. In Fig. \ref{fig:impol_corr} of the Appendix, we present the ISP subtracted Stokes $q$ – $u$ planes for the imaging polarimetry results. For the first epoch, we measure $p =(0.73\pm0.25)\,\%$ and $p =(0.49\pm0.24)\,\%$ for the $V$ and $R$ band respectively, and a polarisation angle of $\chi =(-25.99\pm5.9)\,^{\circ}$ and $\chi =(-38.35\pm12.47)\,^{\circ}$. For the second epoch, we measure $p =(0.34\pm0.55)\,\%$ and $p =(0.30\pm0.51)\,\%$ for the $V$ and $R$ band respectively, and a polarisation angle of $\chi =(-30.66\pm31.00)\,^{\circ}$ and $\chi =(-8.92\pm31.06)\,^{\circ}$. The polarisation values in the first epoch (with a robust S/N $\sim 300$) suggest a mildly aspherical configuration (ellipticity of $b/a \sim 0.85$; \citealt{Hoflich1991}). There seems to be an evolution towards a more spherical configuration although the S/N of the second epoch is not optimal ($\sim 150$) and we cannot draw robust conclusions from those measurements.

\section{Discussion} \label{sec:discussion}

In the previous sections, we presented the optical and NIR properties of SN~2022lxg. We now attempt to piece them together to place SN~2022lxg in the context of other luminous Type II SNe, with a view to shedding light on its progenitor system.

\subsection{The nature of SN~2022lxg}\label{subsec:22lxg}

\subsubsection{CSM interaction} \label{subsubsec:CSM}

There are several features in the data of SN~2022lxg that point to the fact that interaction between the ejecta and the CSM is the primary energy source for a large part of its evolution. Concerning the early phase, the very fast rise and the luminous ($V \lesssim -18.5$\,mag) peak of Type II SNe are usually attributed to such an ejecta/CSM interaction, where the kinetic energy of the ejecta is thermalised by the interaction shock and then radiated (see e.g. \citealt{Moriya2012,Andrews2018};~\pjpnospc). The flash-ionisation lines are also a clear spectroscopic indication \citep{Gal-Yam2014,Shivvers2015,Yaron2017,Dessart2017,Bruch2021,Bruch2023,Jacobson-Galan2024}, along with the blue featureless continua (confirmed by the early blue colours), as shocks produced during the interaction of the fast moving ejecta with the CSM heat the material. The broad-boxy profiles of H$\alpha$ is another signature \citep{Dessart2022}. Furthermore, the blue pseudo-continuum observed (blue-wards of 5700\AA in this case) after $\gtrsim +45$\,d, has been attributed to a forest of blended \ion{Fe}{II} lines provided by fluorescence in the inner wind or post-shock gas \citep{Foley2007,Chugai2009,Smith2009,Pastorello2015,Perley2022}, and its presence here suggests that CSM interaction is ongoing.

\citet{Dessart2023} considered a solar-metallicity 15 $\msun$ red supergiant exploding into circumstellar material, and provide a grid of synthetic spectra resulting from different initial parameters such as the CSM density and the progenitor's mass-loss rate. Although both the progenitor and explosion properties are likely to be different for SN~2022lxg, as the authors note, this would introduce only moderate variations and the overall conclusions presented should hold at a qualitative level. We attempted to find a match within this grid by focussing on our earliest spectrum ($+2.20$\,d), that contains the flash features.
We found one model that provided an adequate match to these lines `\textit{mdot1em3early\_nb5}', a standard RSG star embedded in a steady-state wind with a mass-loss rate of $\rm 1 \times 10^{-3}\,\msun\, \rm yr^{-1}$, 1.25 days after the shock breakout (with a CSM velocity of $50\,\rm km\,s^{-1}$). However, the model lacks the broad wings that we observe in the data, and shows P-Cygni profiles that we do not detect. We show the comparison in Fig. \ref{fig:Keck}.

\begin{figure*}
        \centering
        \begin{subfigure}[b]{1\textwidth}
            \centering
        \includegraphics[width=0.35 \textwidth]{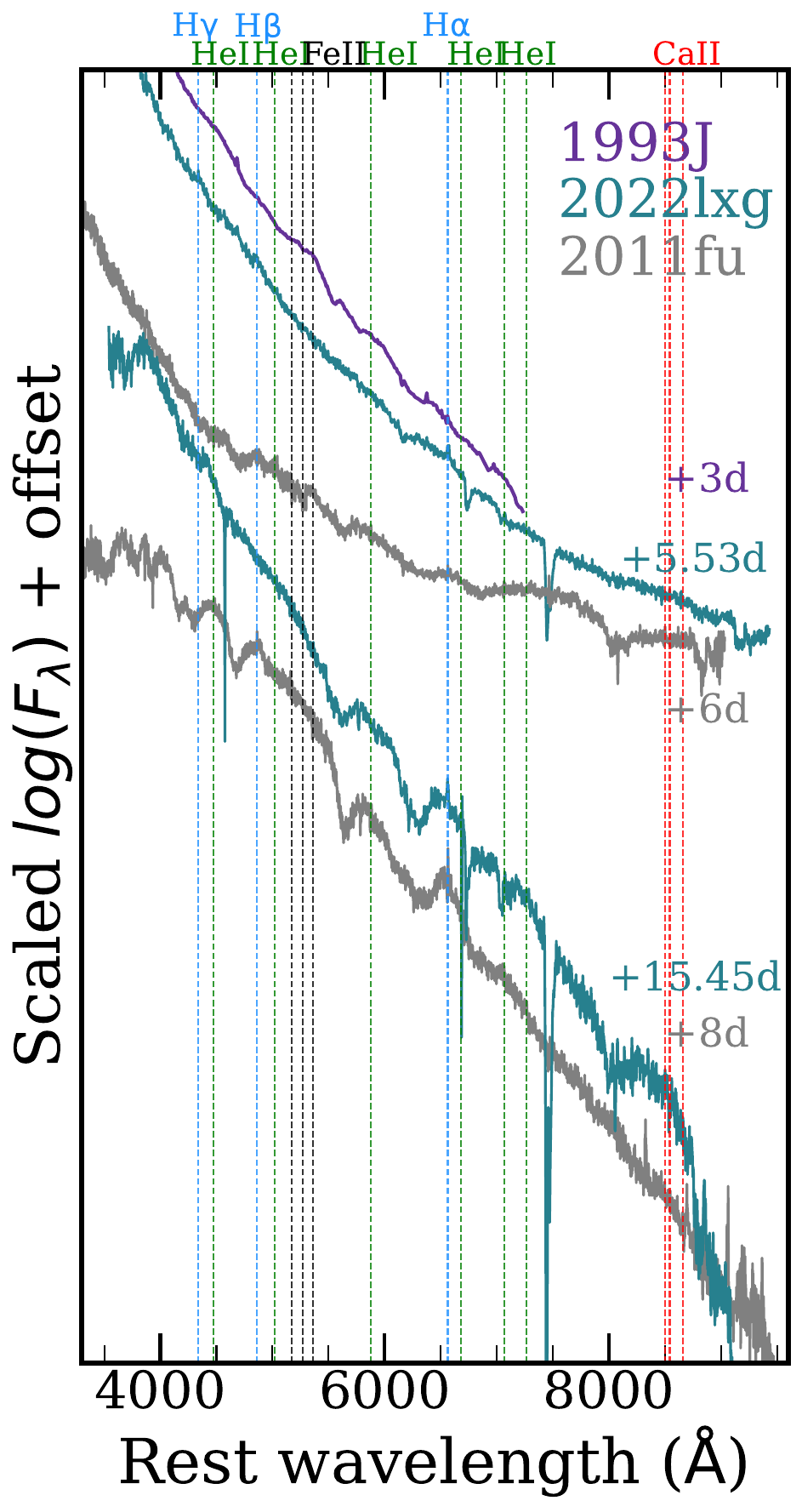}
        \includegraphics[width=0.32 \textwidth]{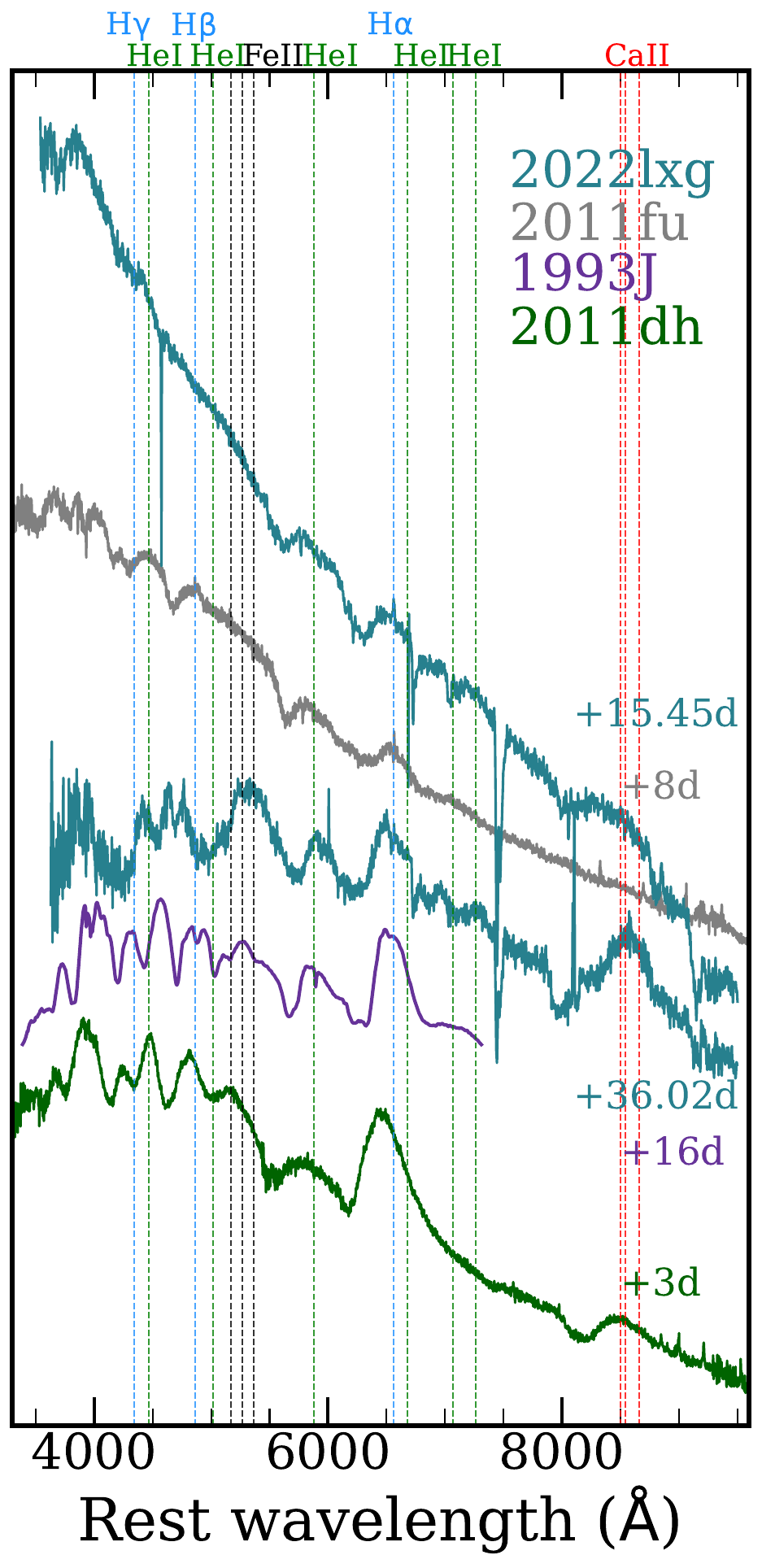}
        \includegraphics[width=0.32 \textwidth]{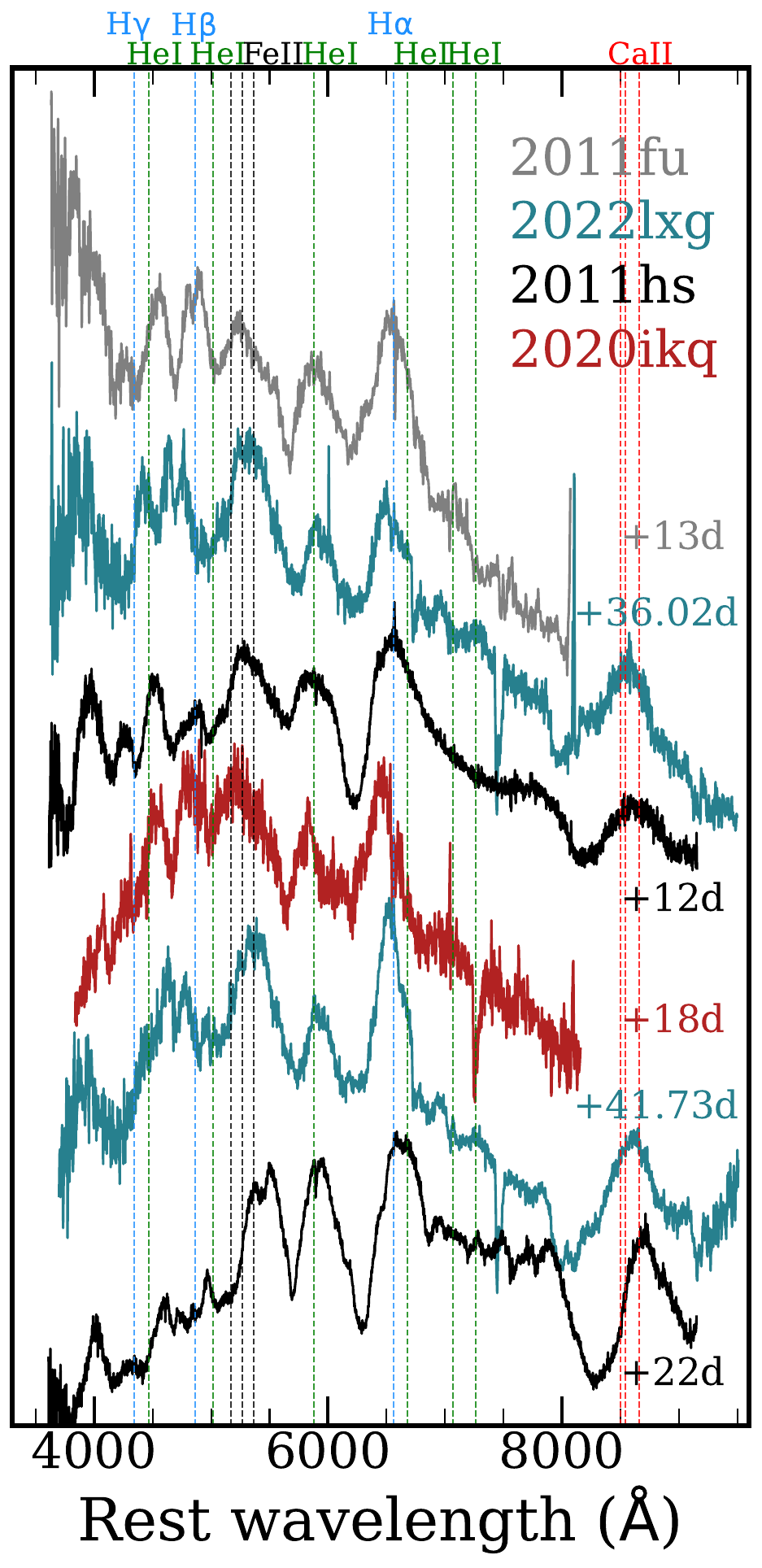}
        \end{subfigure}
        \caption{Spectral comparison of SN~2022lxg to other Type IIb SNe, throughout different phases of its evolution. Left panel: $\lesssim16\,\rm d$, middle panel: between $\sim (16-36)\,\rm d$, right panel: $\gtrsim36\,\rm d$. The y-axis in the left and middle panel is in log-scale while in the right panel it is linear. Emission lines are marked with vertical dashed lines. There is a clear resemblance between spectra of Type IIb SNe and those of SN~2022lxg, throughout different phases of its evolution.}
        \label{fig:spec_comp_IIb}
    \end{figure*}

\begin{figure*}
\centering
\includegraphics[width=0.45 \textwidth]{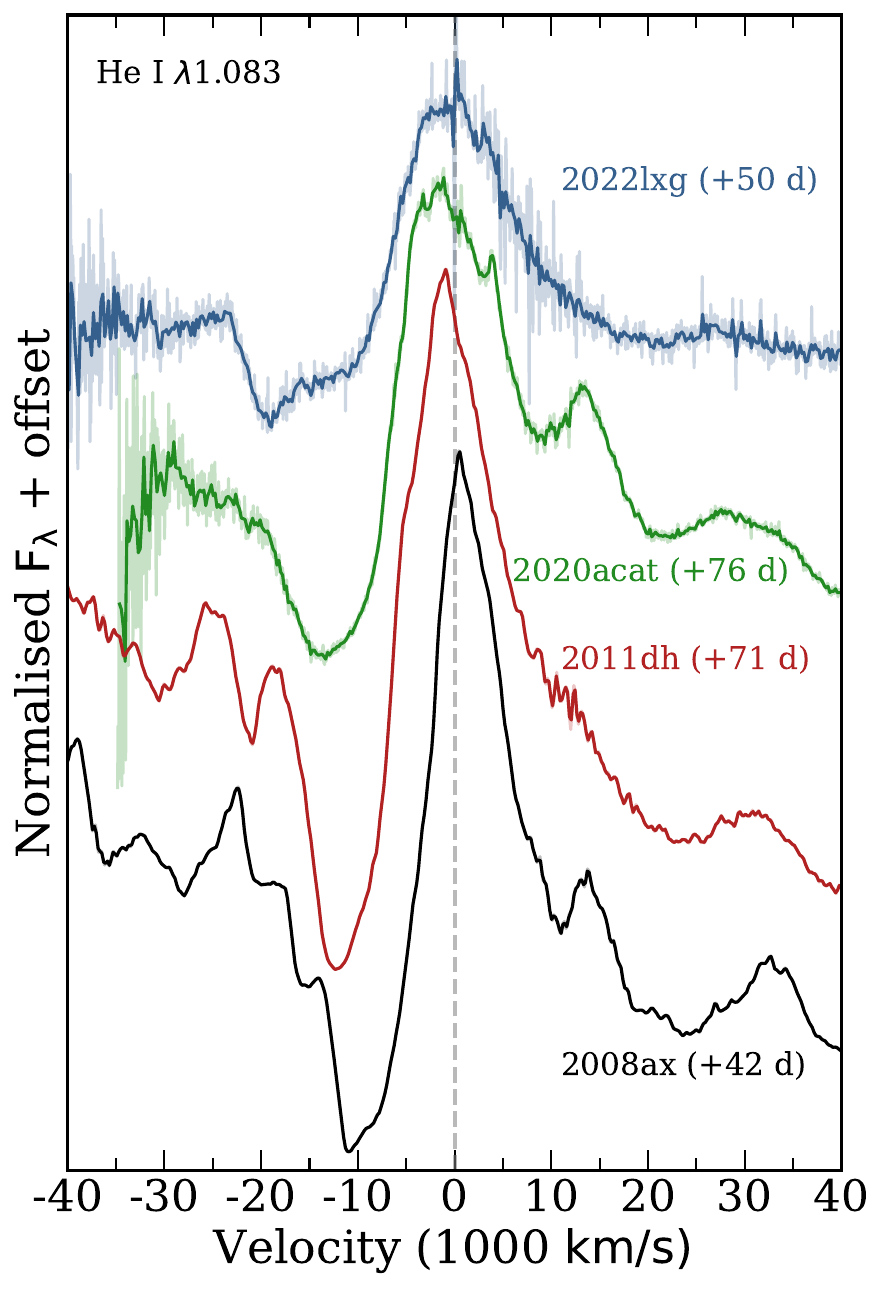}
\includegraphics[width=0.45 \textwidth]{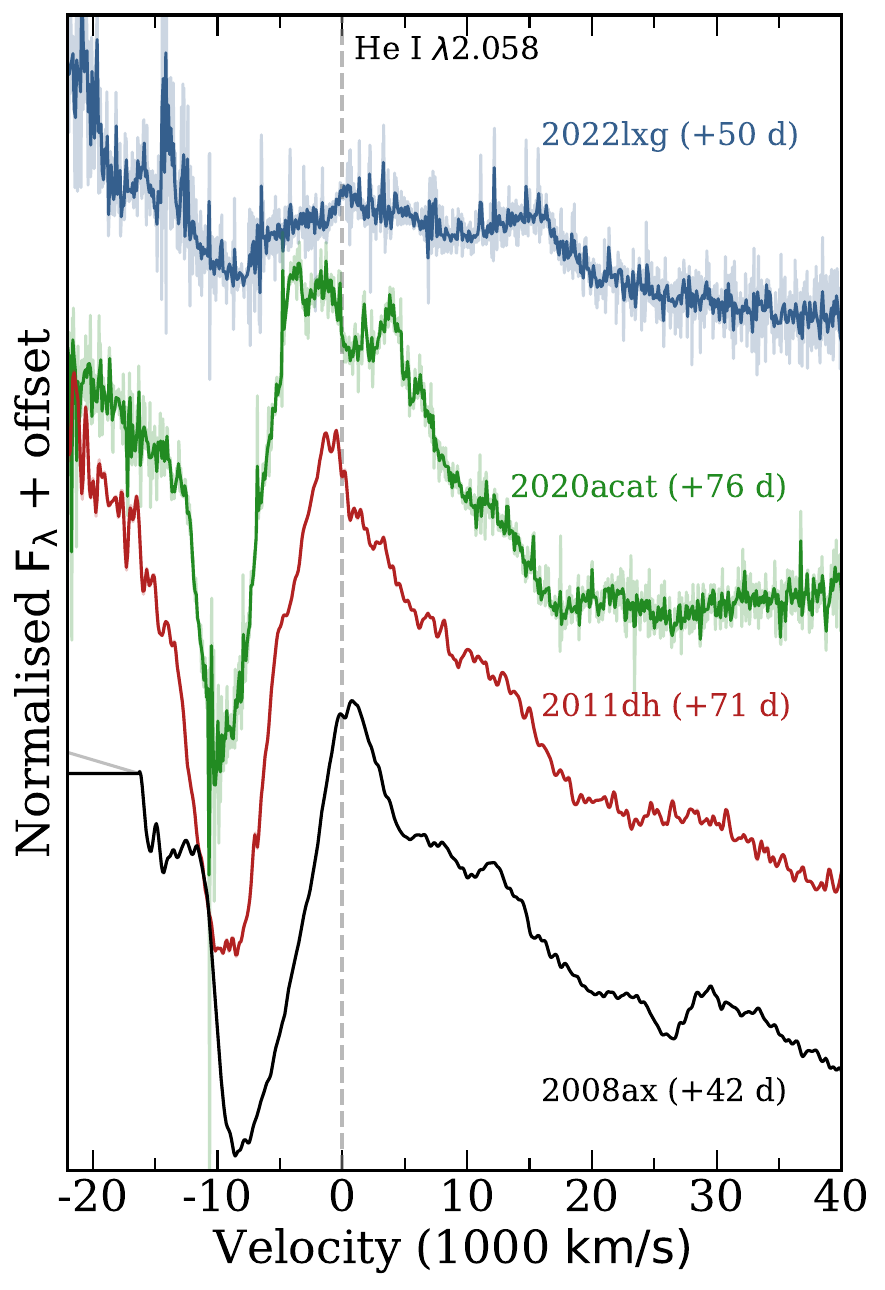}
\caption{SN~2022lxg \ion{He}{I} NIR lines in velocity space and comparison with other Type IIb SNe; SN~2020acat \citep{Medler2023}, SN~2011dh \citep{Ergon2015}, SN~2008ax \citep{Roming2009}. The spectra are binned to 5 \AA\, for visual purposes and the original spectra are plotted with lighter colours in the background. The grey vertical line shows the central wavelength. Left panel: \ion{He}{I} 1.0830 $\mu$m line. The spectra are normalised at $\sim 20\,000\,\rm km\,\rm s^{-1}$ red-wards of the central wavelength. Right panel: \ion{He}{I} 2.0581 $\mu$m line. The spectra are normalised at $\sim 35\,000\,\rm km\,\rm s^{-1}$ red-wards of the central wavelength.}
\label{fig:NIR_vel_comp}
\end{figure*}

During the phases between $+15$ to $+40$\,d, the colours redden quickly. However, around $+40$\,d several important changes seem to happen. The photosphere starts receding (see bottom panel of Fig. \ref{fig:BB}) and we see the cooling stop abruptly and the colours reach a plateau before slowly become bluer until $+65$\,d when cooling seems to take over again through to the end of our spectroscopic coverage.  Spectroscopically, the H$\alpha$ luminosity and FWHM start to decline (first and third panel of Fig. \ref{fig:ha_prop}) and the \ion{He}{I} and \ion{Fe}{II} lines start to fade as well. This is probably the result of the photosphere receding as the ejecta expand. As the temperature drops and the optical depth in the ejecta decreases, we start to see increasing emission from the CSM component and less from the ejecta. It is no surprise then, that this is when we see interaction dominate once again the colours (plateauing) and the spectra, most evident in the latter as H$\alpha$ becomes the most prominent line in the spectrum. Even though the luminosity of H$\alpha$ starts to decline after $+40$\,d, it declines slower than the continuum, which explains the sudden rise in the pEW (bottom panel of Fig. \ref{fig:ha_prop_LII}). Also the velocity of H$\alpha$ decreases due to interaction taking over. We also observe similar behaviour in the NIR hydrogen lines, where Pa$\beta$ clearly emerges in the $+40$\,d spectrum, but within ten days it becomes narrower and stronger with respect to the continuum.

\subsubsection{Similarities with transitional Type II SNe} \label{subsubsec:IIb_SNe}

As discussed in Sect. \ref{subsubsec:spec_evol}, during the first $\sim$ 15 days, we see a blue continuum, featureless or with very shallow broad lines, a result of the shock heated and shock ionised CSM. After that, the continuum starts cooling, hydrogen recombines and we start to see spectral lines likely originating in the ejecta. Along with the Balmer series, we also see strong \ion{He}{I} $\lambda5876$ and \ion{Fe}{II} ($\lambda\lambda$\,5169, 5267, 5363), appearing at the same phases and with similar shapes and FWHM ($\sim$10\,000 -- 15\,000\,km\,s$^{-1}$) as H$\alpha$. This is evidence that all these lines are coming from the ejecta and not from the interaction with the CSM, which means that the ejecta contain both hydrogen and helium. Furthermore, the evolution of those lines and the emergence of the narrower, strong H$\alpha$ at later times ($\gtrsim +35$\,d) together with slow moving ($\sim$150\,km\,s$^{-1}$)
helium suggests that the progenitor has undergone a degree of stripping of its outer layers.

This is reminiscent of a Type IIb SN (c.f. Sect. \ref{subsubsec:spec_evol}) so we investigate whether there is further resemblance to transitional Type II SNe. In Fig. \ref{fig:spec_comp_IIb} we compare SN~2022lxg to Type IIb SNe, at different phases. We find a clear similarity with various SNe IIb, both in the early phases ($\sim +15$\,d) and more so in the subsequent ones (between $+15$ to $+35$\,d), where the ejecta lines have clearly emerged. As the optical depth of the ejecta drops and the photosphere recedes around $+40$\,d (see bottom panel of Fig. \ref{fig:BB}), the interaction with the CSM becomes dominant ($\gtrsim +35$\,d onwards) leading to a spectroscopic change from a SN IIb, to spectra that resemble more an interacting SN II.

There are questions arising from the apparent similarity with Type IIb SNe. We do not see two peaks in the light curves typical in many SNe IIb, with the second peak occurring around three weeks after the first (SN~1993J; \citealt{Filippenko1993,Wheeler1993,Richmond1994,Woosley1994}, SN~2011dh; \citealt{Arcavi2011,Bersten2012,Ergon2015}, SN~2011fu; \citealt{Morales-Garoffolo2015}, SN~2011hs; \citealt{Bufano2014}, SN~2013df; \citealt{Morales-Garoffolo2014}, although there are SN IIb that do not show two distinct peaks (e.g. SN~2008ax; \citealt{Roming2009}, SN~2020acat; \citealt{Medler2022}). A lack of the first shock breakout cooling peak, implies that the progenitor must have been relatively compact. The first luminous peak is thought to result from the shock breakout cooling on the stellar surface, while the secondary peak is assumed to be powered by $^{56}$Ni-heating. The $^{56}$Ni mass that we infer for SN~2022lxg is more than 10$\times$ lower (Sect. \ref{subsec:phot_analysis}) than the mean value of $0.066 \pm 0.006 \msun$ inferred from a sample of 20 SNe IIb \citep{Rodriguez2023}. However, there are increasing indications that SNe IIb with a relatively small amount of $^{56}$Ni exist as a population of under-luminous SNe IIb (\citealt{Ouchi2021} and see discussion in \citealt{Maeda2023a}) and/or as rapidly evolving transients \citep{Ho2023}. In the latter study, they present a population of rapidly evolving SNe IIb without two light curve peaks.

Thus, a plausible scenario might be that SN~2022lxg is intrinsically similar to these faint and rapidly evolving SNe IIb, but that its properties are modified by the presence of CSM. The lack of two distinct peaks in the light curves of SN~2022lxg might be due to a combination of the following reasons: i) the explosion of a compact progenitor leading to a non-evident shock breakout cooling peak, and/or ii) the low amount of $^{56}$Ni leading to a non-prominent secondary peak (with the latter also contributing to the fast evolution of the SN); iii) the interaction of the ejecta with the CSM gives rise to the luminous peak (as well as the fast rise compared to typical SNe IIb) and further blends the two distinct peaks.
We note that the secondary peak can sometimes be fainter than the first one, an effect that is more evident in the bluer bands (e.g. SN~1993J, SN~2011fu). It is possible that the \say{plateau} we see in the $g$-band during the peak (see Sect. \ref{subsubsec:bb_lc} and inset of Fig. \ref{fig:photometry}), and the corresponding lack of it in the redder bands, is a manifestation of the above blended with the luminosity provided by the strong interaction. Furthermore, during the $g$-band \say{plateau} epochs, we note a small drop in brightness and a subsequent rise. That is more evident in the $r$-band. This might indeed be the manifestation of the shock breakout cooling, smoothed by the power provided by the early interaction.

We note that the \ion{He}{I} $\lambda7065$ line, routinely observed in Type IIb SNe is never robustly detected in SN~2022lxg at any epoch, although a very weak, blueshifted feature is present in the spectra from $\sim$54\,d. It is possible that the CSM interaction from $\sim +35$\,d onwards suppresses the emergence of other lines and marks the transition to spectra resembling those of interacting SNe II.

We compare the \ion{He}{I} 1.0830 and 2.0581 $\mu$m lines of SN~2022lxg at +50\,d, the latest near-IR epoch in our dataset, to other type IIb SNe in Fig. \ref{fig:NIR_vel_comp}. 
SN~2022lxg has the broadest \ion{He}{I} 1.0830 $\mu$m line with multiple absorption components (see also Fig. \ref{fig:NIR_vel}), that also appear to be present in SNe~2011dh and 2008ax. At the spectral resolution of the comparison objects, it is not possible to tell whether there is any unshocked, slow moving material ahead of the ejecta as is the case for SN~2022lxg. At this epoch, the \ion{He}{I} 1.0830 $\mu$m line is more centrally peaked in the comparison objects than in SN~2022lxg, which appears to already have the beginnings of a flat-top; this may be linked to its faster evolution. Indeed, both SNe~2008ax and 2020acat went on to develop flat-topped profiles at epochs $>100$\,d \citep{Roming2009,Medler2023}. 
The \ion{He}{I} 2.0581 $\mu$m line of SN~2022lxg is strikingly different compared to the other SNe, being broad and shallow. It is possible that it is blended with Br$\gamma$, while the \ion{H}{I} lines being less prominent in the comparison SNe. The broad multi-component profiles further underscore the effects of ongoing CSM interaction \citep{Dessart2022}.

\subsubsection{Influence of the host environment} \label{subsubsec:LSNII}

As previously noted, SN~2022lxg shows photometric and spectroscopic similarities to the \pjp sample. Differences between these objects might be explained by variations in the amount and configuration of the CSM that lead to features appearing or disappearing at different times. Here we investigate whether this group of SNe also share similar environmental properties. In particular, we focus on the general lack of metal lines in the spectra (\S \ref{subsubsec:spec_evol}). Although it is possible that these may be suppressed by ongoing CSM interaction, one would have to tailor this effect across all objects in this group and across all epochs. The large projected host offset ($\sim$4.6\,kpc) of SN~2022lxg, and blue colours prompt us to consider whether low implied metallicity plays a role. This is further underscored by previous studies, albeit on Type IIP SNe, that noted a correlation between low metallicity hosts (or explosion sites) and SN properties \citep[e.g.,][]{Polshaw2016,Taddia2016,Gutierrez2018}.

\begin{figure}
  \centering
  \includegraphics[width=0.5 \textwidth]{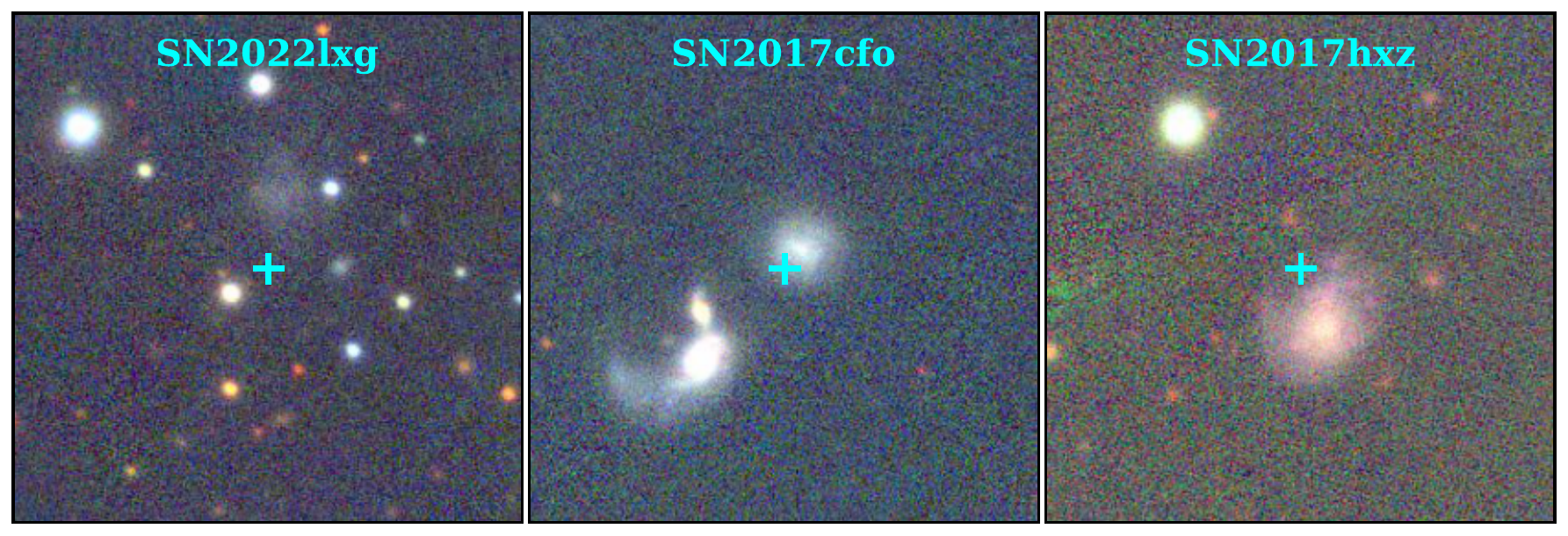}
  \includegraphics[width=0.5 \textwidth]{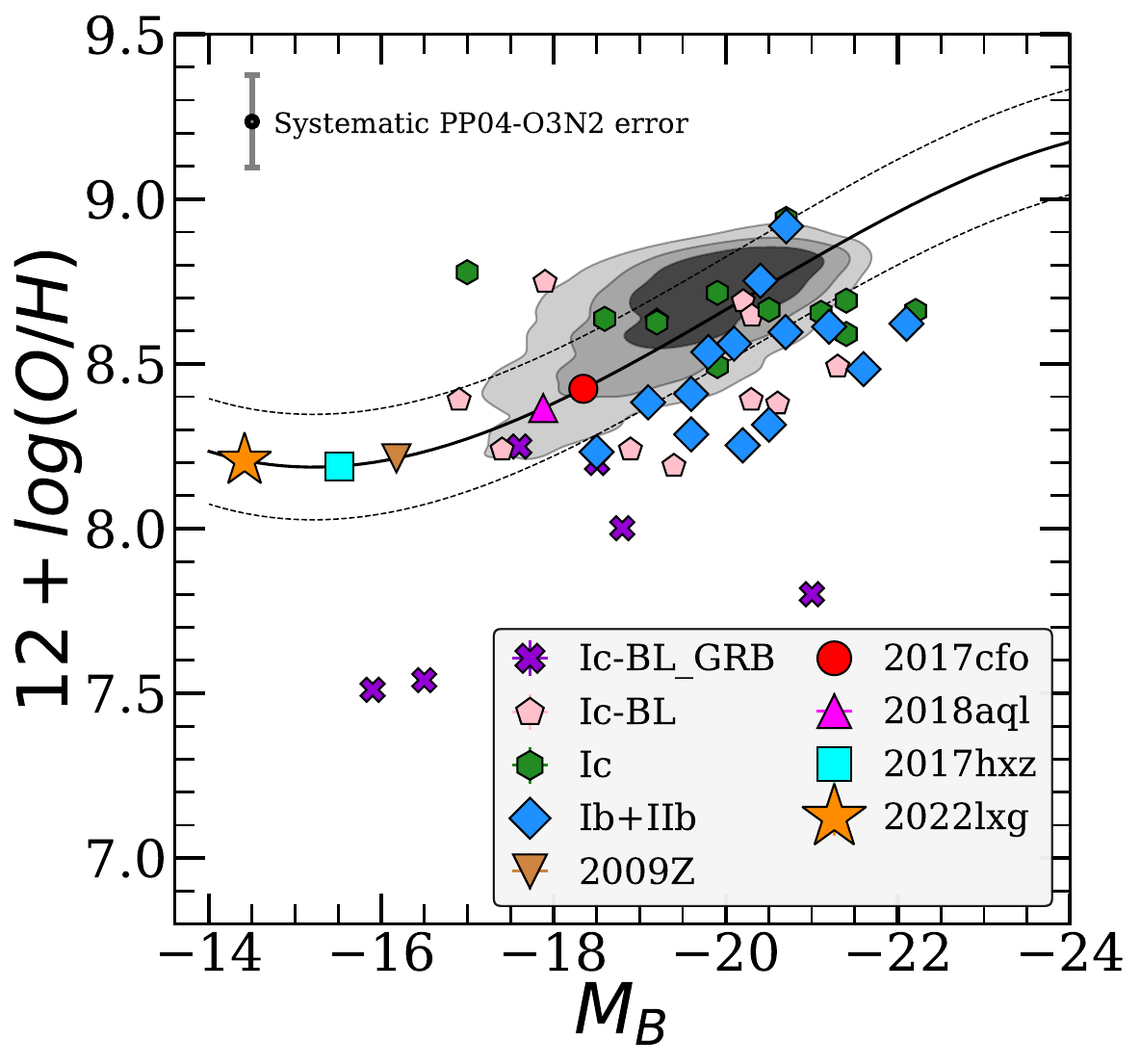}
\caption{Metallicity of host galaxies as a function of host absolute $B$ magnitude. We have used the $L-Z$ \citet{Tremonti2004} relation to estimate the metallicity of the (potential) host of SN~2022lxg, as well as some SNe from the \pjp sample along with SN~2009Z, a Type IIb SN in a LSB galaxy. Contours show 50\%, 75\%, and 90\% containment of the kernel density estimate of the SDSS DR2 sample presented in \citet{Tremonti2004}. The metallicities where converted to the scale of PP04-O3N2 (solid line) including 1$\sigma$ uncertainties (dashed lines), using the empirical calibrations of \citet{Kewley2008}. The representative systematic error of 0.14 dex for the PP04-O3N2 scale is shown in the top left corner. We also include the SNe from the sample of \citet{Modjaz2011} in the comparison whose oxygen abundance was measured via strong-line diagnostics. The metallicity of the LSNe II hosts are consistently low compared to the sample, with the fastest events (SN~2022lxg and SN~2017hxz) having the lowest host metallicities. The tope panels show a mosaic of $gri$ coloured cutouts of the hosts of SN~2022lxg and two SNe from the \pjp sample. A cyan cross denotes the position of the SN. From left to right we show: SN~2022lxg, SN~2017cfo that is spectroscopically the most similar to SN~2022lxg, and SN~2017hxz that is photometrically the most similar to SN~2022lxg.}
\label{fig:metals}
\end{figure}

In order to investigate a potential correlation with the metallicity of their hosts, we use archival photometry and the $L-Z$ \citet{Tremonti2004} relation, in order to make some rough metallicity estimates. For SN~2022lxg, we use the archival PS $g$-band magnitude of the potential host (see Sect. \ref{subsec:host}), and after converting it to the $B$-band ($\rm M_{B}=-14.41\pm0.10$\,mag, using the \citealt{Jordi2006} colour transformations), we find 12 + log(O/H) = 7.90$\pm$0.03, that is $Z=(0.16\pm0.01) \Zsun$ (assuming 12 + log(O/H) = 8.69 for the Sun; \citealt{Prieto2001}). The estimated metallicity is very low compared to the SDSS DR2 sample presented in \citet{Tremonti2004}. We only find archival $g$-band photometry for three out of six SNe of the \pjp sample, namely SN~2017cfo (that is spectroscopically the most similar to SN~2022lxg), SN~2017hxz (that is photometrically the most similar to SN~2022lxg) and SN~2018aql. We show the results in Fig. \ref{fig:metals}. We also include SN~2009Z, a Type IIb SN in a low surface brightness (LSB) galaxy \citep{Zinn2012}. Furthermore, we include the SESNe from the sample of \citet{Modjaz2011} in the comparison whose oxygen abundance was measured via strong-line diagnostics in the actual position of the SNe. Following \citet{Modjaz2011}, we convert the oxygen abundance estimates derived with the empirical $L-Z$ \citet{Tremonti2004} relation to the \citet{Pettini2004} strong-line diagnostic, using the empirical calibrations of \citet{Kewley2008}. All the LSNe II have low metallicites compared to the bulk of the SDSS galaxies, with the faster ones (SN~2022lxg and SN~2017hxz) showing remarkably low metallicites. However, as demonstrated by \citet{Modjaz2011}, the nuclear metallicity when derived from the SN host luminosity is not a good proxy for the local oxygen abundance of the environments of SNe: the local metallicities are often lower than the inferred central ones, something evident in Fig. \ref{fig:metals}. Hence, the true metallicities at the locations of those SNe might be even lower. 

In spite of all the caveats in estimating metallicities from imaging, taken at face-value, it appears that the hosts of SNe such as SN~2022lxg are of low metallicity, which would explain some of the properties of this group. Finally, the apparent similarities of 22lxg-like transients with SNe IIb is in line with the low metallicity environment as studies of SN host environments that found SNe IIb to occur in distinctly different settings compared to other CC-SNe, typically in metal-poor regions with relatively low star formation rates \citep{Galbany2018} or showing a significant excess in dwarf galaxy hosts \citep{Arcavi2010}. However, we note that other studies, based on comparable sample sizes \citep[e.g.,][]{Pessi2023a} find that SNe IIb occur in higher metallicity environments compared to regular Type II SNe.

\subsection{CSM configuration and a potential progenitor}\label{subsec:prog}

We attempt here to make some rough estimates on the CSM mass and the mass-loss history of SN~2022lxg by making some basic assumptions; first, we assume homologous expansion and an ejecta velocity equal to 20\,000\,\,km\,s$^{-1}$ (c.f. Sect. \ref{subsec:spec_analysis}). In order to cover a range of possible steady-state wind velocities of progenitor stars, we assume a value of 75\,\,km\,s$^{-1}$ typical value for a yellow supergiant (YSG) \citep{Humphreys2010,Humphreys2023}, with $\pm$ 25\,km\,s$^{-1}$ serving as our uncertainty estimates, leading to a CSM velocity range typical for RSG and YSG stars \citep{Dessart2023}. The duration of the flash-ionisation features ($\sim$ 8.24 days in our case, see Sect. \ref{subsubsec:ff}) can indicate the extent of the CSM, assuming that the lines disappear when the SN ejecta sweep up the CSM where they originate. Based on our ejecta velocity estimate, that would be ${\rm\sim1.42\times10^{15}}$\,cm, which is where the peak of the blackbody radius lies (${\rm\sim1.37\times10^{15}}$\,cm). Assuming the aforementioned CSM velocities, that would suggest that the mass-loss started $\sim 6.0^{+3.0}_{-1.5}$~ years before the explosion. If we also assume a mass-loss rate ($\rm \frac{dM}{dt}=\mdot$), then we can infer the CSM mass. The models of \citet{Dessart2023} that we compare to due to their resemblance to SN~2022lxg (see Sect. \ref{subsubsec:CSM}) have $\rm \mdot=1\times10^{-3}\,\msun\,yr^{-1}$. If we use the relations between the mass-loss rate and duration of the flash-ionisation features ($\rm {t_{IIn}=376.2\mdot}$) presented in \citet{Jacobson-Galan2024}, we get a ${\rm \mdot\simeq2.2\times10^{-2}\,\msun\,yr^{-1}}$, while if we use our model fits (see Sect. \ref{subsubsec:mosfit}), we can estimate the mass-loss rate at the inner radius of the CSM as $\rm \mdot=4\pi v_{CSM} R_{0}^{2} \rho_{0}$, returning a ${\rm \mdot\simeq4.3\times10^{-3}\,\msun/yr}$. We note however, that this calculation assumes spherical symmetry. These three different values for the mass-loss rate lead to a CSM mass ($\rm M_{CSM}$) of $0.006\msun$, $0.132\msun$, and $0.026\msun$ respectively, with a mean value of $\rm \overline{M}_{CSM}\simeq0.055\msun$. We stress however, that these estimates result from assumptions that may not hold over the timescales probed by our dataset. Nevertheless, they appear to be in line with values found in other studies.

\begin{figure}
  \centering
  \includegraphics[width=0.45\textwidth]{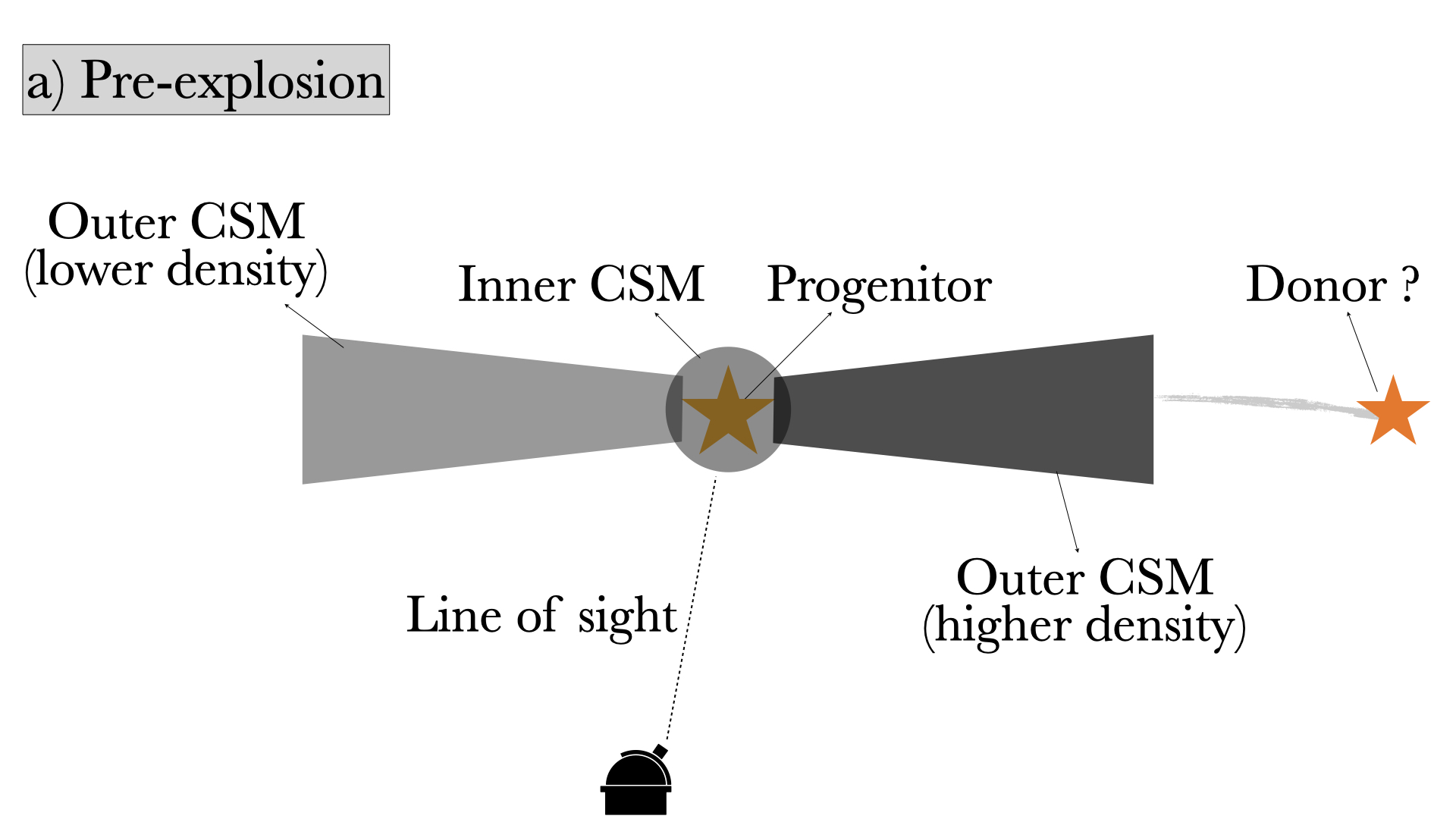}
  \hrule height 0.4pt width 0.5\textwidth
  \includegraphics[width=0.45\textwidth]{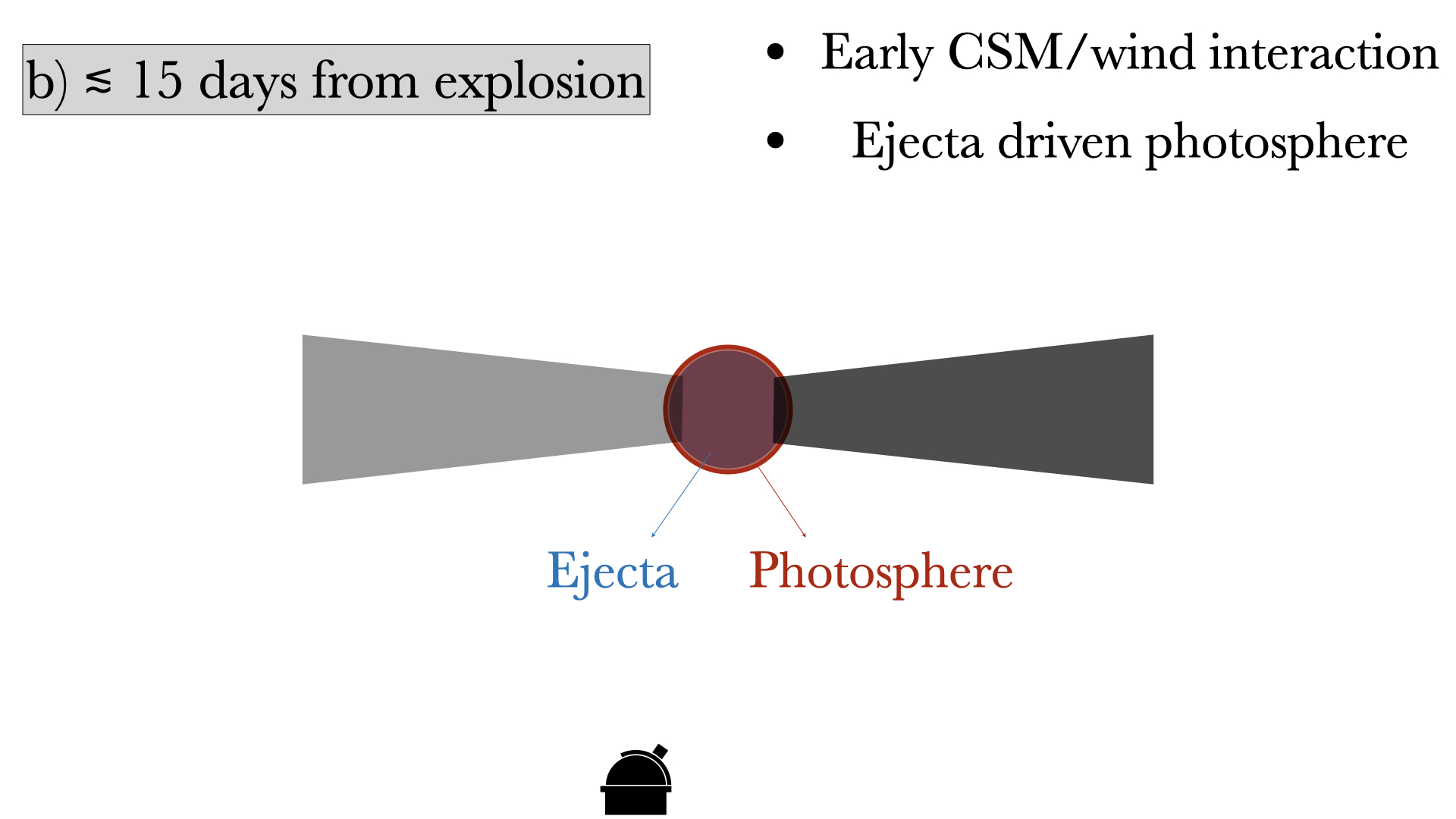}
  \hrule height 0.4pt width 0.5\textwidth
  \includegraphics[width=0.45\textwidth]{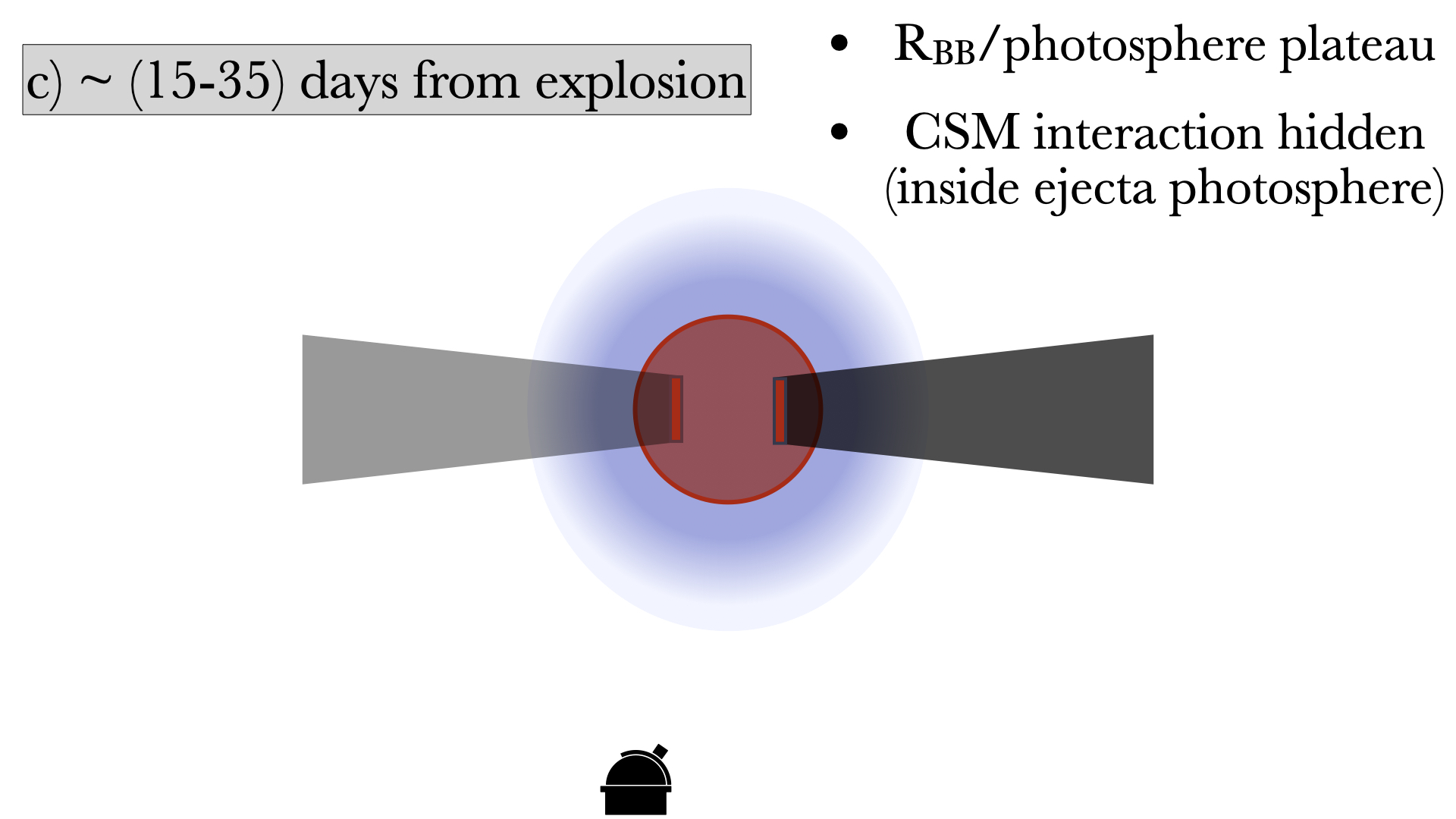}
  \hrule height 0.4pt width 0.5\textwidth
  \includegraphics[width=0.45\textwidth]{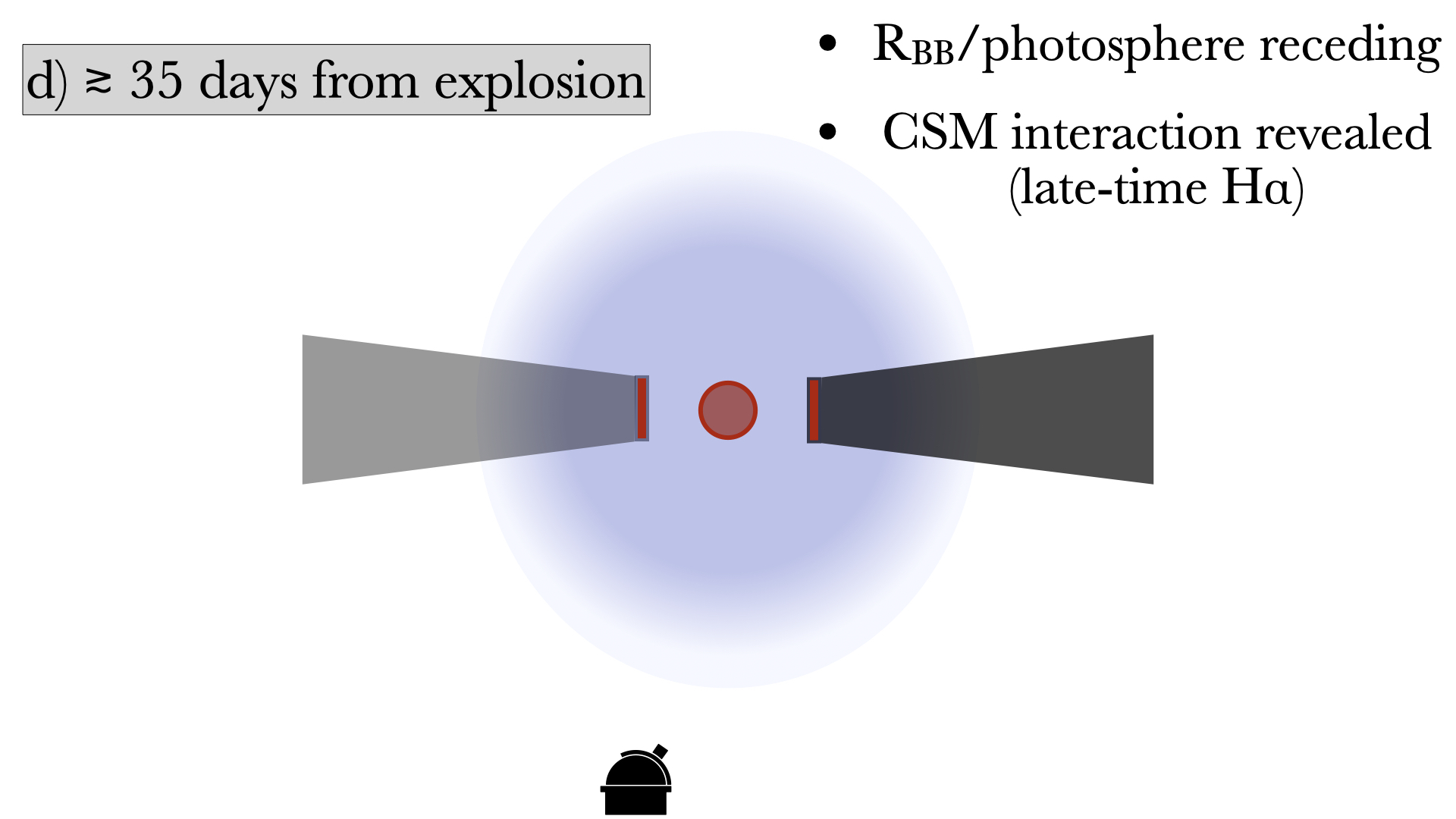}
  \hrule height 0.4pt width 0.5\textwidth
  \vspace{0.2cm}
\caption{An example of a possible configuration that gave rise to SN~2022lxg; we envisage a disc-like CSM (with a potentially azimuthally asymmetric density distribution) with the line-of-sight at low inclination. (a) Before the explosion; we suggest that the CSM closest to the progenitor is denser, and could have been formed by material transferred from a donor combined with stellar winds. (b) the fast-moving ejecta crash into the inner dense CSM giving rise to the observed flash-ionisation features and the luminous peak. Within $\sim$ 10--15\,d this denser CSM is completely swept-up by the ejecta. (c) Between 10--35\,d, the ejecta encounter the outer, hydrogen-rich, disc-like CSM and encompass it. Broad lines from the free-expanding, cooling ejecta emerge while the radius of the composite photoshpere plateaus. (d) After $\sim$35\,d, the photoshpere recedes, the optical depth of the ejecta drops, and the power comes from the interaction. Narrower H$\alpha$ emerges and dominates the spectrum.
}
\label{fig:cartoon}
\end{figure}

However, apart from the interaction that gives rise to the flash-ionisation features, another phase of interaction appears to be taking place 35--40 days post-explosion (Sect. \ref{subsubsec:CSM}). As the optical depth of the ejecta drops and the photosphere recedes, the interaction with the CSM becomes more evident. One possibility might be that there is a change in the CSM density structure. Such a two-component CSM with different density structures has been previously invoked to explain the properties of SN~2018ivc, a Type IIb/IIL SN; that undergoes an extreme case of case C binary mass transfer resulting in a SN intermediate between SNe Ib/c and SNe IIP \citep{Maeda2023}. The transitional properties of SN~2022lxg may well be a variant of such a scenario.

Another option for creating a two-component CSM would be a pre-explosion mass loss episode i.e., with an elevated mass-loss rate rather than a steady-state wind, that could be responsible for late signs of interaction. Making the same assumptions as before, but this time using the starting time of the late interaction signs ($\sim+40$\,d), we can probe where the inner layer of this potential eruption ejected CSM lies. We find that the material would lie at ${\rm\sim6.9\times10^{15}}$\,cm and assuming a conservative eruption velocity of $2000\pm1000$\,km\,s$^{-1}$ (observed velocities show even higher values, e.g. 4 SLSN-I show velocities spanning 3\,000$-$4\,400\,km\,s$^{-1}$; \citealt{Gkini2024}, or the 1843 eruption of $\eta$ Carinae $\sim 3000-6000$\,km\,s$^{-1}$; \citealt{Smith2008}), the eruption should have occurred 
$\sim 1.1^{+1.1}_{-0.4}$~ years before the explosion. 
Observations of precursor emission to core-collapse SNe are a direct means of probing the enhanced mass-loss of the progenitor star during its final moments, especially if it is a violent eruption \citep{Strotjohann2021}. We requested forced photometry from ZTF, ATLAS, and ASAS-SN \citep{Hart2023} for the pre-explosion epochs of SN~2022lxg in order to search for potential precursor emission. Forced photometry was combined into large \say{seasonal} bins, defined by the longest unbroken runs of observations in a given observing season. The fluxes were combined using an inverse-variance weighting scheme, with sigma clipping applied to remove discrepant individual measurements in some surveys. New upper limits were recomputed based on the stacked fluxes and their uncertainties, with 5$\sigma$ limits being used in practice. The results are presented in Fig. \ref{fig:historic}. No significant pre-explosion emission is detected for SN~2022lxg. The forced photometry rules out long-lasting precursors with absolute magnitude $\rm M\lesssim-11$~mag that might have happened four years before the explosion. Data prior to that are not deep enough to rule out outbursts similar to those of other SNe. All the above considered, an eruption or sudden outburst seems less likely although cannot be entirely ruled out.

As previously noted, the rather low photospheric expansion velocity derived from the blackbody fits ($\sim7\,000$ km\,s$^{-1}$) is much lower than ejecta velocity derived from the spectroscopic lines and the light curve model fits. Assuming a spherically-symmetric CSM configuration and given the linear increase in the blackbody radius, the photosphere should lie nearly at the interacting region, that is, at the edge of the ejecta. This discrepancy between $v_{\rm ph}$ and $v_{\rm ej}$, can potentially point to asphericity or line-of-sight (LOS) effects. Assuming a disc-like CSM around the progenitor if our LOS is along the low-density region and does not really intersect the dense CSM, we would expect broad lines (essentially unshocked ejecta) while the power can be provided by the off-LOS interaction (where the ejecta are decelerated, and $v_{\rm ph}$ can be much lower than the original unshocked ejecta velocity. Furthermore the imaging polarimetry at $+17$\,d and $+36$\,d (epochs at the start and the end of the cooling phase respectively) reveals that SN~2022lxg is intrinsically polarised to a $p\sim(0.5-1.0) \%$ level (see Fig. \ref{fig:impol}). This points to the fact that there is some asphericity in the system. 
As the ejecta expand and the photosphere recedes, the temperature drops and the optical depth is decreasing so we start seeing more and more emission from the CSM component and less from the ejecta. The fact that H$\alpha$ gradually starts getting centred after the crucial phase of $35-40$\,d (while it is blueshifted until this point), means that even if the CSM is aspherical, it probably has a symmetric structure around the explosion. This could also point to an asymmetric (and potentially disc-like) CSM around the progenitor. Numerous examples of asymmetric CSM abound in the literature \citep[e.g.,][]{Hoffman2007,Chornock2011,Smith2015,Mauerhan2015a,Mauerhan2017,Pursiainen2022,Reynolds2025a}.

It has been suggested that strong late-time H$\alpha$ profiles observed in some Type II SNe, arise due to late-time interaction with a disc-like CSM, when the photosphere recedes. Such profiles are sometimes redshifted (e.g. PTF11iqb; \citealt{Smith2015}) or blueshifted (e.g. SN~1998S; \citealt{Leonard1999}). \citet{Smith2015} suggest (their Fig. 10) that such disc-like CSM might arise from binaries with an azimuthally asymmetric density distribution around the disc/torus, with higher densities on the receding side. Such a disc/torus configuration was observed in the mass-transferring eclipsing binary RY Scuti \citep{Smith2002}. If one of the stars explodes as a SN in such a system, a high-inclination observer would observe blueshifted or redshifted late-time H$\alpha$, depending on whether the LOS is towards the higher density or the lower density part of the CSM disc respectively. An observer at a low inclination (like the one we suggest that SN~2022lxg is viewed from) would see a narrower and more symmetric line profile. This is in agreement with the late-time centring of the H$\alpha$ profile. Following \citep{Smith2015}, we present a schematic visualisation of a plausible physical evolution of SN~2022lxg in Fig. \ref{fig:cartoon}.

In summary, taking all the clues into account, we find that a plausible scenario for explaining the salient properties of SN~2022lxg and related objects could be to use our findings from Sect. \ref{subsubsec:LSNII} as our starting point. Previous studies of massive binary evolution have shown an inverse correlation between the size of massive stars and metallicity \citep[e.g.,][]{Brott2011,Georgy2013,Schootemeijer2019}.
Thus if we consider mass transfer in a close binary, required to create a CSM with the properties as described above, then it is likely that mass transfer via Roche lobe overflow occurs only late in the evolution (case C) as the progenitor is too compact during earlier evolutionary stages. The resulting CSM would then naturally have a preferred axial symmetry. With this framework, the rapid rise to peak brightness, the colour evolution, and the spectral behaviour can be qualitatively explained. Furthermore, the apparent rarity of such objects also follows, although larger sample sizes are warranted.

\section{Conclusions} \label{sec:conclusion}

We have presented the analysis of the optical and NIR properties of SN~2022lxg, a peculiar, bright ($\rm M_{g}=-19.41$\,mag at peak), and fast evolving SN at $\sim$ 96.6 Mpc. Based on our analysis we find the following:

\begin{enumerate}[label={\arabic*.}]
\item The most probable host galaxy is a faint ($\rm M_{r}=-15.2$) and diffuse source (WISEA J191523.71+481938.5) 4.58 kpc NW from the location of the SN.
\item The SN is luminous and peaks at $\rm M_{g}=-19.41$\,mag. The rise and the decline are relatively fast, 7.6 days from explosion (for which we put very tight constraints) to peak, and a decline of 3.48 $\pm$ 0.26\,mag\,$\rm (50\,d)^{-1}$ (both in the $g$-band). The SN slows down around the peak epochs, where we might see a hint of the shock breakout cooling in the light curves, blended with the power from the interaction (responsible for the luminous peak).
\item The spectral evolution can be divided into three phases:
\begin{itemize}
    \item Until 10 days post explosion, the SN showed very blue continua (T$>15\,000$\,K) with flash-ionisation features of hydrogen and \ion{He}{II} until $\sim$ $+8$\,d.
    \item Between roughly $+10$ to $+35$\,d, the SN cools and broad ($\sim2\times10^{4}\rm\,km\,s^{-1}$) lines appear, identified as hydrogen, helium, a strong iron complex around 5300 \AA\, and the \ion{Ca}{II} NIR triplet. At this phase there are striking similarities with early spectra of SNe IIb.
    \item After $\sim$ $+35$\,d, a previously blueshifted H$\alpha$ gets centred to the rest wavelength, becomes narrower, and its pseudo-equivalent width rises sharply, dominating the spectra until $+80$\,d, when our spectroscopic follow up stops as the SN is already too faint (m$_{\rm g}\sim20$\,mag).
\end{itemize}
\item Using various standard methods and light curve model fits, we find that SN~2022lxg must have a low amount of $^{56}$Ni ($\lesssim0.013\msun$). Several signs point to interaction between the ejecta and circumstellar material (CSM) such as the luminous peak (despite the very low $\rm ^{56}Ni$ mass), the fast rise, the late ($\gtrsim +35$\,d) spectral transitioning, and narrow ($\sim 100$\,km\,s$^{-1}$) \ion{He}{I} 1.0830 $\mu$m P-Cygni lines revealing unshocked helium CSM and suggesting partial stripping of the progenitor.
\item A discrepancy between the photospheric and the ejecta velocity (7\,000 compared to 20\,000 $\rm\,km\,s^{-1}$) might potentially point to line-of-sight effects and asphericity. This is further confirmed by two epochs of imaging polarimetry in the $V$ and $R$ filters, that show intrinsic polarisation of $p\sim(0.5-1.0) \%$. 
\item SN~2022lxg shares many similar properties with a sample of LSNe II, such as higher luminosity, fast decline, blue colours, lack of persistent narrow lines, broad H$\alpha$ emission, \ion{He}{I} 5876\,\AA\, emission, and weak or non-existent H$\alpha$ absorption and metal lines. There is tentative evidence that the properties of these SNe might be correlated with the low metallicity of their host galaxies, however the numbers are still very low in order to draw statistically robust conclusions. 
\item We can reconcile most of the observed properties of SN~2022lxg by invoking late mass transfer in a close binary system that gives rise to a disc-like CSM oriented such that our line-of-sight does not directly traverse it. As the optical depth of the ejecta drops and the photosphere recedes, the interaction with the CSM becomes dominant, leading to a spectroscopic change from a SN IIb, to spectra that resemble more akin to an interacting SN II.

\end{enumerate}

\begin{acknowledgements}
We thank the anonymous referee for the helpful comments that improved the manuscript. We thank T. Nagao for many interesting discussions and providing the imaging polarimetry data presented in \S \ref{subsec:pola_analysis}. R.K. also acknowledges discussions with L. Dessart. P.C., R.K., and T.L.K acknowledge support via the Research Council of Finland (grant 340613). C.P.G. acknowledges financial support from the Secretary of Universities and Research (Government of Catalonia) and by the Horizon 2020 Research and Innovation Programme of the European Union under the Marie Sk\l{}odowska-Curie and the Beatriu de Pin\'os 2021 BP 00168 programme, from the Spanish Ministerio de Ciencia e Innovaci\'on (MCIN) and the Agencia Estatal de Investigaci\'on (AEI) 10.13039/501100011033 under the PID2020-115253GA-I00 HOSTFLOWS project, and the program Unidad de Excelencia Mar\'ia de Maeztu CEX2020-001058-M. M.P. acknowledges support from a UK Research and Innovation Fellowship (MR/T020784/1). T.L.K acknowledges support from the Turku University Foundation (grant no. 081810). S.S. is partially supported by LBNL Subcontract 7707915. K.M. acknowledges support from the Japan Society for the Promotion of Science (JSPS) KAKENHI grant JP24KK0070 and 24H01810. The work is partly supported by the JSPS Open Partnership Bilateral Joint Research Projects between Japan and Finland (JPJSBP120229923). T.K. acknowledges support from the Research Council of Finland project 360274. Y.-Z.C. is supported by the National Natural Science Foundation of China (NSFC, Grant No. 12303054), the National Key Research and Development Program of China (Grant No. 2024YFA1611603), the Yunnan Fundamental Research Projects (Grant No. 202401AU070063), and the International Centre of Supernovae, Yunnan Key Laboratory (No. 202302AN360001). E.K. acknowledges financial support from the Emil Aaltonen foundation. H.K. was funded by the Research Council of Finland projects 324504, 328898, and 353019. S.M. and T.M.R acknowledge support from the Research Council of Finland project 350458. A.R. acknowledges financial support from the GRAWITA Large Program Grant (PI P. D’Avanzo) and from the PRIN-INAF 2022 "Shedding light on the nature of gap transients: from the observations to the models. T.M.R is part of the Cosmic Dawn Center (DAWN), which is funded by the Danish National Research Foundation under grant DNRF140. M.D.S is funded by the Independent Research Fund Denmark (IRFD, grant number 10.46540/2032-00022B) and by an Aarhus University Research Foundation Nova project (AUFF-E-2023-9-28). Based on observations obtained with the Samuel Oschin Telescope 48-inch and the 60-inch Telescope at the Palomar Observatory as part of the Zwicky Transient Facility project. ZTF is supported by the National Science Foundation under Grants No. AST-1440341 and AST-2034437 and a collaboration including Caltech, IPAC, the Weizmann Institute of Science, the Oskar Klein Center at Stockholm University, the University of Maryland, Deutsches Elektronen-Synchrotron and Humboldt University, the TANGO Consortium of Taiwan, the University of Wisconsin at Milwaukee, Trinity College Dublin, Lawrence Livermore National Laboratories, IN2P3, University of Warwick, Ruhr University Bochum, Northwestern University and former partners the University of Washington, Los Alamos National Laboratories, and Lawrence Berkeley National Laboratories. Operations are conducted by COO, IPAC, and UW. SED Machine is based upon work supported by the National Science Foundation under Grant No. 1106171. The ZTF forced-photometry service was funded under the Heising-Simons Foundation grant \#12540303 (PI: Graham). The Gordon and Betty Moore Foundation, through both the Data-Driven Investigator Program and a dedicated grant, provided critical funding for SkyPortal. Based on observations made with the Nordic Optical Telescope, owned in collaboration by the University of Turku and Aarhus University, and operated jointly by Aarhus University, the University of Turku and the University of Oslo, representing Denmark, Finland, and Norway, the University of Iceland and Stockholm University at the Observatorio del Roque de los Muchachos, La Palma, Spain, of the Instituto de Astrofisica de Canarias. The data presented here were obtained with ALFOSC, which is provided by the Instituto de Astrofisica de Andalucia (IAA) under a joint agreement with the University of Copenhagen and NOT. Observations from the Nordic Optical Telescope were obtained through the NUTS2 collaboration which is supported in part by the Instrument Centre for Danish Astrophysics (IDA), and the Finnish Centre for Astronomy with ESO (FINCA) via Academy of Finland grant nr 306531, through the proposal P65-005 (PI: T. Nagao), and through the fast-track proposal P66-415 (PI: P. Charalampopoulos).

\end{acknowledgements}

\bibliographystyle{aa}
\bibliography{bib.bib}

\begin{thebibliography}{170}
\expandafter\ifx\csname natexlab\endcsname\relax\def\natexlab#1{#1}\fi

\bibitem[{Aghanim {et~al.}(2020)Aghanim, Akrami, Ashdown, Aumont, Baccigalupi,
  Ballardini, Banday, Barreiro, Bartolo, Basak, Battye, Benabed, Bernard,
  Bersanelli, Bielewicz, Bock, Bond, Borrill, Bouchet, Boulanger, Bucher,
  Burigana, Butler, Calabrese, Cardoso, Carron, Challinor, Chiang, Chluba,
  Colombo, Combet, Contreras, Crill, Cuttaia, {De Bernardis}, {De Zotti},
  Delabrouille, Delouis, {Di Valentino}, Diego, Dor{\'{e}}, Douspis, Ducout,
  Dupac, Dusini, Efstathiou, Elsner, En{\ss}lin, Eriksen, Fantaye, Farhang,
  Fergusson, Fernandez-Cobos, Finelli, Forastieri, Frailis, Fraisse,
  Franceschi, Frolov, Galeotta, Galli, Ganga, G{\'{e}}nova-Santos, Gerbino,
  Ghosh, Gonz{\'{a}}lez-Nuevo, G{\'{o}}rski, Gratton, Gruppuso, Gudmundsson,
  Hamann, Handley, Hansen, Herranz, Hildebrandt, Hivon, Huang, Jaffe, Jones,
  Karakci, Keih{\"{a}}nen, Keskitalo, Kiiveri, Kim, Kisner, Knox,
  Krachmalnicoff, Kunz, Kurki-Suonio, Lagache, Lamarre, Lasenby, Lattanzi,
  Lawrence, {Le Jeune}, Lemos, Lesgourgues, Levrier, Lewis, Liguori, Lilje,
  Lilley, Lindholm, L{\'{o}}pez-Caniego, Lubin, Ma, Maci{\'{a}}s-P{\'{e}}rez,
  Maggio, Maino, Mandolesi, Mangilli, Marcos-Caballero, Maris, Martin,
  Martinelli, Mart{\'{i}}nez-Gonz{\'{a}}lez, Matarrese, Mauri, McEwen,
  Meinhold, Melchiorri, Mennella, Migliaccio, Millea, Mitra,
  Miville-Desch{\^{e}}nes, Molinari, Montier, Morgante, Moss, Natoli,
  N{\o}rgaard-Nielsen, Pagano, Paoletti, Partridge, Patanchon, Peiris,
  Perrotta, Pettorino, Piacentini, Polastri, Polenta, Puget, Rachen, Reinecke,
  Remazeilles, Renzi, Rocha, Rosset, Roudier, Rubi{\~{n}}o-Mart{\'{i}}n,
  Ruiz-Granados, Salvati, Sandri, Savelainen, Scott, Shellard, Sirignano,
  Sirri, Spencer, Sunyaev, Suur-Uski, Tauber, Tavagnacco, Tenti, Toffolatti,
  Tomasi, Trombetti, Valenziano, Valiviita, {Van Tent}, Vibert, Vielva, Villa,
  Vittorio, Wandelt, Wehus, White, White, Zacchei, \& Zonca}]{Aghanim2020}
Aghanim, N., Akrami, Y., Ashdown, M., {et~al.} 2020, A\&A, 641, A6

\bibitem[{{Allende Prieto} {et~al.}(2001){Allende Prieto}, Lambert, \&
  Asplund}]{Prieto2001}
{Allende Prieto}, C., Lambert, D.~L., \& Asplund, M. 2001, ApJ, 556, L63

\bibitem[{Anderson {et~al.}(2014{\natexlab{a}})Anderson, Dessart, Gutierrez,
  Hamuy, Morrell, Phillips, Folatelli, Stritzinger, Freedman,
  Gonz{\'{a}}lez-gait{\'{a}}n, Mccarthy, Suntzeff, \&
  Thomas-osip}]{Anderson2014a}
Anderson, J.~P., Dessart, L., Gutierrez, C.~P., {et~al.} 2014{\natexlab{a}},
  MNRAS, 441, 671

\bibitem[{Anderson {et~al.}(2014{\natexlab{b}})Anderson,
  Gonz{\'{a}}lez-Gait{\'{a}}n, Hamuy, Guti{\'{e}}rrez, Stritzinger, {Olivares
  E.}, Phillips, Schulze, Antezana, Bolt, Campillay, Castell{\'{o}}n,
  Contreras, {De Jaeger}, Folatelli, F{\"{o}}rster, Freedman, Gonz{\'{a}}lez,
  Hsiao, Krzemi{\'{n}}ski, Krisciunas, Maza, McCarthy, Morrell, Persson, Roth,
  Salgado, Suntzeff, \& Thomas-Osip}]{Anderson2014}
Anderson, J.~P., Gonz{\'{a}}lez-Gait{\'{a}}n, S., Hamuy, M., {et~al.}
  2014{\natexlab{b}}, ApJ, 786, 67

\bibitem[{Andrews {et~al.}(2025)Andrews, Shrestha, Bostroem, {Dong 董},
  Pearson, Fausnaugh, Sand, Valenti, Ravi, Hoang, Hosseinzadeh, Ilyin, Janzen,
  Lundquist, Meza, Smith, Jha, Andrews, Farah, {Padilla Gonzalez}, Howell,
  McCully, Newsome, Pellegrino, Terreran, Wiggins, Hsu, Christy, Franz, Wang,
  Liu, \& Chen}]{Andrews2025}
Andrews, J.~E., Shrestha, M., Bostroem, K.~A., {et~al.} 2025, ApJ, 980, 37

\bibitem[{Andrews \& Smith(2018)}]{Andrews2018}
Andrews, J.~E. \& Smith, N. 2018, MNRAS, 477, 74

\bibitem[{Arcavi {et~al.}(2010)Arcavi, Gal-Yam, Kasliwal, Quimby, Ofek,
  Kulkarni, Nugent, Cenko, Bloom, Sullivan, Howell, Poznanski, Filippenko, Law,
  Hook, {\"{O}}nsson, Blake, Cooke, Dekany, Rahmer, Hale, Smith, Zolkower,
  Velur, Walters, Henning, Bui, McKenna, \& Jacobsen}]{Arcavi2010}
Arcavi, I., Gal-Yam, A., Kasliwal, M.~M., {et~al.} 2010, ApJ, 721, 777

\bibitem[{Arcavi {et~al.}(2011)Arcavi, Gal-Yam, Yaron, Sternberg, Rabinak,
  Waxman, Kasliwal, Quimby, Ofek, Horesh, Kulkarni, Filippenko, Silverman,
  Cenko, Li, Bloom, Sullivan, Nugent, Poznanski, Gorbikov, Fulton, Howell,
  Bersier, Riou, Lamotte-Bailey, Griga, Cohen, Hachinger, Polishook, Xu,
  Ben-Ami, Manulis, Walker, Maguire, Pan, Matheson, Mazzali, Pian, Fox,
  Gehrels, Law, James, Marchant, Smith, Mottram, Barnsley, Kandrashoff, \&
  Clubb}]{Arcavi2011}
Arcavi, I., Gal-Yam, A., Yaron, O., {et~al.} 2011, ApJ, 742, L18

\bibitem[{Arnett(1982)}]{Arnett1982}
Arnett, W.~D. 1982, ApJ, 253, 785

\bibitem[{Ashall(2022)}]{Ashall2022}
Ashall, C. 2022, Transient Name Serv. Classif. Report, No. 2022-1602,
  2022-1602, 1

\bibitem[{Barbary {et~al.}(2009)Barbary, Dawson, Tokita, Aldering, Amanullah,
  Connolly, Doi, Faccioli, Fadeyev, Fruchter, Goldhaber, Goobar, Gude, Huang,
  Ihara, Konishi, Kowalski, Lidman, Meyers, Morokuma, Nugent, Perlmutter,
  Rubin, Schlegel, Spadafora, Suzuki, Swift, Takanashi, Thomas, \&
  Yasuda}]{Barbary2009}
Barbary, K., Dawson, K.~S., Tokita, K., {et~al.} 2009, ApJ, 690, 1358

\bibitem[{Barnsley {et~al.}(2012)Barnsley, Smith, \& Steele}]{Barnsley2012}
Barnsley, R., Smith, R., \& Steele, I. 2012, Astron. Nachrichten, 333, 101

\bibitem[{Bellm {et~al.}(2019)Bellm, Kulkarni, Graham, Dekany, Smith, Riddle,
  Masci, Helou, Prince, Adams, Barbarino, Barlow, Bauer, Beck, Belicki, Biswas,
  Blagorodnova, Bodewits, Bolin, Brinnel, Brooke, Bue, Bulla, Burruss, Cenko,
  Chang, Connolly, Coughlin, Cromer, Cunningham, De, Delacroix, Desai, Duev,
  Eadie, Farnham, Feeney, Feindt, Flynn, Franckowiak, Frederick, Fremling,
  Gal-Yam, Gezari, Giomi, Goldstein, Golkhou, Goobar, Groom, Hacopians, Hale,
  Henning, Ho, Hover, Howell, Hung, Huppenkothen, Imel, Ip, Ivezi{\'{c}},
  Jackson, Jones, Juric, Kasliwal, Kaspi, Kaye, Kelley, Kowalski, Kramer,
  Kupfer, Landry, Laher, Lee, Lin, Lin, Lunnan, Giomi, Mahabal, Mao, Miller,
  Monkewitz, Murphy, Ngeow, Nordin, Nugent, Ofek, Patterson, Penprase, Porter,
  Rauch, Rebbapragada, Reiley, Rigault, Rodriguez, van Roestel, Rusholme, van
  Santen, Schulze, Shupe, Singer, Soumagnac, Stein, Surace, Sollerman, Szkody,
  Taddia, Terek, {Van Sistine}, van Velzen, Vestrand, Walters, Ward, Ye, Yu,
  Yan, \& Zolkower}]{Bellm2019a}
Bellm, E.~C., Kulkarni, S.~R., Graham, M.~J., {et~al.} 2019, PASP, 131, 018002

\bibitem[{Benetti {et~al.}(2016)Benetti, Chugai, Utrobin, Cappellaro, Patat,
  Pastorello, Turatto, Cupani, Neuh{\"{a}}user, Caldwell, Pignata, \&
  Tomasella}]{Benetti2016}
Benetti, S., Chugai, N.~N., Utrobin, V.~P., {et~al.} 2016, MNRAS, 456, 3296

\bibitem[{Bersten {et~al.}(2012)Bersten, Benvenuto, Nomoto, Ergon, Folatelli,
  Sollerman, Benetti, Botticella, Fraser, Kotak, Maeda, Ochner, \&
  Tomasella}]{Bersten2012}
Bersten, M.~C., Benvenuto, O.~G., Nomoto, K., {et~al.} 2012, ApJ, 757

\bibitem[{Blagorodnova {et~al.}(2018)Blagorodnova, Neill, Walters, Kulkarni,
  Fremling, Ben-Ami, Dekany, Fucik, Konidaris, Nash, Ngeow, Ofek, O'Sullivan,
  Quimby, Ritter, \& Vyhmeister}]{Blagorodnova2018a}
Blagorodnova, N., Neill, J.~D., Walters, R., {et~al.} 2018, PASP, 130, 035003

\bibitem[{Blondin \& Tonry(2007)}]{Blondin2007}
Blondin, S. \& Tonry, J.~L. 2007, ApJ, 666, 1024

\bibitem[{Bostroem {et~al.}(2020)Bostroem, Valenti, Sand, Andrews, {Van Dyk},
  Galbany, Pooley, Amaro, Smith, Yang, Anupama, Arcavi, Baron, Brown, Burke,
  Cartier, Hiramatsu, Dastidar, DerKacy, Dong, Egami, Ertel, Filippenko, Fox,
  Haislip, Hosseinzadeh, Howell, Gangopadhyay, Jha, Kouprianov, Kumar,
  Lundquist, Milisavljevic, McCully, Milne, Misra, Reichart, Sahu, Sai, Singh,
  Smith, Vinko, Wang, Wang, Wheeler, Williams, Wyatt, Zhang, \&
  Zhang}]{Bostroem2020}
Bostroem, K.~A., Valenti, S., Sand, D.~J., {et~al.} 2020, ApJ, 895, 31

\bibitem[{Bradley {et~al.}(2024)Bradley, Sipőcz, Robitaille, Tollerud,
  Vin{\'{i}}cius, Deil, Barbary, Wilson, Busko, Donath, G{\"{u}}nther, Cara,
  Lim, Me{\ss}linger, Conseil, Burnett, Bostroem, Droettboom, Bray, {Andersen
  Bratholm}, Ginsburg, Jamieson, Barentsen, Craig, Morris, Perrin, Rathi,
  Pascual, Georgiev, Bradley, Sipőcz, Robitaille, Tollerud, Vin{\'{i}}cius,
  Deil, Barbary, Wilson, Busko, Donath, G{\"{u}}nther, Cara, Lim,
  Me{\ss}linger, Conseil, Burnett, Bostroem, Droettboom, Bray, {Andersen
  Bratholm}, Ginsburg, Jamieson, Barentsen, Craig, Morris, Perrin, Rathi,
  Pascual, \& Georgiev}]{Bradley2024}
Bradley, L., Sipőcz, B., Robitaille, T., {et~al.} 2024, Zenodo

\bibitem[{Branch {et~al.}(1981)Branch, Falk, Uomoto, Wills, McCall, \&
  Rybski}]{Branch1981}
Branch, D., Falk, S.~W., Uomoto, A.~K., {et~al.} 1981, ApJ, 244, 780

\bibitem[{Brott {et~al.}(2011)Brott, {De Mink}, Cantiello, Langer, {De Koter},
  Evans, Hunter, Trundle, Vink, Brott, {De Mink}, Cantiello, Langer, {De
  Koter}, Evans, Hunter, Trundle, \& Vink}]{Brott2011}
Brott, I., {De Mink}, S.~E., Cantiello, M., {et~al.} 2011, A\&A, 530
  [\eprint[arXiv]{1102.0530}]

\bibitem[{Bruch {et~al.}(2021)Bruch, Gal-Yam, Schulze, Yaron, Yang, Soumagnac,
  Rigault, Strotjohann, Ofek, Sollerman, Masci, Barbarino, Ho, Fremling,
  Perley, Nordin, Cenko, Adams, Adreoni, Bellm, Blagorodnova, Bulla, Burdge,
  De, Dhawan, Drake, Duev, Dugas, Graham, Graham, Irani, Jencson,
  Karamehmetoglu, Kasliwal, Kim, Kulkarni, Kupfer, Liang, Mahabal, Miller,
  Prince, Riddle, Sharma, Smith, Taddia, Taggart, Walters, \& Yan}]{Bruch2021}
Bruch, R.~J., Gal-Yam, A., Schulze, S., {et~al.} 2021, ApJ, 912, 46

\bibitem[{Bruch {et~al.}(2023)Bruch, Gal-Yam, Yaron, Chen, Strotjohann, Irani,
  Zimmerman, Schulze, Yang, Kim, Bulla, Sollerman, Rigault, Ofek, Soumagnac,
  Masci, Fremling, Perley, Nordin, Cenko, Ho, Adams, Adreoni, Bellm,
  Blagorodnova, Burdge, De, Dekany, Dhawan, Drake, Duev, Graham, Graham,
  Jencson, Karamehmetoglu, Kasliwal, Kulkarni, Miller, Neill, Prince, Riddle,
  Rusholme, Sharma, Smith, Sravan, Taggart, Walters, \& Yan}]{Bruch2023}
Bruch, R.~J., Gal-Yam, A., Yaron, O., {et~al.} 2023, ApJ, 952, 119

\bibitem[{Bufano {et~al.}(2014)Bufano, Pignata, Bersten, Mazzali, Ryder,
  Margutti, Milisavljevic, Morelli, Benetti, Cappellaro, Gonzalez-Gaitan,
  Romero-Ca{\~{n}}izales, Stritzinger, Walker, Anderson, Contreras, de~Jaeger,
  F{\"{o}}rster, Gutierrez, Hamuy, Hsiao, Morrell, {Olivares E.}, Paillas,
  Parker, Pian, Pickering, Sanders, Stockdale, Turatto, Valenti, Fesen, Maza,
  Nomoto, Phillips, \& Soderberg}]{Bufano2014}
Bufano, F., Pignata, G., Bersten, M., {et~al.} 2014, MNRAS, 439, 1807

\bibitem[{Cappellaro {et~al.}(1997)Cappellaro, Mazzali, Benetti, Danziger,
  Turatto, {Della Valle}, \& Patat}]{Cappellaro1997}
Cappellaro, E., Mazzali, P.~A., Benetti, S., {et~al.} 1997, A\&A, 328, 203

\bibitem[{Cardelli {et~al.}(1989)Cardelli, Clayton, \& Mathis}]{Cardelli1989}
Cardelli, J.~A., Clayton, G.~C., \& Mathis, J.~S. 1989, ApJ, 345, 245

\bibitem[{Chatzopoulos {et~al.}(2012)Chatzopoulos, {Craig Wheeler}, \&
  Vinko}]{Chatzopoulos2012}
Chatzopoulos, E., {Craig Wheeler}, J., \& Vinko, J. 2012, ApJ, 746, 121

\bibitem[{Chatzopoulos {et~al.}(2013)Chatzopoulos, Wheeler, Vinko, Horvath, \&
  Nagy}]{Chatzopoulos2013}
Chatzopoulos, E., Wheeler, J.~C., Vinko, J., Horvath, Z.~L., \& Nagy, A. 2013,
  ApJ, 773, 76

\bibitem[{Chevalier \& Irwin(2011)}]{Chevalier2011}
Chevalier, R.~A. \& Irwin, C.~M. 2011, ApJ, 729, L6

\bibitem[{Chornock {et~al.}(2011)Chornock, Filippenko, Li, Marion, Foley,
  Modjaz, Rafelski, Becker, {De Vries}, Garnavich, Jorgenson, Lynch, Malec,
  Moran, Murphy, Rudy, Russell, Silverman, Steele, Stockton, Wolfe, \&
  Woodward}]{Chornock2011}
Chornock, R., Filippenko, A.~V., Li, W., {et~al.} 2011, ApJ, 739, 41

\bibitem[{Chugai(2001)}]{Chugai2001}
Chugai, N.~N. 2001, MNRAS, 326, 1448

\bibitem[{Chugai(2009)}]{Chugai2009}
Chugai, N.~N. 2009, MNRAS, 400, 866

\bibitem[{Claeys {et~al.}(2011)Claeys, {De Mink}, Pols, Eldridge, \&
  Baes}]{Claeys2011}
Claeys, J.~S., {De Mink}, S.~E., Pols, O.~R., Eldridge, J.~J., \& Baes, M.
  2011, A\&A, 528, A131

\bibitem[{Davis {et~al.}(2019)Davis, Hsiao, Ashall, Hoeflich, Phillips, Marion,
  Kirshner, Morrell, Sand, Burns, Contreras, Stritzinger, Anderson, Baron,
  Diamond, Guti{\'{e}}rrez, Hamuy, Holmbo, Kasliwal, Krisciunas, Kumar, Lu,
  Pessi, Piro, Prieto, Shahbandeh, \& Suntzeff}]{Davis2019}
Davis, S., Hsiao, E.~Y., Ashall, C., {et~al.} 2019, ApJ, 887, 4

\bibitem[{de~Jaeger {et~al.}(2018)de~Jaeger, Anderson, Galbany,
  Gonz{\'{a}}lez-Gait{\'{a}}n, Hamuy, Phillips, Stritzinger, Contreras,
  Folatelli, Guti{\'{e}}rrez, Hsiao, Morrell, Suntzeff, Dessart, \&
  Filippenko}]{DeJaeger2018}
de~Jaeger, T., Anderson, J.~P., Galbany, L., {et~al.} 2018, MNRAS, 476, 4592

\bibitem[{de~Vaucouleurs {et~al.}(1981)de~Vaucouleurs, de~Vaucouleurs, Buta,
  Ables, \& Hewitt}]{DeVaucouleurs1981}
de~Vaucouleurs, G., de~Vaucouleurs, A., Buta, R., Ables, H.~D., \& Hewitt,
  A.~V. 1981, PASP, 93, 36

\bibitem[{Dekany {et~al.}(2020)Dekany, Smith, Riddle, Feeney, Porter, Hale,
  Zolkower, Belicki, Kaye, Henning, Walters, Cromer, Delacroix, Rodriguez,
  Reiley, Mao, Hover, Murphy, Burruss, Baker, Kowalski, Reif, Mueller, Bellm,
  Graham, \& Kulkarni}]{Dekany2020}
Dekany, R., Smith, R.~M., Riddle, R., {et~al.} 2020, PASP, 132, 038001

\bibitem[{D'Elia {et~al.}(2015)D'Elia, Pian, Melandri, D'Avanzo, {Della Valle},
  Mazzali, Piranomonte, Tagliaferri, Antonelli, Bufano, Covino, Fugazza,
  Malesani, M{\o}ller, \& Palazzi}]{DElia2015}
D'Elia, V., Pian, E., Melandri, A., {et~al.} 2015, A\&A, 577, A116

\bibitem[{Dessart \& Hillier(2022)}]{Dessart2022}
Dessart, L. \& Hillier, D.~J. 2022, A\&A, 660, L9

\bibitem[{Dessart {et~al.}(2017)Dessart, Hillier, \& Audit}]{Dessart2017}
Dessart, L., Hillier, D.~J., \& Audit, E. 2017, A\&A, 605, A83

\bibitem[{Dessart \& Jacobson-Gal{\'{a}}n(2023)}]{Dessart2023}
Dessart, L. \& Jacobson-Gal{\'{a}}n, W.~V. 2023, A\&A, 677, A105

\bibitem[{Ercolino {et~al.}(2024)Ercolino, Jin, Langer, \&
  Dessart}]{Ercolino2024}
Ercolino, A., Jin, H., Langer, N., \& Dessart, L. 2024, A\&A, 685, A58

\bibitem[{Ergon {et~al.}(2015)Ergon, Jerkstrand, Sollerman, Elias-Rosa,
  Fransson, Fraser, Pastorello, Kotak, Taubenberger, Tomasella, Valenti,
  Benetti, Helou, Kasliwal, Maund, Smartt, \& Spyromilio}]{Ergon2015}
Ergon, M., Jerkstrand, A., Sollerman, J., {et~al.} 2015, A\&A, 580, A142

\bibitem[{Faran {et~al.}(2014)Faran, Poznanski, Filippenko, Chornock, Foley,
  Ganeshalingam, Leonard, Li, Modjaz, Serduke, \& Silverman}]{Faran2014}
Faran, T., Poznanski, D., Filippenko, A.~V., {et~al.} 2014, MNRAS, 445, 554

\bibitem[{Fassia {et~al.}(2000)Fassia, Meikle, Vacca, Kemp, Walton, Pollacco,
  Smartt, Oscoz, Arag{\'{o}}n-Salamanca, Bennett, Hawarden, Alonso, Alcalde,
  Pedrosa, Telting, Arevalo, Deeg, Garz{\'{o}}n, G{\'{o}}mez-Rold{\'{a}}n,
  G{\'{o}}mez, Guti{\'{e}}rrez, L{\'{o}}pez, Rozas, Serra-Ricart, \&
  Zapatero-Osorio}]{Fassia2000}
Fassia, A., Meikle, W.~P., Vacca, W.~D., {et~al.} 2000, MNRAS, 318, 1093

\bibitem[{Filippenko {et~al.}(1993)Filippenko, Matheson, \&
  Ho}]{Filippenko1993}
Filippenko, A.~V., Matheson, T., \& Ho, L.~C. 1993, ApJ, 415, L103

\bibitem[{Foley {et~al.}(2007)Foley, Smith, Ganeshalingam, Li, Chornock, \&
  Filippenko}]{Foley2007}
Foley, R.~J., Smith, N., Ganeshalingam, M., {et~al.} 2007, ApJ, 657, L105

\bibitem[{Fransson {et~al.}(2005)Fransson, Challis, Chevalier, Filippenko,
  Kirshner, Kozma, Leonard, Matheson, Baron, Garnavich, Jha, Leibundgut,
  Lundqvist, Pun, Wang, \& Wheeler}]{Fransson2005}
Fransson, C., Challis, P.~M., Chevalier, R.~A., {et~al.} 2005, ApJ, 622, 991

\bibitem[{Fremling {et~al.}(2016)Fremling, Sollerman, Taddia, Ergon, Fraser,
  Karamehmetoglu, Valenti, Jerkstrand, Arcavi, Bufano, {Elias Rosa},
  Filippenko, Fox, Gal-Yam, Howell, Kotak, Mazzali, Milisavljevic, Nugent,
  Nyholm, Pian, \& Smartt}]{Fremling2016}
Fremling, C., Sollerman, J., Taddia, F., {et~al.} 2016, A\&A, 593
  [\eprint[arXiv]{1606.03074}]

\bibitem[{Gal-Yam {et~al.}(2014)Gal-Yam, Arcavi, Ofek, Ben-Ami, Cenko,
  Kasliwal, Cao, Yaron, Tal, Silverman, Horesh, {De Cia}, Taddia, Sollerman,
  Perley, Vreeswijk, Kulkarni, Nugent, Filippenko, \& Wheeler}]{Gal-Yam2014}
Gal-Yam, A., Arcavi, I., Ofek, E.~O., {et~al.} 2014, Nature, 509, 471

\bibitem[{Galbany {et~al.}(2018)Galbany, Anderson, S{\'{a}}nchez, Kuncarayakti,
  Pedraz, Gonz{\'{a}}lez-Gait{\'{a}}n, Stanishev, Dom{\'{i}}nguez, Moreno-Raya,
  Wood-Vasey, Mour{\~{a}}o, Ponder, Badenes, Moll{\'{a}},
  L{\'{o}}pez-S{\'{a}}nchez, Rosales-Ortega, V{\'{i}}lchez,
  Garc{\'{i}}a-Benito, \& Marino}]{Galbany2018}
Galbany, L., Anderson, J.~P., S{\'{a}}nchez, S.~F., {et~al.} 2018, ApJ, 855,
  107

\bibitem[{Gall {et~al.}(2015)Gall, Polshaw, Kotak, Jerkstrand, Leibundgut,
  Rabinowitz, Sollerman, Sullivan, Smartt, Anderson, Benetti, Baltay, Feindt,
  Fraser, Gonz{\'{a}}lez-Gait{\'{a}}n, Inserra, Maguire, McKinnon, Valenti, \&
  Young}]{Gall2015}
Gall, E.~E., Polshaw, J., Kotak, R., {et~al.} 2015, A\&A, 582, A3

\bibitem[{Georgy {et~al.}(2013)Georgy, Ekstr{\"{o}}m, Eggenberger, Meynet,
  Haemmerl{\'{e}}, Maeder, Granada, Groh, Hirschi, Mowlavi, Yusof, Charbonnel,
  Decressin, \& Barblan}]{Georgy2013}
Georgy, C., Ekstr{\"{o}}m, S., Eggenberger, P., {et~al.} 2013, A\&A, 558, A103

\bibitem[{Gezari {et~al.}(2009)Gezari, Halpern, Grupe, Yuan, Quimby, McKay,
  Chamarro, Sisson, Akerlof, Wheeler, Brown, Cenko, Rau, Djordjevic, \&
  Terndrup}]{Gezari2009}
Gezari, S., Halpern, J.~P., Grupe, D., {et~al.} 2009, ApJ, 690, 1313

\bibitem[{Gkini {et~al.}(2024)Gkini, Fransson, Lunnan, Schulze, Poidevin,
  Sarin, K{\"{o}}nyves-T{\'{o}}th, Sollerman, Omand, Brennan, Hinds, Anderson,
  Bronikowski, Chen, Dekany, Fraser, Fremling, Galbany, Gal-Yam, Gangopadhyay,
  Geier, Gonzalez, Gromadzki, Groom, Guti{\'{e}}rrez, Hiramatsu, Howell, Hu,
  Inserra, Kopsacheili, Lacroix, Masci, Matilainen, McCully, Moore,
  M{\"{u}}ller-Bravo, Nicholl, Pellegrino, P{\'{e}}rez-Fournon, Perley, Pessi,
  Petrushevska, Pignata, Ragosta, Sahu, Singh, Srivastav, Wise, Yan, \&
  Young}]{Gkini2024}
Gkini, A., Fransson, C., Lunnan, R., {et~al.} 2024, arXiv, 18, arXiv:2409.17296

\bibitem[{Gonz{\'{a}}lez-Gait{\'{a}}n
  {et~al.}(2015)Gonz{\'{a}}lez-Gait{\'{a}}n, Tominaga, Molina, Galbany, Bufano,
  Anderson, Gutierrez, F{\"{o}}rster, Pignata, Bersten, Howell, Sullivan,
  Carlberg, {De Jaeger}, Hamuy, Baklanov, \& Blinnikov}]{Gonzalez-Gaitan2015}
Gonz{\'{a}}lez-Gait{\'{a}}n, S., Tominaga, N., Molina, J., {et~al.} 2015,
  MNRAS, 451, 2212

\bibitem[{Graham {et~al.}(2019)Graham, Kulkarni, Bellm, Adams, Barbarino,
  Blagorodnova, Bodewits, Bolin, Brady, Cenko, Chang, Coughlin, {Kishalay De},
  Eadie, Farnham, Feindt, Franckowiak, Fremling, Gezari, Ghosh, Goldstein,
  Golkhou, Goobar, Ho, Huppenkothen, Ivezi{\'{c}}, Jones, Juric, Kaplan,
  Kasliwal, Kelley, Kupfer, Lee, Lin, Lunnan, Mahabal, Miller, Ngeow, Nugent,
  Ofek, Prince, Rauch, {Van Roestel}, Schulze, Singer, Sollerman, Taddia, Yan,
  Ye, Yu, Barlow, Bauer, Beck, Belicki, Biswas, Brinnel, Brooke, Bue, Bulla,
  Burruss, Connolly, Cromer, Cunningham, Dekany, Delacroix, Desai, Duev,
  Feeney, Flynn, Frederick, Gal-Yam, Giomi, Groom, Hacopians, Hale, Helou,
  Henning, Hover, Hillenbrand, Howell, Hung, Imel, Ip, Jackson, Kaspi, Kaye,
  Kowalski, Kramer, Kuhn, Landry, Laher, Mao, Masci, Monkewitz, Murphy, Nordin,
  Patterson, Penprase, Porter, Rebbapragada, Reiley, Riddle, Rigault,
  Rodriguez, Rusholme, {Van Santen}, Shupe, Smith, Soumagnac, Stein, Surace,
  Szkody, Terek, {Van Sistine}, {Van Velzen}, Vestrand, Walters, Ward, Zhang,
  \& Zolkower}]{Graham2019}
Graham, M.~J., Kulkarni, S.~R., Bellm, E.~C., {et~al.} 2019, PASP, 131, 078001

\bibitem[{Guillochon {et~al.}(2018)Guillochon, Nicholl, Villar, Mockler,
  Narayan, Mandel, Berger, \& Williams}]{Guillochon2018}
Guillochon, J., Nicholl, M., Villar, V.~A., {et~al.} 2018, Astrophys. J. Suppl.
  Ser., 236, 6

\bibitem[{Guti{\'{e}}rrez {et~al.}(2014)Guti{\'{e}}rrez, Anderson, Hamuy,
  Gonz{\'{a}}lez-Gait{\'{a}}n, Folatelli, Morrell, Stritzinger, Phillips,
  McCarthy, Suntzeff, \& Thomas-Osip}]{Gutierrez2014}
Guti{\'{e}}rrez, C.~P., Anderson, J.~P., Hamuy, M., {et~al.} 2014, ApJ, 786,
  L15

\bibitem[{Guti{\'{e}}rrez {et~al.}(2017)Guti{\'{e}}rrez, Anderson, Hamuy,
  Morrell, Gonz{\'{a}}lez-Gaitan, Stritzinger, Phillips, Galbany, Folatelli,
  Dessart, Contreras, Valle, Freedman, Hsiao, Krisciunas, Madore, Maza,
  Suntzeff, Prieto, Gonz{\'{a}}lez, Cappellaro, Navarrete, Pizzella, Ruiz,
  Smith, \& Turatto}]{Gutierrez2017}
Guti{\'{e}}rrez, C.~P., Anderson, J.~P., Hamuy, M., {et~al.} 2017, ApJ, 850, 89

\bibitem[{Guti{\'{e}}rrez {et~al.}(2018)Guti{\'{e}}rrez, Anderson, Sullivan,
  Dessart, Gonz{\'{a}}lez-Gaitan, Galbany, Dimitriadis, Arcavi, Bufano, Chen,
  Dennefeld, Gromadzki, Haislip, Hosseinzadeh, Howell, Inserra, Kankare,
  Leloudas, Maguire, McCully, Morrell, Olivares, Pignata, Reichart, Reynolds,
  Smartt, Sollerman, Taddia, Tak{\'{a}}ts, Terreran, Valenti, \&
  Young}]{Gutierrez2018}
Guti{\'{e}}rrez, C.~P., Anderson, J.~P., Sullivan, M., {et~al.} 2018, MNRAS,
  479, 3232

\bibitem[{Guti{\'{e}}rrez {et~al.}(2020)Guti{\'{e}}rrez, Pastorello,
  Jerkstrand, Galbany, Sullivan, Anderson, Taubenberger, Kuncarayakti,
  Gonz{\'{a}}lez-Gait{\'{a}}n, Wiseman, Inserra, Fraser, Maguire, Smartt,
  M{\"{u}}ller-Bravo, Arcavi, Benetti, Bersier, Bose, Bostroem, Burke, Chen,
  Chen, {Della Valle}, Dong, Gal-Yam, Gromadzki, Hiramatsu, Holoien,
  Hosseinzadeh, Howell, Kankare, Kochanek, McCully, Nicholl, Pignata, Prieto,
  Shappee, Taggart, Tomasella, Valenti, \& Young}]{Gutierrez2020}
Guti{\'{e}}rrez, C.~P., Pastorello, A., Jerkstrand, A., {et~al.} 2020, MNRAS,
  499, 974

\bibitem[{G{\"{u}}ver \& {\"{O}}zel(2009)}]{Guver2009}
G{\"{u}}ver, T. \& {\"{O}}zel, F. 2009, MNRAS, 400, 2050

\bibitem[{Hamuy(2003)}]{Hamuy2003}
Hamuy, M. 2003, ApJ, 582, 905

\bibitem[{Hart {et~al.}(2023)Hart, Shappee, Hey, Kochanek, Stanek, Lim, Dobbs,
  Tucker, Jayasinghe, Beacom, Boright, Holoien, Ong, Prieto, Thompson, \&
  Will}]{Hart2023}
Hart, K., Shappee, B.~J., Hey, D., {et~al.} 2023, arXiv, arXiv:2304.03791

\bibitem[{Harutyunyan {et~al.}(2008)Harutyunyan, Pfahler, Pastorello,
  Taubenberger, Turatto, Cappellaro, Benetti, Elias-Rosa, Navasardyan, Valenti,
  Stanishev, Patat, Riello, Pignata, \& Hillebrandt}]{Harutyunyan2008}
Harutyunyan, A.~H., Pfahler, P., Pastorello, A., {et~al.} 2008, A\&A, 488, 383

\bibitem[{Heiles(2000)}]{Heiles2000}
Heiles, C. 2000, AJ, 119, 923

\bibitem[{Ho {et~al.}(2023)Ho, Perley, Gal-Yam, Lunnan, Sollerman, Schulze,
  Das, Dobie, Yao, Fremling, Adams, Anand, Andreoni, Bellm, Bruch, Burdge,
  Castro-Tirado, Dahiwale, De, Dekany, Drake, Duev, Graham, Helou, Kaplan,
  Karambelkar, Kasliwal, Kool, Kulkarni, Mahabal, Medford, Miller, Nordin,
  Ofek, Petitpas, Riddle, Sharma, Smith, Stewart, Taggart, Tartaglia,
  Tzanidakis, \& Winters}]{Ho2023}
Ho, A. Y.~Q., Perley, D.~A., Gal-Yam, A., {et~al.} 2023, ApJ, 949, 120

\bibitem[{Hoffman {et~al.}(2008)Hoffman, Leonard, Chornock, Filippenko, Barth,
  \& Matheson}]{Hoffman2007}
Hoffman, J.~L., Leonard, D.~C., Chornock, R., {et~al.} 2008, ApJ, 688, 1186

\bibitem[{Hoflich(1991)}]{Hoflich1991}
Hoflich, P. 1991, A\&A, 246, 481

\bibitem[{Huber {et~al.}(2015)Huber, Chambers, Flewelling, Willman, Primak,
  Schultz, Gibson, Magnier, Waters, Tonry, Wainscoat, Smith, Wright, Smartt,
  Foley, Jha, Rest, Scolnic, Huber, Chambers, Flewelling, Willman, Primak,
  Schultz, Gibson, Magnier, Waters, Tonry, Wainscoat, Smith, Wright, Smartt,
  Foley, Jha, Rest, \& Scolnic}]{Huber2015}
Huber, M., Chambers, K.~C., Flewelling, H., {et~al.} 2015, ATel, 7153, 1

\bibitem[{Humphreys(2010)}]{Humphreys2010}
Humphreys, R. 2010, in Hot Cool Bridg. Gaps Massive Star Evol., Vol. 425, 247

\bibitem[{Humphreys {et~al.}(2023)Humphreys, Jones, \& Martin}]{Humphreys2023}
Humphreys, R.~M., Jones, T.~J., \& Martin, J.~C. 2023, AJ, 166, 50

\bibitem[{Jacobson-Gal{\'{a}}n {et~al.}(2024)Jacobson-Gal{\'{a}}n, Dessart,
  Davis, Kilpatrick, Margutti, Foley, Chornock, Terreran, Hiramatsu, Newsome,
  {Padilla Gonzalez}, Pellegrino, Howell, Filippenko, Anderson, Angus,
  Auchettl, Bostroem, Brink, Cartier, Coulter, de~Boer, Drout, Earl, Ertini,
  Farah, Farias, Gall, Gao, Gerlach, Guo, Haynie, Hosseinzadeh, Ibik, Jha,
  Jones, Langeroodi, LeBaron, Magnier, Piro, Raimundo, Rest, Rest, Rich,
  Rojas-Bravo, Sears, Taggart, Villar, Wainscoat, Wang, Wasserman, Yan, Yang,
  Zhang, \& Zheng}]{Jacobson-Galan2024}
Jacobson-Gal{\'{a}}n, W.~V., Dessart, L., Davis, K.~W., {et~al.} 2024, ApJ,
  970, 189

\bibitem[{Jordi {et~al.}(2006)Jordi, Grebel, \& Ammon}]{Jordi2006}
Jordi, K., Grebel, E.~K., \& Ammon, K. 2006, A\&A, 460, 339

\bibitem[{Kaiser {et~al.}(2002)Kaiser, Aussel, Burke, Boesgaard, Chambers,
  Chun, Heasley, Hodapp, Hunt, Jedicke, Jewitt, Kudritzki, Luppino, Maberry,
  Magnier, Monet, Onaka, Pickles, Rhoads, Simon, Szalay, Szapudi, Tholen,
  Tonry, Waterson, \& Wick}]{Kaiser2002}
Kaiser, N., Aussel, H., Burke, B.~E., {et~al.} 2002, in Surv. Other Telesc.
  Technol. Discov., Vol. 4836 (SPIE), 154

\bibitem[{Kewley \& Ellison(2008)}]{Kewley2008}
Kewley, L.~J. \& Ellison, S.~L. 2008, ApJ, 681, 1183

\bibitem[{Khazov {et~al.}(2016)Khazov, Yaron, Gal-Yam, Manulis, Rubin,
  Kulkarni, Arcavi, Kasliwal, Ofek, Cao, Perley, Sollerman, Horesh, Sullivan,
  Filippenko, Nugent, Howell, Cenko, Silverman, Ebeling, Taddia, Johansson,
  Laher, Surace, Rebbapragada, Wozniak, \& Matheson}]{Khazov2016}
Khazov, D., Yaron, O., Gal-Yam, A., {et~al.} 2016, ApJ, 818, 3

\bibitem[{Kim {et~al.}(2022)Kim, Rigault, Neill, Briday, Copin, Lezmy, Nicolas,
  Riddle, Sharma, Smith, Sollerman, \& Walters}]{Kim2022}
Kim, Y.~L., Rigault, M., Neill, J.~D., {et~al.} 2022, PASP, 134, 024505

\bibitem[{Kotak {et~al.}(2004)Kotak, Meikle, Adamson, \& Leggett}]{Kotak2004}
Kotak, R., Meikle, W.~P., Adamson, A., \& Leggett, S.~K. 2004, MNRAS, 354, L13

\bibitem[{Kulkarni {et~al.}(1998)Kulkarni, Frail, Wieringa, Ekers, Sadler,
  Wark, Higdon, Phinney, \& Bloom}]{Kulkarni1998}
Kulkarni, S.~R., Frail, D.~A., Wieringa, M.~H., {et~al.} 1998, Nature, 395, 663

\bibitem[{Langer(1998)}]{Langer1998}
Langer, N. 1998, A\&A, 329, 551

\bibitem[{Langer(2012)}]{Langer2012}
Langer, N. 2012, {Presupernova evolution of massive single and binary stars}

\bibitem[{Leloudas {et~al.}(2017)Leloudas, Maund, Gal-Yam, Pursimo, Hsiao,
  Malesani, Patat, {de Ugarte Postigo}, Sollerman, Stritzinger, \&
  Wheeler}]{Leloudas2017}
Leloudas, G., Maund, J.~R., Gal-Yam, A., {et~al.} 2017, ApJ, 837, L14

\bibitem[{Leonard {et~al.}(2000)Leonard, Filippenko, Barth, \&
  Matheson}]{Leonard1999}
Leonard, D.~C., Filippenko, A.~V., Barth, A.~J., \& Matheson, T. 2000, ApJ,
  536, 239

\bibitem[{Maeda {et~al.}(2023{\natexlab{a}})Maeda, Chandra, Moriya, Reguitti,
  Ryder, Matsuoka, Michiyama, Pignata, Hiramatsu, Bostroem, Kundu,
  Kuncarayakti, Bersten, Pooley, Lee, Patnaude, Rodr{\'{i}}guez, \&
  Folatelli}]{Maeda2023a}
Maeda, K., Chandra, P., Moriya, T.~J., {et~al.} 2023{\natexlab{a}}, ApJ, 942,
  17

\bibitem[{Maeda {et~al.}(2023{\natexlab{b}})Maeda, Michiyama, Chandra, Ryder,
  Kuncarayakti, Hiramatsu, \& Imanishi}]{Maeda2023}
Maeda, K., Michiyama, T., Chandra, P., {et~al.} 2023{\natexlab{b}}, ApJ, 945,
  L3

\bibitem[{Mandigo-Stoba {et~al.}(2022)Mandigo-Stoba, Fremling, \&
  Kasliwal}]{Mandigo-Stoba2022}
Mandigo-Stoba, M.~S., Fremling, C., \& Kasliwal, M.~M. 2022, J. Open Source
  Softw., 7, 3612

\bibitem[{Masci {et~al.}(2019)Masci, Laher, Rusholme, Shupe, Groom, Surace,
  Jackson, Monkewitz, Beck, Flynn, Terek, Landry, Hacopians, Desai, Howell,
  Brooke, Imel, Wachter, Ye, Lin, Cenko, Cunningham, Rebbapragada, Bue, Miller,
  Mahabal, Bellm, Patterson, Juri{\'{c}}, Golkhou, Ofek, Walters, Graham,
  Kasliwal, Dekany, Kupfer, Burdge, Cannella, Barlow, {Van Sistine}, Giomi,
  Fremling, Blagorodnova, Levitan, Riddle, Smith, Helou, Prince, \&
  Kulkarni}]{Masci2019}
Masci, F.~J., Laher, R.~R., Rusholme, B., {et~al.} 2019, PASP, 131

\bibitem[{Mauerhan {et~al.}(2015)Mauerhan, Smith, {Van Dyk}, Morzinski, Close,
  Hinz, Males, \& Rodigas}]{Mauerhan2015a}
Mauerhan, J., Smith, N., {Van Dyk}, S.~D., {et~al.} 2015, MNRAS, 450, 2551

\bibitem[{Mauerhan {et~al.}(2017)Mauerhan, {Van Dyk}, Johansson, Hu, Fox, Wang,
  Graham, Filippenko, \& Shivvers}]{Mauerhan2017}
Mauerhan, J.~C., {Van Dyk}, S.~D., Johansson, J., {et~al.} 2017, ApJ, 834, 118

\bibitem[{Medler {et~al.}(2023)Medler, Mazzali, Ashall, Teffs, Shahbandeh, \&
  Shappee}]{Medler2023}
Medler, K., Mazzali, P.~A., Ashall, C., {et~al.} 2023, Mon. Not. R. Astron.
  Soc. Lett. Vol. 518, Issue 1, pp.L40-L44, 518, L40

\bibitem[{Medler {et~al.}(2022)Medler, Mazzali, Teffs, Ashall, Anderson,
  Arcavi, Benetti, Bostroem, Burke, Cai, Charalampopoulos, Elias-Rosa, Ergon,
  Galbany, Gromadzki, Hiramatsu, Howell, Inserra, Lundqvist, McCully,
  {M{\"{u}}ller -Bravo}, Newsome, Nicholl, Gonzalez, Paraskeva, Pastorello,
  Pellegrino, Pessi, Reguitti, Reynolds, Roy, Terreran, Tomasella, \&
  Young}]{Medler2022}
Medler, K., Mazzali, P.~A., Teffs, J., {et~al.} 2022, MNRAS, 513, 5540

\bibitem[{Meynet {et~al.}(1994)Meynet, Maeder, Schaller, Schaerer, \&
  Charbonnel}]{Meynet1994}
Meynet, G., Maeder, A., Schaller, G., Schaerer, D., \& Charbonnel, C. 1994,
  \aaps, 103, 97

\bibitem[{Miller {et~al.}(2009)Miller, Chornock, Perley, Ganeshalingam, Li,
  Butler, Bloom, Smith, Modjaz, Poznanski, Filippenko, Griffith, Shiode, \&
  Silverman}]{Miller2009}
Miller, A.~A., Chornock, R., Perley, D.~A., {et~al.} 2009, ApJ, 690, 1303

\bibitem[{Modjaz {et~al.}(2011)Modjaz, Kewley, Bloom, Filippenko, Perley, \&
  Silverman}]{Modjaz2011}
Modjaz, M., Kewley, L., Bloom, J.~S., {et~al.} 2011, ApJ, 731, L4

\bibitem[{Morales-Garoffolo {et~al.}(2014)Morales-Garoffolo, Elias-Rosa,
  Benetti, Taubenberger, Cappellaro, Pastorello, Klauser, Valenti, Howerton,
  Ochner, Schramm, Siviero, Tartaglia, \& Tomasella}]{Morales-Garoffolo2014}
Morales-Garoffolo, A., Elias-Rosa, N., Benetti, S., {et~al.} 2014, Mon. Not. R.
  Astron. Soc. Vol. 445, Issue 2, p.1647-1662, 445, 1647

\bibitem[{Morales-Garoffolo {et~al.}(2015)Morales-Garoffolo, Elias-Rosa,
  Bersten, Jerkstrand, Taubenberger, Benetti, Cappellaro, Kotak, Pastorello,
  Bufano, Dom{\'{i}}nguez, Ergon, Fraser, Gao, Garc{\'{i}}a, Howell, Isern,
  Smartt, Tomasella, \& Valenti}]{Morales-Garoffolo2015}
Morales-Garoffolo, A., Elias-Rosa, N., Bersten, M., {et~al.} 2015, MNRAS, 454,
  95

\bibitem[{Moriya \& Tominaga(2012)}]{Moriya2012}
Moriya, T.~J. \& Tominaga, N. 2012, ApJ, 747, 118

\bibitem[{Nadyozhin(1994)}]{Nadyozhin1994}
Nadyozhin, D.~K. 1994, Astrophys. J. Suppl. Ser., 92, 527

\bibitem[{Nagy(2018)}]{Nagy2018}
Nagy, A.~P. 2018, ApJ, 862, 143

\bibitem[{Nicholl(2018)}]{Nicholl2018a}
Nicholl, M. 2018, Res. Notes AAS, 2, 230

\bibitem[{Nomoto {et~al.}(1995)Nomoto, Iwamoto, \& Suzuki}]{Nomoto1995}
Nomoto, K., Iwamoto, K., \& Suzuki, T. 1995, Phys. Rep., 256, 173

\bibitem[{Nomoto {et~al.}(1993)Nomoto, Suzuki, Shigeyama, Kumagai, Yamaoka, \&
  Saio}]{Nomoto1993}
Nomoto, K., Suzuki, T., Shigeyama, T., {et~al.} 1993, Nature, 364, 507

\bibitem[{Oke {et~al.}(1995)Oke, Cohen, Carr, Cromer, Dingizian, Harris,
  Labrecque, Lucinio, Schaal, Epps, \& Miller}]{Oke1995}
Oke, J.~B., Cohen, J.~G., Carr, M., {et~al.} 1995, PASP, 107, 375

\bibitem[{Oke \& Gunn(1983)}]{Oke1983}
Oke, J.~B. \& Gunn, J.~E. 1983, ApJ, 266, 713

\bibitem[{Ouchi {et~al.}(2021)Ouchi, Maeda, Anderson, \& Sawada}]{Ouchi2021}
Ouchi, R., Maeda, K., Anderson, J.~P., \& Sawada, R. 2021, ApJ, 922, 141

\bibitem[{Pastorello {et~al.}(2015)Pastorello, Prieto, Elias-Rosa, Bersier,
  Hosseinzadeh, Morales-Garoffolo, Noebauer, Taubenberger, Tomasella, Kochanek,
  Falco, Basu, Beacom, Benetti, Brimacombe, Cappellaro, Danilet, Dong,
  Fernandez, Goss, Granata, Harutyunyan, Holoien, Ishida, Kiyota, Krannich,
  Nicholls, Ochner, Pojma{\'{n}}ski, Shappee, Simonian, Stanek, Starrfield,
  Szczygiel, Tartaglia, Terreran, Thompson, Turatto, Wagner, Wiethoff, Wilber,
  \& Wo{\'{z}}niak}]{Pastorello2015}
Pastorello, A., Prieto, J.~L., Elias-Rosa, N., {et~al.} 2015, MNRAS, 453, 3649

\bibitem[{Patat {et~al.}(1994)Patat, Barbon, Cappellaro, Turatto, Patat,
  Barbon, Cappellaro, \& Turatto}]{Patat1994}
Patat, F., Barbon, R., Cappellaro, E., {et~al.} 1994, A\&A, 282, 731

\bibitem[{Patat {et~al.}(2001)Patat, Cappellaro, Danziger, Mazzali, Sollerman,
  Augusteijn, Brewer, Doublier, Gonzalez, Hainaut, Lidman, Leibundgut, Nomoto,
  Nakamura, Spyromilio, Rizzi, Turatto, Walsh, Galama, van Paradijs,
  Kouveliotou, Vreeswijk, Frontera, Masetti, Palazzi, \& Pian}]{Patat2001}
Patat, F., Cappellaro, E., Danziger, J., {et~al.} 2001, ApJ, 555, 900

\bibitem[{Perley {et~al.}(2019)Perley, Mazzali, Yan, Cenko, Gezari, Taggart,
  Blagorodnova, Fremling, Mockler, Singh, Tominaga, Tanaka, Watson, Ahumada,
  Anupama, Ashall, Becerra, Bersier, Bhalerao, Bloom, Butler, Copperwheat,
  Coughlin, De, Drake, Duev, Frederick, Gonz{\'{a}}lez, Goobar, Heida, Ho,
  Horst, Hung, Itoh, Jencson, Kasliwal, Kawai, Khanam, Kulkarni, Kumar, Kumar,
  Kutyrev, Lee, Maeda, Mahabal, Murata, Neill, Ngeow, Penprase, Pian, Quimby,
  Ramirez-Ruiz, Richer, Roma{\'{n}}-Z{\'{u}}{\~{n}}iga, Sahu, Srivastav, Socia,
  Sollerman, Tachibana, Taddia, Tinyanont, Troja, Ward, Wee, \&
  Yu}]{Perley2019}
Perley, D.~A., Mazzali, P.~A., Yan, L., {et~al.} 2019, MNRAS, 484, 1031

\bibitem[{Perley {et~al.}(2022)Perley, Sollerman, Schulze, Yao, Fremling,
  Gal-Yam, Ho, Yang, Kool, Irani, Yan, Andreoni, Baade, Bellm, Brink, Chen,
  Cikota, Coughlin, Dahiwale, Dekany, Duev, Filippenko, Hoeflich, Kasliwal,
  Kulkarni, Lunnan, Masci, Maund, Medford, Riddle, Rosnet, Shupe, Strotjohann,
  Tzanidakis, \& Zheng}]{Perley2022}
Perley, D.~A., Sollerman, J., Schulze, S., {et~al.} 2022, ApJ, 927, 180

\bibitem[{Pessi {et~al.}(2023{\natexlab{a}})Pessi, Anderson, Folatelli,
  Dessart, Gonz{\'{a}}lez-Gait{\'{a}}n, M{\"{o}}ller, Guti{\'{e}}rrez, Mattila,
  Reynolds, Charalampopoulos, Filippenko, Galbany, Gal-Yam, Gromadzki,
  Hiramatsu, Howell, Inserra, Kankare, Lunnan, Martinez, McCully, Meza,
  M{\"{u}}ller-Bravo, Nicholl, Pellegrino, Pignata, Sollerman, Tucker, Wang, \&
  Young}]{Pessi2023}
Pessi, P.~J., Anderson, J.~P., Folatelli, G., {et~al.} 2023{\natexlab{a}},
  MNRAS, 523, 5315

\bibitem[{Pessi {et~al.}(2023{\natexlab{b}})Pessi, Prieto, Anderson, Galbany,
  Lyman, Kochanek, Dong, Forster, Gonz{\'{a}}lez-D{\'{i}}az, Gonzalez-Gaitan,
  Guti{\'{e}}rrez, Holoien, James, Jim{\'{e}}nez-Palau, Johnston, Kuncarayakti,
  Rosales-Ortega, S{\'{a}}nchez, Schulze, \& Shappee}]{Pessi2023a}
Pessi, T., Prieto, J.~L., Anderson, J.~P., {et~al.} 2023{\natexlab{b}}, A\&A,
  677, A28

\bibitem[{Pettini \& Pagel(2004)}]{Pettini2004}
Pettini, M. \& Pagel, B.~E. 2004, MNRAS, 348, L59

\bibitem[{Piascik {et~al.}(2014)Piascik, Steele, Bates, Mottram, Smith,
  Barnsley, \& Bolton}]{Piascik2014}
Piascik, A.~S., Steele, I.~A., Bates, S.~D., {et~al.} 2014, in Ground-based
  Airborne Instrum. Astron. V, Vol. 9147 (SPIE), 91478H

\bibitem[{Plaszczynski {et~al.}(2014)Plaszczynski, Montier, Levrier, \&
  Tristram}]{Plaszczynski2014}
Plaszczynski, S., Montier, L., Levrier, F., \& Tristram, M. 2014, MNRAS, 439,
  4048

\bibitem[{Podsiadlowski(1992)}]{Podsiadlowski1992}
Podsiadlowski, P. 1992, PASP, 104, 717

\bibitem[{Polshaw {et~al.}(2016)Polshaw, Kotak, Dessart, Fraser, Gal-Yam,
  Inserra, Sim, Smartt, Sollerman, Baltay, Rabinowitz, Benetti, Botticella,
  Campbell, Chen, Galbany, McKinnon, Nicholl, Smith, Sullivan, Tak{\'{a}}ts,
  Valenti, \& Young}]{Polshaw2016}
Polshaw, J., Kotak, R., Dessart, L., {et~al.} 2016, A\&A, 588, A1

\bibitem[{Prochaska {et~al.}(2020)Prochaska, Hennawi, Westfall, Cooke, Wang,
  Hsyu, Davies, Farina, \& Pelliccia}]{Prochaska2020}
Prochaska, J., Hennawi, J., Westfall, K., {et~al.} 2020, J. Open Source Softw.,
  5, 2308

\bibitem[{Pursiainen {et~al.}(2023)Pursiainen, Leloudas, Cikota, Bulla,
  Inserra, Patat, Wheeler, Aamer, Gal-Yam, Maund, Nicholl, Schulze, Sollerman,
  \& Yang}]{Pursiainen2023}
Pursiainen, M., Leloudas, G., Cikota, A., {et~al.} 2023, A\&A, 674, A81

\bibitem[{Pursiainen {et~al.}(2022)Pursiainen, Leloudas, Paraskeva, Cikota,
  Anderson, Angus, Brennan, Bulla, Camacho-I{\~{n}}iguez, Charalampopoulos,
  Chen, {Delgado Manche{\~{n}}o}, Fraser, Frohmaier, Galbany, Guti{\'{e}}rrez,
  Gromadzki, Inserra, Maund, M{\"{u}}ller-Bravo, {Mu{\~{n}}oz Torres}, Nicholl,
  Onori, Patat, Pessi, Roy, Spyromilio, Wiseman, \& Young}]{Pursiainen2022}
Pursiainen, M., Leloudas, G., Paraskeva, E., {et~al.} 2022, A\&A, 666, A30

\bibitem[{Rasmussen \& Williams(2004)}]{Rasmussen2004}
Rasmussen, C.~E. \& Williams, C. K.~I. 2004, {Gaussian processes for machine
  learning.}

\bibitem[{Reguitti {et~al.}(2024)Reguitti, Dastidar, Pignata, Maeda, Moriya,
  Kuncarayakti, Rodr{\'{i}}guez, Bersten, Anderson, Charalampopoulos, Fraser,
  Gromadzki, Young, Benetti, Cai, Elias-Rosa, Lundqvist, Carini, Cosentino,
  Galbany, Gonzalez-Ba{\~{n}}uelos, Guti{\'{e}}rrez, Kopsacheili, {Pineda
  Garc{\'{i}}a}, \& Ramirez}]{Reguitti2024}
Reguitti, A., Dastidar, R., Pignata, G., {et~al.} 2024, A\&A, 692,
  arXiv:2409.16890

\bibitem[{Reynolds {et~al.}(2020)Reynolds, Fraser, Mattila, Ergon, Dessart,
  Lundqvist, Dong, Elias-Rosa, Galbany, Gutierrez, Kangas, Kankare, Kotak,
  Kuncarayakti, Pastorello, Rodriguez, Smartt, Stritzinger, Tomasella, Chen,
  Harmanen, Hosseinzadeh, Howell, Inserra, Nicholl, Nielsen, Smith, Somero,
  Tronsgaard, \& Young}]{Reynolds2020}
Reynolds, T.~M., Fraser, M., Mattila, S., {et~al.} 2020, MNRAS, 493, 1761

\bibitem[{Reynolds {et~al.}(2025{\natexlab{a}})Reynolds, Nagao, Gottumukkala,
  Guti{\'{e}}rrez, Kangas, Kravtsov, Kuncarayakti, Maeda, Elias-Rosa, Fraser,
  Kotak, Mattila, Pastorello, Pessi, Cai, Fynbo, Kawabata, Lundqvist,
  Matilainen, Moran, Reguitti, Taguchi, \& Yamanaka}]{Reynolds2025}
Reynolds, T.~M., Nagao, T., Gottumukkala, R., {et~al.} 2025{\natexlab{a}},
  arXiv (Submitted to A\&A), arXiv:2501.13619

\bibitem[{Reynolds {et~al.}(2025{\natexlab{b}})Reynolds, Nagao, Maeda,
  Elias-Rosa, Fraser, Guti{\'{e}}rrez, Kangas, Kuncarayakti, Mattila, \&
  Pessi}]{Reynolds2025a}
Reynolds, T.~M., Nagao, T., Maeda, K., {et~al.} 2025{\natexlab{b}}, arXiv
  (Submitted to A\&A), arXiv:2501.13621

\bibitem[{Richmond {et~al.}(1994)Richmond, Treffers, Filippenko, Paik,
  Leibundgut, Schulman, \& Cox}]{Richmond1994}
Richmond, M.~W., Treffers, R.~R., Filippenko, A.~V., {et~al.} 1994, AJ, 107,
  1022

\bibitem[{Rigault {et~al.}(2019)Rigault, Neill, Blagorodnova, Dugas, Feeney,
  Walters, Brinnel, Copin, Fremling, Nordin, \& Sollerman}]{Rigault2019}
Rigault, M., Neill, J.~D., Blagorodnova, N., {et~al.} 2019, A\&A, 627, A115

\bibitem[{Rodrigo {et~al.}(2024)Rodrigo, Cruz, Aguilar, Aller, Solano,
  G{\'{a}}lvez-Ortiz, Jim{\'{e}}nez-Esteban, Mas-Buitrago, Bayo,
  Cort{\'{e}}s-Contreras, Murillo-Ojeda, Bonoli, Cenarro, Dupke,
  L{\'{o}}pez-Sanjuan, Mar{\'{i}}n-Franch, {De Oliveira}, Moles, Taylor,
  Varela, \& Rami{\'{o}}}]{Rodrigo2024}
Rodrigo, C., Cruz, P., Aguilar, J.~F., {et~al.} 2024, A\&A, 689

\bibitem[{Rodrigo \& Solano(2020)}]{Rodrigo2020}
Rodrigo, C. \& Solano, E. 2020, XIV.0 Sci. Meet. Spanish Astron. Soc. p. 182,
  182

\bibitem[{Rodrigo {et~al.}(2012)Rodrigo, Solano, \& Bayo}]{Rodrigo2012}
Rodrigo, C., Solano, E., \& Bayo, A. 2012, {SVO Filter Profile Service Version
  1.0}, Tech. rep., International Virtual Observatory Alliance

\bibitem[{Rodr{\'{i}}guez {et~al.}(2023)Rodr{\'{i}}guez, Maoz, \&
  Nakar}]{Rodriguez2023}
Rodr{\'{i}}guez, {\'{O}}., Maoz, D., \& Nakar, E. 2023, ApJ, 955, 71

\bibitem[{Roming {et~al.}(2009)Roming, Pritchard, Brown, Holland, Immler,
  Stockdale, Weiler, Panagia, {Van Dyk}, Hoversten, Milne, Oates, Russell, \&
  Vandrevala}]{Roming2009}
Roming, P.~W., Pritchard, T.~A., Brown, P.~J., {et~al.} 2009, ApJ, 704, L118

\bibitem[{Sahu {et~al.}(2009)Sahu, Tanaka, Anupama, Gurugubelli, \&
  Nomoto}]{Sahu2008}
Sahu, D.~K., Tanaka, M., Anupama, G.~C., Gurugubelli, U.~K., \& Nomoto, K.
  2009, ApJ, 697, 676

\bibitem[{Schlafly \& Finkbeiner(2011)}]{Schlafly2010}
Schlafly, E.~F. \& Finkbeiner, D.~P. 2011, ApJ, 737, 103

\bibitem[{Schlegel(1990)}]{Schlegel1990}
Schlegel, E.~M. 1990, MNRAS, 244, 269

\bibitem[{Schootemeijer {et~al.}(2019)Schootemeijer, Langer, Grin, \&
  Wang}]{Schootemeijer2019}
Schootemeijer, A., Langer, N., Grin, N.~J., \& Wang, C. 2019, A\&A, 625, A132

\bibitem[{Serkowski {et~al.}(1975)Serkowski, Mathewson, \&
  Ford}]{Serkowski1975}
Serkowski, K., Mathewson, D.~L., \& Ford, V.~L. 1975, ApJ, 196, 261

\bibitem[{Shappee {et~al.}(2014)Shappee, Prieto, Stanek, Kochanek, Holoien,
  Jencson, Basu, Beacom, Szczygiel, Pojmanski, Brimacombe, Dubberley, Elphick,
  Foale, Hawkins, Mullins, Rosing, Ross, \& Walker}]{Shappee2014}
Shappee, B., Prieto, J., Stanek, K.~Z., {et~al.} 2014, Am. Astron. Soc., 223,
  236.03

\bibitem[{Shingles {et~al.}(2021)Shingles, Smith, Young, Smartt, Tonry,
  Denneau, Heinze, Weiland, Flewelling, Stalder, Clocchiatti, F{\"{o}}rster,
  Pignata, Rest, Anderson, Stubbs, Erasmus, Shingles, Smith, Young, Smartt,
  Tonry, Denneau, Heinze, Weiland, Flewelling, Stalder, Clocchiatti,
  F{\"{o}}rster, Pignata, Rest, Anderson, Stubbs, \& Erasmus}]{Shingles2021}
Shingles, L., Smith, K.~W., Young, D.~R., {et~al.} 2021, TNSAN, 7, 1

\bibitem[{Shivvers {et~al.}(2015)Shivvers, Groh, Mauerhan, Fox, Leonard, \&
  Filippenko}]{Shivvers2015}
Shivvers, I., Groh, J.~H., Mauerhan, J.~C., {et~al.} 2015, ApJ, 806, 213

\bibitem[{Simmons \& Stewart(1985)}]{Simmons1985}
Simmons, J. F.~L. \& Stewart, B.~G. 1985, A\&A, 142, 100

\bibitem[{Smith {et~al.}(2002{\natexlab{a}})Smith, Tucker, Kent, Richmond,
  Fukugita, Ichikawa, Ichikawa, Jorgensen, Uomoto, Gunn, Hamabe, Watanabe,
  Tolea, Henden, Annis, Pier, McKay, Brinkmann, Chen, Holtzman, Shimasaku, \&
  York}]{Smith2002}
Smith, J.~A., Tucker, D.~L., Kent, S., {et~al.} 2002{\natexlab{a}}, AJ, 123,
  2121

\bibitem[{Smith {et~al.}(2020)Smith, Smartt, Young, Tonry, Denneau, Flewelling,
  Heinze, Weiland, Stalder, Rest, Stubbs, Anderson, Chen, Clark, Do,
  F{\"{o}}rster, Fulton, Gillanders, McBrien, O'neill, Srivastav, \&
  Wright}]{Smith2020}
Smith, K.~W., Smartt, S.~J., Young, D.~R., {et~al.} 2020, PASP, 132, 1

\bibitem[{Smith(2008)}]{Smith2008}
Smith, N. 2008, Nature, 455, 201

\bibitem[{Smith {et~al.}(2002{\natexlab{b}})Smith, Gehrz, Stahl, Balick, \&
  Kaufer}]{Smith2002a}
Smith, N., Gehrz, R.~D., Stahl, O., Balick, B., \& Kaufer, A.
  2002{\natexlab{b}}, ApJ, 578, 464

\bibitem[{Smith {et~al.}(2015)Smith, Mauerhan, Cenko, Kasliwal, Silverman,
  Filippenko, Gal-Yam, Clubb, Graham, Leonard, Horst, Williams, Andrews,
  Kulkarni, Nugent, Sullivan, Maguire, Xu, \& Ben-Ami}]{Smith2015}
Smith, N., Mauerhan, J.~C., Cenko, S.~B., {et~al.} 2015, MNRAS, 449, 1876

\bibitem[{Smith {et~al.}(2009)Smith, Silverman, Chornock, Filippenko, Wang, Li,
  Ganeshalingam, Foley, Rex, \& Steele}]{Smith2009}
Smith, N., Silverman, J.~M., Chornock, R., {et~al.} 2009, ApJ, 695, 1334

\bibitem[{Speagle(2020)}]{Speagle2020}
Speagle, J.~S. 2020, MNRAS, 493, 3132

\bibitem[{Stanek {et~al.}(2022)Stanek, Stanek, \& Z.}]{Stanek2022}
Stanek, K.~Z., Stanek, \& Z., K. 2022, TNSTR, 2022-1548, 1

\bibitem[{Steele(2004)}]{Steele2004}
Steele, I.~A. 2004, Astron. Nachrichten, 325, 519

\bibitem[{Stritzinger {et~al.}(2012)Stritzinger, Taddia, Fransson, Fox,
  Morrell, Phillips, Sollerman, Anderson, Boldt, Brown, Campillay, Castellon,
  Contreras, Folatelli, Habergham, Hamuy, Hjorth, James, Krzeminski, Mattila,
  Persson, \& Roth}]{Stritzinger2012}
Stritzinger, M., Taddia, F., Fransson, C., {et~al.} 2012, ApJ, 756, 173

\bibitem[{Strotjohann {et~al.}(2021)Strotjohann, Ofek, Gal-Yam, Bruch, Schulze,
  Shaviv, Sollerman, Filippenko, Yaron, Fremling, Nordin, Kool, Perley, Ho,
  Yang, Yao, Soumagnac, Graham, Barbarino, Tartaglia, De, Goldstein, Cook,
  Brink, Taggart, Yan, Lunnan, Kasliwal, Kulkarni, Nugent, Masci, Rosnet,
  Adams, Andreoni, Bagdasaryan, Bellm, Burdge, Duev, Dugas, Frederick,
  Goldwasser, Hankins, Irani, Karambelkar, Kupfer, Liang, Neill, Porter,
  Riddle, Sharma, Short, Taddia, Tzanidakis, van Roestel, Walters, \&
  Zhuang}]{Strotjohann2021}
Strotjohann, N.~L., Ofek, E.~O., Gal-Yam, A., {et~al.} 2021, ApJ, 907, 99

\bibitem[{Taddia {et~al.}(2016)Taddia, Moquist, Sollerman, Rubin, Leloudas,
  Gal-Yam, Arcavi, Cao, Filippenko, Graham, Mazzali, Nugent, Pan, Silverman,
  Xu, \& Yaron}]{Taddia2016}
Taddia, F., Moquist, P., Sollerman, J., {et~al.} 2016, A\&A, 587, L7

\bibitem[{Tinyanont {et~al.}(2024)Tinyanont, Foley, Taggart, Davis, LeBaron,
  Andrews, Bustamante-Rosell, Camacho-Neves, Chornock, Coulter, Galbany, Jha,
  Kilpatrick, Kwok, Larison, Pierel, Siebert, Aldering, Auchettl, Bloom,
  Dhawan, Filippenko, French, Gagliano, Grayling, Howell, Jacobson-Gal{\'{a}}n,
  Jones, {Le Saux}, Macias, Mandel, McCully, {Padilla Gonzalez}, Rest, Rho,
  Rojas-Bravo, Skrutskie, Thorp, Wang, \& Ward}]{Tinyanont2024}
Tinyanont, S., Foley, R.~J., Taggart, K., {et~al.} 2024, PASP, 136
  [\eprint[arXiv]{2309.07102}]

\bibitem[{Tonry {et~al.}(2018)Tonry, Denneau, Heinze, Stalder, Smith, Smartt,
  Stubbs, Weiland, \& Rest}]{Tonry2018}
Tonry, J.~L., Denneau, L., Heinze, A.~N., {et~al.} 2018, PASP, 130, 064505

\bibitem[{Tremonti {et~al.}(2004)Tremonti, Heckman, Kauffmann, Brinchmann,
  Charlot, White, Seibert, Peng, Schlegel, Uomoto, Fukugita, \&
  Brinkmann}]{Tremonti2004}
Tremonti, C.~A., Heckman, T.~M., Kauffmann, G., {et~al.} 2004, ApJ, 613, 898

\bibitem[{Trundle {et~al.}(2008)Trundle, Kotak, Vink, \& Meikle}]{Trundle2008}
Trundle, C., Kotak, R., Vink, J.~S., \& Meikle, W.~P. 2008, A\&A, 483, L47

\bibitem[{Valenti {et~al.}(2016)Valenti, Howell, Stritzinger, Graham,
  Hosseinzadeh, Arcavi, Bildsten, Jerkstrand, McCully, Pastorello, Piro, Sand,
  Smartt, Terreran, Baltay, Benetti, Brown, Filippenko, Fraser, Rabinowitz,
  Sullivan, \& Yuan}]{Valenti2016}
Valenti, S., Howell, D.~A., Stritzinger, M.~D., {et~al.} 2016, MNRAS, 459, 3939

\bibitem[{Wang {et~al.}(1997)Wang, Wheeler, \& H{\"{o}}flich}]{Wang1997}
Wang, L., Wheeler, J.~C., \& H{\"{o}}flich, P. 1997, ApJ, 476, L27

\bibitem[{Wheeler {et~al.}(1993)Wheeler, Barker, Benjamin, Boisseau,
  Clocchiatti, de~Vaucouleurs, Gaffney, Harkness, Khokhlov, Lester, Smith,
  Smith, \& Tomkin}]{Wheeler1993}
Wheeler, J.~C., Barker, E., Benjamin, R., {et~al.} 1993, ApJ, 417, L71

\bibitem[{Woosley {et~al.}(1999)Woosley, Eastman, \& Schmidt}]{Woosley1999}
Woosley, S.~E., Eastman, R.~G., \& Schmidt, B.~P. 1999, ApJ, 516, 788

\bibitem[{Woosley {et~al.}(1994)Woosley, Eastman, Weaver, \&
  Pinto}]{Woosley1994}
Woosley, S.~E., Eastman, R.~G., Weaver, T.~A., \& Pinto, P.~A. 1994, ApJ, 429,
  300

\bibitem[{Woosley {et~al.}(1989)Woosley, Hartmann, \& Pinto}]{Woosley1989}
Woosley, S.~E., Hartmann, D., \& Pinto, P.~A. 1989, ApJ, 346, 395

\bibitem[{Woosley {et~al.}(2002)Woosley, Heger, \& Weaver}]{Woosley2002}
Woosley, S.~E., Heger, A., \& Weaver, T.~A. 2002, {The evolution and explosion
  of massive stars}

\bibitem[{Yaron \& Gal-Yam(2012)}]{Yaron2012}
Yaron, O. \& Gal-Yam, A. 2012, PASP, 124, 668

\bibitem[{Yaron {et~al.}(2017)Yaron, Perley, Gal-Yam, Groh, Horesh, Ofek,
  Kulkarni, Sollerman, Fransson, Rubin, Szabo, Sapir, Taddia, Cenko, Valenti,
  Arcavi, Howell, Kasliwal, Vreeswijk, Khazov, Fox, Cao, Gnat, Kelly, Nugent,
  Filippenko, Laher, Wozniak, Lee, Rebbapragada, Maguire, Sullivan, \&
  Soumagnac}]{Yaron2017}
Yaron, O., Perley, D.~A., Gal-Yam, A., {et~al.} 2017, Nat. Phys., 13, 510

\bibitem[{Yoon {et~al.}(2017)Yoon, Dessart, \& Clocchiatti}]{Yoon2017}
Yoon, S.-C., Dessart, L., \& Clocchiatti, A. 2017, ApJ, 840, 10

\bibitem[{Zinn {et~al.}(2012)Zinn, Stritzinger, Braithwaite, Gallazzi, Grunden,
  Bomans, Morrell, \& Bach}]{Zinn2012}
Zinn, P.~C., Stritzinger, M., Braithwaite, J., {et~al.} 2012, A\&A, 538, A30

\end{thebibliography}

\appendix{}
\twocolumn
\centering

\section{Tables} \label{apdx:phot_tab}

\begin{table}[h]
    \def\arraystretch{1.1}%
    \setlength\tabcolsep{3pt}
    \centering
    \fontsize{9}{11}\selectfont
    \caption{Photometry of SN~2022lxg.}
   
    \begin{tabular}{c c c c c c}
    \hline
    \hline
        Date & MJD & Phase (d) & Telescope & Band & Magnitude (mag) \\
    \hline
2022-05-28 & 59727.43 & -3.86 & ZTF & g & 21.15 (99.00) \\ 
2022-06-01 & 59731.39 & 0.02 & ZTF & g & 21.02 (0.20) \\ 
2022-06-03 & 59733.32 & 1.91 & ZTF & g & 16.08 (0.01) \\ 
2022-06-05 & 59735.43 & 3.97 & ZTF & g & 15.78 (0.00) \\ 
2022-06-07 & 59737.41 & 5.91 & ZTF & g & 15.73 (0.00) \\ 
2022-06-09 & 59739.39 & 7.85 & ZTF & g & 15.76 (0.00) \\ 
2022-06-11 & 59741.24 & 9.66 & SEDM & g & 15.73 (0.01) \\ 
2022-06-11 & 59741.43 & 9.85 & ZTF & g & 15.73 (0.00) \\ 
2022-06-13 & 59743.43 & 11.80 & ZTF & g & 15.77 (0.01) \\ 
2022-06-18 & 59748.28 & 16.55 & ZTF & g & 15.87 (0.01) \\ 
2022-06-20 & 59750.32 & 18.55 & ZTF & g & 15.94 (0.01) \\ 
2022-06-22 & 59752.32 & 20.51 & ZTF & g & 15.98 (0.01) \\ 
2022-06-25 & 59755.39 & 23.51 & ZTF & g & 16.18 (0.01) \\ 
2022-06-27 & 59757.32 & 25.40 & ZTF & g & 16.32 (0.01) \\ 
2022-06-27 & 59757.35 & 25.43 & ZTF & g & 16.32 (0.01) \\ 
2022-06-28 & 59758.27 & 26.33 & SEDM & g & 16.41 (0.01) \\ 
2022-06-29 & 59759.37 & 27.41 & ZTF & g & 16.50 (0.01) \\ 
2022-07-03 & 59763.39 & 31.35 & ZTF & g & 16.96 (0.01) \\ 
2022-07-05 & 59765.27 & 33.19 & SEDM & g & 17.17 (0.01) \\ 
2022-07-05 & 59765.39 & 33.30 & ZTF & g & 17.20 (0.01) \\ 
2022-07-08 & 59768.37 & 36.22 & ZTF & g & 17.63 (0.01) \\ 
2022-07-08 & 59768.43 & 36.28 & ZTF & g & 17.65 (0.01) \\ 
2022-07-09 & 59769.39 & 37.22 & ZTF & g & 17.77 (0.01) \\ 
2022-07-10 & 59770.33 & 38.14 & SEDM & g & 17.85 (0.01) \\ 
2022-07-10 & 59770.34 & 38.15 & ZTF & g & 17.89 (0.02) \\ 
2022-07-10 & 59770.39 & 38.20 & ZTF & g & 17.88 (0.01) \\ 
2022-07-10 & 59770.46 & 38.27 & ZTF & g & 17.90 (0.01) \\ 
2022-07-13 & 59773.08 & 40.83 & LT & g & 18.06 (0.07) \\ 
2022-07-13 & 59773.08 & 40.83 & LT & g & 18.11 (0.06) \\ 
2022-07-16 & 59776.20 & 43.89 & SEDM & g & 18.31 (0.01) \\ 
2022-07-17 & 59777.07 & 44.74 & LT & g & 18.36 (0.05) \\ 
2022-07-17 & 59777.07 & 44.74 & LT & g & 18.37 (0.05) \\ 
2022-07-20 & 59780.44 & 48.04 & SEDM & g & 18.61 (0.02) \\ 
2022-07-21 & 59781.16 & 48.74 & LT & g & 18.62 (0.04) \\ 
2022-07-21 & 59781.16 & 48.74 & LT & g & 18.63 (0.04) \\ 
2022-07-22 & 59782.30 & 49.86 & ZTF & g & 18.76 (0.03) \\ 
2022-07-24 & 59784.37 & 51.89 & ZTF & g & 18.85 (0.03) \\ 
2022-07-26 & 59786.32 & 53.80 & ZTF & g & 19.00 (0.03) \\ 
2022-07-26 & 59786.39 & 53.87 & SEDM & g & 18.99 (0.02) \\ 
2022-07-27 & 59787.21 & 54.67 & LT & g & 19.07 (0.07) \\ 
2022-07-27 & 59787.21 & 54.67 & LT & g & 19.06 (0.07) \\ 
2022-07-27 & 59787.21 & 54.67 & LT & g & 18.96 (0.07) \\ 
2022-07-27 & 59787.39 & 54.84 & SEDM & g & 19.15 (0.02) \\ 
2022-07-28 & 59788.48 & 55.91 & SEDM & g & 19.20 (0.03) \\ 
2022-07-30 & 59790.23 & 57.62 & SEDM & g & 19.21 (0.02) \\ 
2022-07-30 & 59790.30 & 57.69 & ZTF & g & 19.22 (0.04) \\ 
2022-08-02 & 59793.27 & 60.60 & ZTF & g & 19.41 (0.04) \\ 
2022-08-04 & 59795.39 & 62.68 & SEDM & g & 19.53 (0.02) \\ 
2022-08-06 & 59797.39 & 64.64 & ZTF & g & 19.58 (0.05) \\ 
2022-08-11 & 59802.31 & 69.45 & ZTF & g & 19.79 (0.12) \\ 
2022-08-14 & 59805.25 & 72.33 & ZTF & g & 20.15 (0.21) \\ 
    \hline
    \hline
    \end{tabular}
    \label{tab:phot}
\end{table}

\begin{table}[h]
    \def\arraystretch{1.1}%
    \setlength\tabcolsep{3pt}
    \centering
    \fontsize{9}{11}\selectfont
    \caption*{Table \ref{tab:phot} continued.}
   
    \begin{tabular}{c c c c c c}
    \hline
    \hline
        Date & MJD & Phase (d) & Telescope & Band & Mag (Error) \\
    \hline
2022-08-16 & 59807.36 & 74.40 & SEDM & g & 20.36 (0.08) \\ 
2022-08-18 & 59809.34 & 76.33 & ZTF & g & 20.41 (0.15) \\ 
2022-08-22 & 59813.29 & 80.20 & ZTF & g & 20.37 (0.10) \\ 
2022-08-24 & 59815.25 & 82.12 & ZTF & g & 20.75 (0.15) \\ 
2022-08-26 & 59817.18 & 84.01 & ZTF & g & 20.72 (0.12) \\ 
2022-08-28 & 59819.18 & 85.97 & ZTF & g & 21.30 (0.21) \\ 
2022-08-30 & 59821.18 & 87.93 & ZTF & g & 21.26 (0.20) \\ 
2022-05-28 & 59727.50 & -3.79 & ATLAS & c & 20.31 (99.00) \\ 
2022-06-01 & 59731.50 & 0.12 & ATLAS & c & 19.12 (0.09) \\ 
2022-06-05 & 59735.46 & 4.00 & ATLAS & c & 15.73 (0.01) \\ 
2022-06-09 & 59739.51 & 7.97 & ATLAS & c & 15.66 (0.01) \\ 
2022-06-21 & 59751.43 & 19.64 & ATLAS & c & 15.79 (0.01) \\ 
2022-07-23 & 59783.42 & 50.96 & ATLAS & c & 18.44 (0.04) \\ 
2022-07-27 & 59787.42 & 54.87 & ATLAS & c & 18.95 (0.08) \\ 
2022-07-31 & 59791.43 & 58.80 & ATLAS & c & 19.09 (0.08) \\ 
2022-08-20 & 59811.38 & 78.33 & ATLAS & c & 20.20 (0.20) \\ 
2022-05-28 & 59727.39 & -3.90 & ZTF & r & 21.01 (99.00) \\ 
2022-06-01 & 59731.44 & 0.07 & ZTF & r & 20.19 (0.11) \\ 
2022-06-03 & 59733.35 & 1.94 & ZTF & r & 16.31 (0.01) \\ 
2022-06-04 & 59734.21 & 2.78 & SEDM & r & 16.08 (0.01) \\ 
2022-06-05 & 59735.39 & 3.93 & ZTF & r & 16.01 (0.01) \\ 
2022-06-07 & 59737.30 & 5.80 & ZTF & r & 15.90 (0.00) \\ 
2022-06-09 & 59739.43 & 7.89 & ZTF & r & 15.90 (0.00) \\ 
2022-06-10 & 59740.21 & 8.65 & SEDM & r & 15.87 (0.01) \\ 
2022-06-11 & 59741.21 & 9.63 & SEDM & r & 15.86 (0.01) \\ 
2022-06-13 & 59743.30 & 11.68 & ZTF & r & 15.80 (0.01) \\ 
2022-06-18 & 59748.34 & 16.61 & ZTF & r & 15.83 (0.00) \\ 
2022-06-19 & 59749.25 & 17.50 & SEDM & r & 15.89 (0.01) \\ 
2022-06-20 & 59750.34 & 18.57 & ZTF & r & 15.92 (0.01) \\ 
2022-06-28 & 59758.25 & 26.31 & SEDM & r & 16.30 (0.01) \\ 
2022-06-28 & 59758.27 & 26.33 & SEDM & r & 16.31 (0.00) \\ 
2022-07-01 & 59761.37 & 29.37 & ZTF & r & 16.47 (0.01) \\ 
2022-07-03 & 59763.32 & 31.28 & ZTF & r & 16.62 (0.01) \\ 
2022-07-05 & 59765.25 & 33.17 & SEDM & r & 16.83 (0.01) \\ 
2022-07-05 & 59765.28 & 33.20 & SEDM & r & 16.78 (0.01) \\ 
2022-07-05 & 59765.34 & 33.26 & ZTF & r & 16.75 (0.01) \\ 
2022-07-10 & 59770.31 & 38.12 & SEDM & r & 17.27 (0.01) \\ 
2022-07-10 & 59770.33 & 38.14 & SEDM & r & 17.27 (0.01) \\ 
2022-07-12 & 59772.19 & 39.96 & SEDM & r & 17.43 (0.02) \\ 
2022-07-13 & 59773.08 & 40.83 & LT & r & 17.53 (0.05) \\ 
2022-07-13 & 59773.08 & 40.83 & LT & r & 17.51 (0.05) \\ 
2022-07-13 & 59773.30 & 41.05 & ZTF & r & 17.40 (0.02) \\ 
2022-07-13 & 59773.37 & 41.12 & ZTF & r & 17.44 (0.02) \\ 
2022-07-14 & 59774.37 & 42.10 & ZTF & r & 17.47 (0.02) \\ 
2022-07-15 & 59775.26 & 42.97 & ZTF & r & 17.54 (0.02) \\ 
2022-07-15 & 59775.31 & 43.02 & ZTF & r & 17.49 (0.02) \\ 
2022-07-15 & 59775.40 & 43.11 & ZTF & r & 17.51 (0.02) \\ 
2022-07-16 & 59776.18 & 43.87 & SEDM & r & 17.69 (0.01) \\ 
2022-07-16 & 59776.20 & 43.89 & SEDM & r & 17.69 (0.01) \\ 
2022-07-16 & 59776.30 & 43.99 & ZTF & r & 17.60 (0.01) \\ 
2022-07-17 & 59777.08 & 44.75 & LT & r & 17.77 (0.05) \\ 
2022-07-17 & 59777.08 & 44.75 & LT & r & 17.77 (0.05) \\ 
2022-07-18 & 59778.37 & 46.01 & ZTF & r & 17.76 (0.02) \\ 
2022-07-19 & 59779.37 & 46.99 & ZTF & r & 17.85 (0.02) \\ 
    \hline
    \hline
    \end{tabular}
\end{table}

\begin{table}[h]
    \def\arraystretch{1.1}%
    \setlength\tabcolsep{3pt}
    \centering
    \fontsize{9}{11}\selectfont
    \caption*{Table \ref{tab:phot} continued.}
   
    \begin{tabular}{c c c c c c}
    \hline
    \hline
        Date & MJD & Phase (d) & Telescope & Band & Mag (Error) \\
    \hline
2022-07-20 & 59780.42 & 48.02 & SEDM & r & 17.99 (0.02) \\ 
2022-07-20 & 59780.44 & 48.04 & SEDM & r & 17.96 (0.01) \\ 
2022-07-21 & 59781.16 & 48.74 & LT & r & 18.01 (0.04) \\ 
2022-07-21 & 59781.16 & 48.74 & LT & r & 18.04 (0.04) \\ 
2022-07-22 & 59782.36 & 49.92 & ZTF & r & 17.99 (0.02) \\ 
2022-07-23 & 59783.30 & 50.84 & SEDM & r & 18.16 (0.01) \\ 
2022-07-24 & 59784.30 & 51.82 & ZTF & r & 18.18 (0.02) \\ 
2022-07-26 & 59786.29 & 53.77 & ZTF & r & 18.25 (0.03) \\ 
2022-07-26 & 59786.37 & 53.85 & SEDM & r & 18.43 (0.02) \\ 
2022-07-26 & 59786.40 & 53.88 & SEDM & r & 18.42 (0.01) \\ 
2022-07-27 & 59787.21 & 54.67 & LT & r & 18.39 (0.05) \\ 
2022-07-27 & 59787.21 & 54.67 & LT & r & 18.38 (0.05) \\ 
2022-07-27 & 59787.21 & 54.67 & LT & r & 18.44 (0.05) \\ 
2022-07-27 & 59787.21 & 54.67 & LT & r & 18.44 (0.05) \\ 
2022-07-27 & 59787.21 & 54.67 & LT & r & 18.41 (0.04) \\ 
2022-07-27 & 59787.36 & 54.81 & SEDM & r & 18.47 (0.02) \\ 
2022-07-27 & 59787.39 & 54.84 & SEDM & r & 18.47 (0.01) \\ 
2022-07-28 & 59788.29 & 55.73 & ZTF & r & 18.43 (0.02) \\ 
2022-07-28 & 59788.45 & 55.88 & SEDM & r & 18.58 (0.03) \\ 
2022-07-28 & 59788.48 & 55.91 & SEDM & r & 18.54 (0.02) \\ 
2022-07-30 & 59790.20 & 57.60 & SEDM & r & 18.62 (0.03) \\ 
2022-07-30 & 59790.29 & 57.68 & ZTF & r & 18.52 (0.03) \\ 
2022-08-02 & 59793.23 & 60.56 & ZTF & r & 18.70 (0.02) \\ 
2022-08-04 & 59795.36 & 62.65 & SEDM & r & 19.02 (0.03) \\ 
2022-08-06 & 59797.35 & 64.60 & ZTF & r & 18.94 (0.03) \\ 
2022-08-09 & 59800.41 & 67.59 & ZTF & r & 19.12 (0.05) \\ 
2022-08-11 & 59802.25 & 69.39 & ZTF & r & 18.89 (0.11) \\ 
2022-08-16 & 59807.29 & 74.33 & ZTF & r & 19.51 (0.06) \\ 
2022-08-16 & 59807.36 & 74.40 & SEDM & r & 19.63 (0.04) \\ 
2022-08-18 & 59809.26 & 76.26 & ZTF & r & 19.71 (0.06) \\ 
2022-08-22 & 59813.19 & 80.10 & ZTF & r & 19.99 (0.08) \\ 
2022-08-24 & 59815.28 & 82.15 & ZTF & r & 19.97 (0.08) \\ 
2022-08-26 & 59817.22 & 84.05 & ZTF & r & 20.14 (0.08) \\ 
2022-08-28 & 59819.20 & 85.99 & ZTF & r & 20.26 (0.09) \\ 
2022-08-30 & 59821.25 & 88.00 & ZTF & r & 20.45 (0.12) \\ 
2022-09-01 & 59823.27 & 89.97 & ZTF & r & 20.74 (0.17) \\ 
2022-09-05 & 59827.31 & 93.93 & ZTF & r & 20.73 (0.21) \\ 
2022-09-17 & 59839.29 & 105.66 & SEDM & r & 21.47 (0.14) \\ 
2023-02-23 & 59998.25 & 261.29 & NOT & r & 23.12 (99.00) \\ 
2023-04-17 & 60051.19 & 313.12 & NOT & r & 24.15 (0.12) \\ 
2022-06-03 & 59733.50 & 2.08 & ATLAS & o & 16.35 (0.01) \\ 
2022-06-07 & 59737.50 & 6.00 & ATLAS & o & 16.03 (0.01) \\ 
2022-06-11 & 59741.48 & 9.90 & ATLAS & o & 15.95 (0.01) \\ 
2022-06-15 & 59745.48 & 13.81 & ATLAS & o & 15.91 (0.01) \\ 
2022-06-17 & 59747.47 & 15.76 & ATLAS & o & 15.89 (0.01) \\ 
2022-06-19 & 59749.57 & 17.82 & ATLAS & o & 15.97 (0.01) \\ 
2022-06-23 & 59753.51 & 21.67 & ATLAS & o & 16.11 (0.01) \\ 
2022-06-27 & 59757.38 & 25.46 & ATLAS & o & 16.32 (0.01) \\ 
2022-07-01 & 59761.43 & 29.43 & ATLAS & o & 16.59 (0.01) \\ 
2022-07-05 & 59765.49 & 33.40 & ATLAS & o & 16.90 (0.01) \\ 
2022-07-07 & 59767.41 & 35.28 & ATLAS & o & 17.01 (0.03) \\ 
2022-07-09 & 59769.44 & 37.27 & ATLAS & o & 17.21 (0.02) \\ 
2022-07-11 & 59771.45 & 39.24 & ATLAS & o & 17.58 (0.16) \\ 
2022-07-13 & 59773.39 & 41.14 & ATLAS & o & 17.63 (0.05) \\ 
2022-07-17 & 59777.48 & 45.14 & ATLAS & o & 17.88 (0.04) \\ 
    \hline
    \hline
    \end{tabular}
\end{table}

\begin{table}[h]
    \def\arraystretch{1.1}%
    \setlength\tabcolsep{3pt}
    \centering
    \fontsize{9}{11}\selectfont
    \caption*{Table \ref{tab:phot} continued.}
   
    \begin{tabular}{c c c c c c}
    \hline
    \hline
        Date & MJD & Phase (d) & Telescope & Band & Mag (Error) \\
    \hline
2022-07-19 & 59779.48 & 47.10 & ATLAS & o & 17.98 (0.03) \\ 
2022-07-25 & 59785.44 & 52.94 & ATLAS & o & 18.50 (0.05) \\ 
2022-07-29 & 59789.37 & 56.78 & ATLAS & o & 18.78 (0.06) \\ 
2022-08-02 & 59793.39 & 60.72 & ATLAS & o & 19.00 (0.06) \\ 
2022-08-08 & 59799.41 & 66.61 & ATLAS & o & 19.42 (0.16) \\ 
2022-08-10 & 59801.40 & 68.56 & ATLAS & o & 19.31 (0.36) \\ 
2022-08-12 & 59803.40 & 70.52 & ATLAS & o & 19.28 (0.22) \\ 
2022-08-14 & 59805.36 & 72.44 & ATLAS & o & 19.61 (0.20) \\ 
2022-08-16 & 59807.37 & 74.41 & ATLAS & o & 19.37 (0.13) \\ 
2022-08-18 & 59809.40 & 76.39 & ATLAS & o & 19.69 (0.15) \\ 
2022-08-22 & 59813.38 & 80.29 & ATLAS & o & 20.16 (0.25) \\ 
2022-08-26 & 59817.38 & 84.21 & ATLAS & o & 20.42 (0.27) \\ 
2022-05-31 & 59730.45 & -0.90 & ZTF & i & 20.90 (99.00) \\ 
2022-06-04 & 59734.43 & 2.99 & ZTF & i & 16.29 (0.01) \\ 
2022-06-07 & 59737.27 & 5.77 & ZTF & i & 16.12 (0.01) \\ 
2022-06-11 & 59741.24 & 9.66 & SEDM & i & 15.99 (0.01) \\ 
2022-06-13 & 59743.34 & 11.72 & ZTF & i & 16.00 (0.01) \\ 
2022-06-17 & 59747.41 & 15.70 & ZTF & i & 15.89 (0.01) \\ 
2022-06-20 & 59750.46 & 18.69 & ZTF & i & 16.00 (0.01) \\ 
2022-06-23 & 59753.37 & 21.54 & ZTF & i & 16.02 (0.01) \\ 
2022-06-26 & 59756.33 & 24.43 & ZTF & i & 16.23 (0.01) \\ 
2022-06-28 & 59758.28 & 26.34 & SEDM & i & 16.29 (0.01) \\ 
2022-07-02 & 59762.41 & 30.39 & ZTF & i & 16.58 (0.01) \\ 
2022-07-05 & 59765.28 & 33.20 & SEDM & i & 16.76 (0.01) \\ 
2022-07-06 & 59766.45 & 34.34 & ZTF & i & 16.92 (0.01) \\ 
2022-07-10 & 59770.33 & 38.14 & SEDM & i & 17.27 (0.01) \\ 
2022-07-13 & 59773.08 & 40.83 & LT & i & 17.61 (0.07) \\ 
2022-07-13 & 59773.08 & 40.83 & LT & i & 17.64 (0.07) \\ 
2022-07-16 & 59776.20 & 43.89 & SEDM & i & 17.71 (0.01) \\ 
2022-07-17 & 59777.08 & 44.75 & LT & i & 17.92 (0.06) \\ 
2022-07-17 & 59777.08 & 44.75 & LT & i & 17.92 (0.05) \\ 
2022-07-20 & 59780.44 & 48.04 & SEDM & i & 18.04 (0.02) \\ 
2022-07-21 & 59781.16 & 48.74 & LT & i & 18.20 (0.06) \\ 
2022-07-21 & 59781.16 & 48.74 & LT & i & 18.20 (0.06) \\ 
2022-07-22 & 59782.35 & 49.91 & ZTF & i & 18.21 (0.03) \\ 
2022-07-23 & 59783.30 & 50.84 & SEDM & i & 18.21 (0.02) \\ 
2022-07-26 & 59786.40 & 53.88 & SEDM & i & 18.37 (0.01) \\ 
2022-07-27 & 59787.21 & 54.67 & LT & i & 18.56 (0.08) \\ 
2022-07-27 & 59787.21 & 54.67 & LT & i & 18.53 (0.08) \\ 
2022-07-27 & 59787.21 & 54.67 & LT & i & 18.53 (0.07) \\ 
2022-07-27 & 59787.39 & 54.84 & SEDM & i & 18.57 (0.01) \\ 
2022-07-29 & 59789.32 & 56.73 & ZTF & i & 18.68 (0.04) \\ 
2022-08-01 & 59792.24 & 59.59 & ZTF & i & 18.83 (0.07) \\ 
2022-08-06 & 59797.43 & 64.67 & ZTF & i & 19.19 (0.07) \\ 
2022-08-10 & 59801.37 & 68.53 & ZTF & i & 19.23 (0.09) \\ 
2022-08-14 & 59805.36 & 72.44 & ZTF & i & 19.92 (0.18) \\ 
2022-08-16 & 59807.36 & 74.40 & SEDM & i & 19.90 (0.05) \\ 
2022-08-17 & 59808.30 & 75.32 & ZTF & i & 19.62 (0.11) \\ 
2022-08-20 & 59811.30 & 78.25 & ZTF & i & 19.98 (0.12) \\ 
2022-08-23 & 59814.34 & 81.23 & ZTF & i & 20.09 (0.15) \\ 
2022-07-13 & 59773.09 & 40.84 & LT & z & 17.38 (0.09) \\ 
2022-07-13 & 59773.09 & 40.84 & LT & z & 17.39 (0.08) \\ 
2022-07-17 & 59777.08 & 44.75 & LT & z & 17.59 (0.07) \\ 
2022-07-17 & 59777.08 & 44.75 & LT & z & 17.61 (0.07) \\ 
2022-07-21 & 59781.17 & 48.75 & LT & z & 17.81 (0.07) \\ 
    \hline
    \hline
    \end{tabular}
\end{table}

\begin{table}[h]
    \def\arraystretch{1.1}%
    \setlength\tabcolsep{3pt}
    \centering
    \fontsize{9}{11}\selectfont
    \caption*{Table \ref{tab:phot} continued.}
   
    \begin{tabular}{c c c c c c}
    \hline
    \hline
        Date & MJD & Phase (d) & Telescope & Band & Mag (Error) \\
    \hline
2022-07-21 & 59781.17 & 48.75 & LT & z & 17.76 (0.08) \\ 
2022-07-27 & 59787.21 & 54.67 & LT & z & 18.01 (0.09) \\ 
2022-07-27 & 59787.22 & 54.68 & LT & z & 18.00 (0.09) \\ 
    \hline
    \hline
    \end{tabular}
\tablefoot{All reported magnitudes are in the AB system and are host subtracted but not dereddened for Milky Way reddening. The table is sorted from bluest to reddest band. A value of 99 in the error refers to an upper-limit.}
\end{table}

\begin{table*} 
\renewcommand{\arraystretch}{1.2}
\setlength\tabcolsep{0.1cm}
\fontsize{10}{11}\selectfont
\begin{center}
\caption{Spectroscopic observations of SN~2022lxg} 
\label{tab:spec_log}
\begin{tabular}{ccrccccc}
\hline
      UT date & 
    MJD & 
    Phase$^{a}$ & 
    Telescope+Instrument &
    Grism/Grating &
        Slit Width &
        Airmass &
    Exposure Time  \\
    (yyyy-mm-dd)   &     &   (days)  &     &   &   (arcsec)  &     &  (s) \\

2022-06-03 & 59733.62 & 2.20 & Keck-I+LRIS & 400/3400+400/8500 & 1.0 & 1.26 & 500 \\
2022-06-04 & 59734.21 & 2.78 & P60+SEDM & - & IFU & 1.84 & 2160 \\
2022-06-07 & 59737.02 & 5.53 & NOT+ALFOSC & GR\#4 & 1.3 & 1.26 & 2400 \\
2022-06-08 & 59738.42 & 6.90 & P200+DBSP & 316/7500+600/4000 & 1.5 & 1.04 & 300 \\
2022-06-08 & 59738.51 & 6.98 & UH88+SNIFS\textsuperscript{\textdagger} & B+R & IFU & 1.14 & 1800 \\
2022-06-09 & 59739.35 & 7.82 & P200+DBSP & 316/7500+600/4000 & 1.5 & 1.08 & 120 \\
2022-06-10 & 59740.21 & 8.66 & P60+SEDM & - & IFU & 1.66 & 2160 \\
2022-06-11 & 59741.25 & 9.67 & P60+SEDM & - & IFU & 1.68 & 2160 \\
2022-06-16 & 59747.15 & 15.45 & NOT+ALFOSC & GR\#4 & 1.0 & 1.08 & 600 \\
2022-06-19 & 59749.25 & 17.51 & P60+SEDM & - & IFU & 1.26 & 2160 \\
2022-06-28 & 59758.25 & 26.31 & P60+SEDM & - & IFU & 1.19 & 1800 \\
2022-07-05 & 59765.25 & 33.17 & P60+SEDM & - & IFU & 1.13 & 1800 \\
2022-07-07 & 59768.16 & 36.02 & NOT+ALFOSC & GR\#4 & 1.0 & 1.21 & 600 \\
2022-07-10 & 59770.31 & 38.12 & P60+SEDM & - & IFU & 1.04 & 1800 \\
2022-07-12 & 59772.09 & 39.87 & LT+SPRAT & Wasatch600 & 1.8	& 1.08 & 1500 \\
2022-07-13 & 59773.99 & 41.73 & NOT+ALFOSC & GR\#4 & 1.0 & 1.09 & 1800 \\
2022-07-16 & 59776.18 & 43.87 & P60+SEDM & - & IFU & 1.27 & 1800 \\
2022-07-18 & 59779.17 & 46.79 & NOT+ALFOSC & GR\#4 & 1.3 & 1.34 & 900 \\
2022-07-20 & 59780.42 & 48.02 & P60+SEDM & - & IFU & 1.22 & 1800 \\
2022-07-21 & 59781.40 & 48.98 & P200+DBSP\textsuperscript{\textdagger} & 316/7500+600/4000 & - & - & - \\
2022-07-23 & 59783.28 & 50.82 & P60+SEDM & - & IFU & 1.04 & 1800 \\
2022-07-26 & 59786.37 & 53.84 & P60+SEDM & - & IFU & 1.12 & 2250 \\
2022-07-27 & 59787.36 & 54.82 & P60+SEDM & - & IFU & 1.13 & 1800 \\
2022-07-27 & 59788.03 & 55.47 & NOT+ALFOSC & GR\#4 & 1.3 & 1.06 & 2400 \\
2022-07-28 & 59788.45 & 55.88 & P60+SEDM & - & IFU & 1.49 & 2250 \\
2022-07-30 & 59790.20 & 57.60 & P60+SEDM & - & IFU & 1.10 & 2250 \\
2022-08-04 & 59795.36 & 62.65 & P60+SEDM & - & IFU & 1.17 & 2250 \\
2022-08-07 & 59798.05 & 65.28 & NOT+ALFOSC & GR\#4 & 1.3 & 1.15 & 2400 \\
2022-08-21 & 59813.04 & 79.96 & NOT+ALFOSC & GR\#4 & 1.3 & 1.17 & 3600 \\

\hline
\end{tabular}
\\[-10pt]
\end{center}
\tablefoot{$^{a}$With respect to the date of explosion ($\mathrm{MJD} = 59\,731.37$) and given in the rest-frame of SN~2022lxg ($z=0.0214$).\\
\textdagger These spectra are not plotted in Fig. \ref{fig:spectra} either due to phase overlap with other higher resolution spectra, or due to low S/N, leading to reduced clarity in the figure.}
\end{table*}

\begin{table}
    \def\arraystretch{1.1}%
    \setlength\tabcolsep{3pt}
    \fontsize{10}{11}\selectfont
    \centering
    \caption{$^{56}$Ni mass estimates for SN~2022lxg. We derive the peak and the tail $^{56}$Ni mass estimates from the luminosities calculated with two different methods, either from the blackbody fits and the Stefan-Boltzmann law, or from the pseudo-bolometric luminosity estimates (see Sect. \ref{subsubsec:Bol} for details).}
    \begin{tabular}{l c}
    \hline
        Method & Value \\
        \hline         
        Arnett, peak, Stf-Bol & 1.210$\pm$ 0.160\,$\msun$ \\
        Arnett, Tail, Stf-Bol &  0.008$\pm$ 0.003\,$\msun$ \\
        Hamuy, tail, Stf-Bol &  0.002$\pm$ 0.001\,$\msun$ \\
        1987A, tail, Stf-Bol &  0.009$\pm$ 0.004\,$\msun$ \\
        Arnett, peak, pseudo-bol &  0.247$\pm$ 0.003\,$\msun$ \\
        Arnett, Tail, pseudo-bol &  0.002$\pm$ 0.001\,$\msun$ \\
        Hamuy, tail, pseudo-bol &  <0.001\,$\msun$ \\
        1987A, tail, pseudo-bol &  0.003$\pm$ 0.001\,$\msun$ \\
         \hline
    \end{tabular}
    \label{tab:nickel}
\end{table}

\begin{table}[h]
    \centering
    \def\arraystretch{1.1}%
    \setlength\tabcolsep{3pt}
    \fontsize{9}{11}\selectfont
    \caption{Imaging polarimetry observations of SN~2022lxg with ALFOSC.}
   
    \begin{tabular}{c c c c c c c c c}
    \hline
    \hline
        Date & MJD & Phase (d) & Band & Exp. time$^{a}$ (s) & q (\%) & u (\%) & p$^{b}$ (\%) & $\chi$ (deg) \\
    \hline
2022-06-18 & 59749.15 & 17.41 & V & 70 & 0.66 (0.24) & $-$0.84 (0.24) & 1.04 (0.24) & -25.92 (6.44) \\ 
2022-07-07 & 59768.22 & 36.07 & V & 80 & 0.43 (0.51) & $-$0.68 (0.56) & 0.65 (0.55) & -28.85 (19.45) \\ 
2022-06-18 & 59749.15 & 17.41 & R & 70 & 0.31 (0.23) & $-$0.76 (0.23) & 0.79 (0.23) & -33.90 (8.03) \\ 
2022-07-07 & 59768.22 & 36.07 & R & 80 & 0.63 (0.50) & $-$0.09 (0.54) & 0.46 (0.50) & -4.01 (22.55) \\ 
\hline
\multicolumn{8}{c}{ISP subtracted}\\
\hline
2022-06-18 & 59749.15 & 17.41 & V & 70 & 0.47 (0.25) & $-$0.61 (0.25) & 0.73 (0.25) & -25.99 (9.12) \\ 
2022-07-07 & 59768.22 & 36.07 & V & 80 & 0.24 (0.51) & $-$0.45 (0.56) & 0.34 (0.55) & -30.66 (31.00) \\ 
2022-06-18 & 59749.15 & 17.41 & R & 70 & 0.12 (0.24) & $-$0.53 (0.24) & 0.49 (0.24) & -38.35 (12.47) \\ 
2022-07-07 & 59768.22 & 36.07 & R & 80 & 0.44 (0.50) & 0.14 (0.54) & 0.30 (0.51) & -8.92 (31.06) \\ 
    \hline
    \hline
    \end{tabular}
\tablefoot{The values in the parentheses are the uncertainties and correspond to 68\% ($1\sigma$). The table is sorted from bluest to reddest band.\\
$^{a}$Per half-wave retarder plate.\\
$^{b}$Corrected for polarisation bias following \citet{Plaszczynski2014}.}
\label{tab:pol_log}
\hfill
\end{table}

\onecolumn
\centering

\section{Supplementary figures} \label{apdx:graphs}

\begin{figure*}[h]
  \centering
  \includegraphics[width=19.5cm]{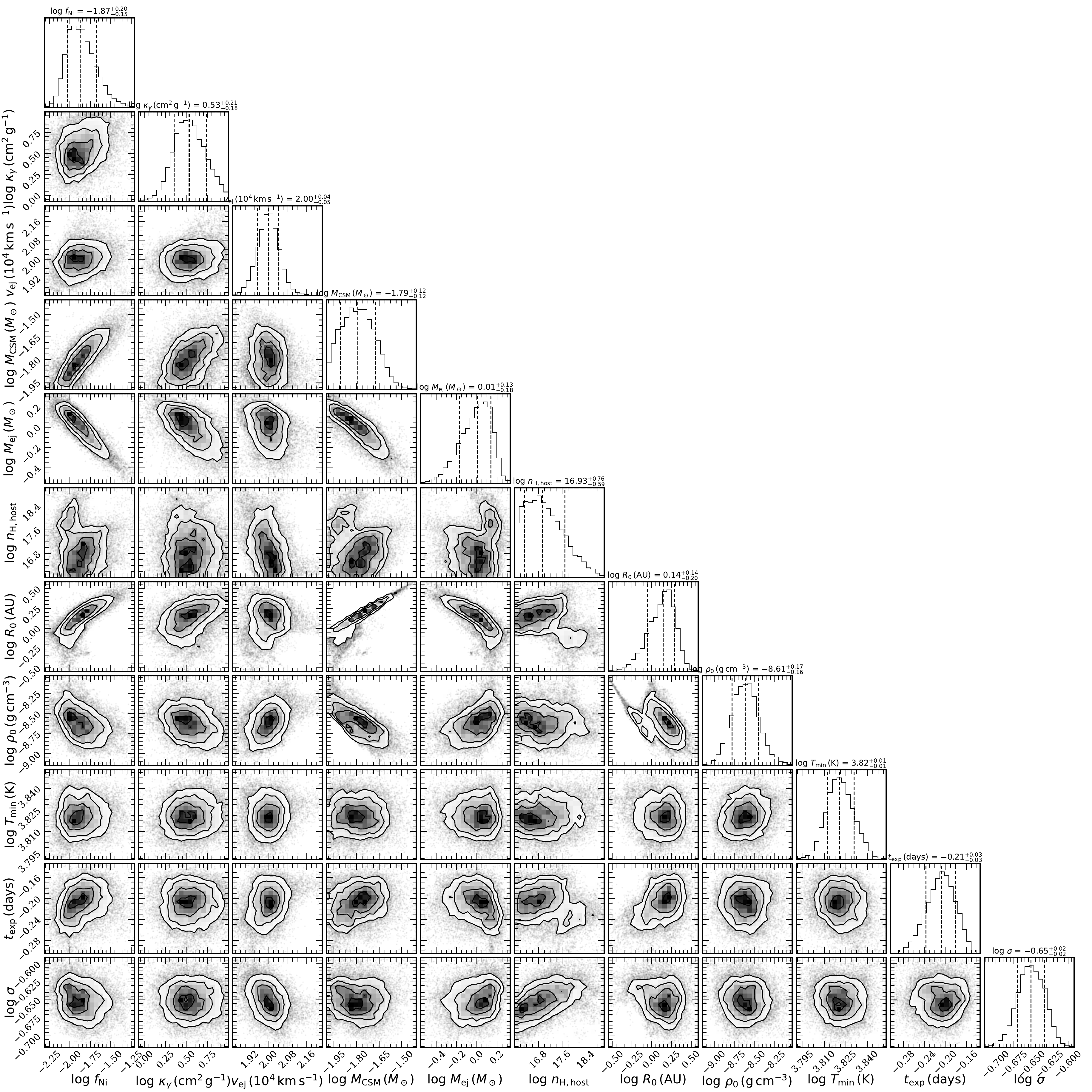}
  \caption{Posterior probability density functions for the free parameters of the model light curves in Fig. \ref{fig:mosfit}.}
  \label{fig:corner}
\end{figure*}

\begin{figure*}
\centering
\includegraphics[width=0.45 \textwidth]{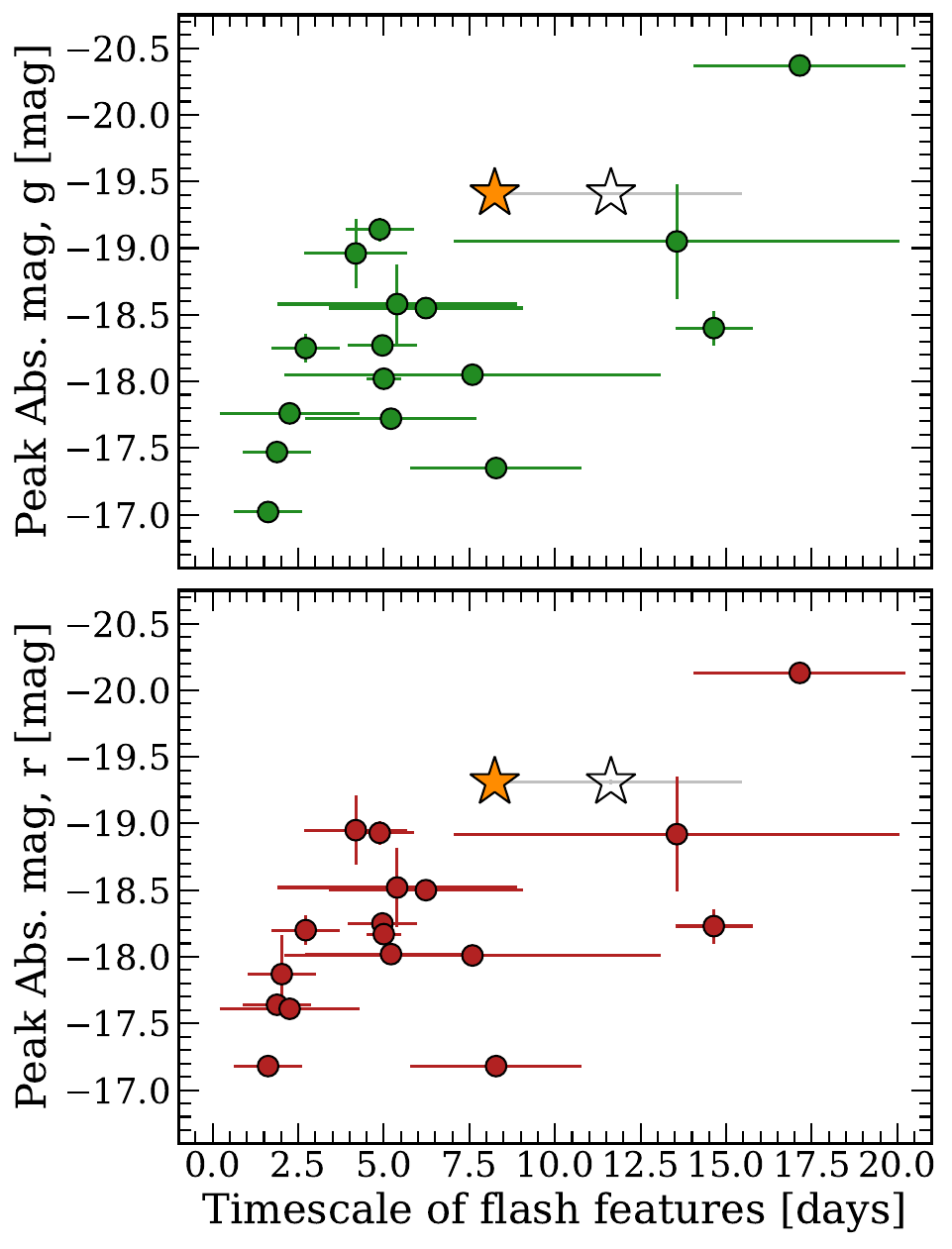}
\includegraphics[width=0.45 \textwidth]{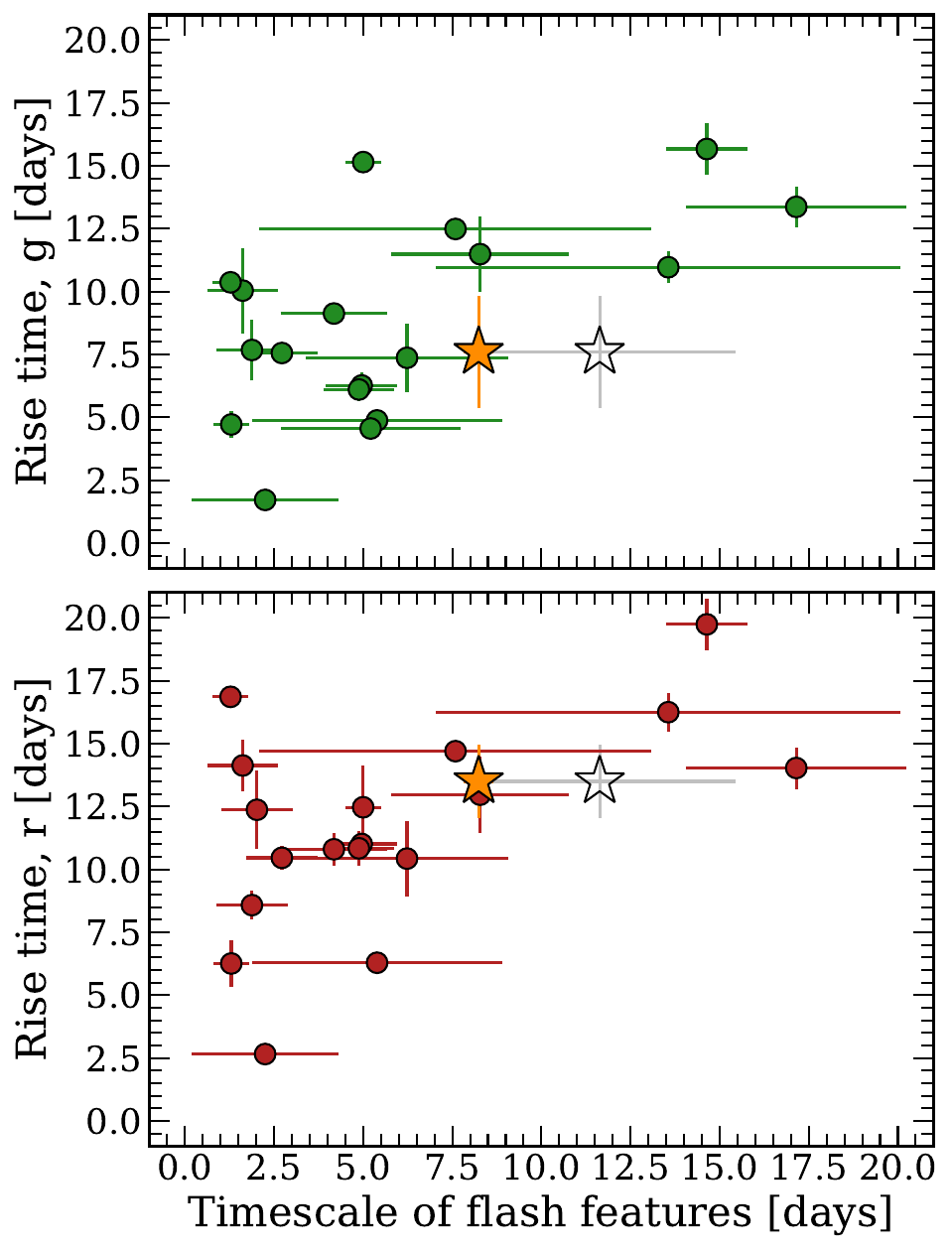}
\caption{Comparison of the timescale of the \ion{He}{II} flash-ionisation feature versus the peak absolute magnitude (left panels) and the rise time (right panels), for various SNe (adapted from \citealt{Bruch2023}). Top panels are in the $g$-band while bottom panels in the $r$-band. SN~2022lxg is shown as a star, where the filled or empty markers denote two potential timescales (8.24 and 11.63 days respectively; see Sect. \ref{subsubsec:ff}. The filled circles are a sample of SNe showing flash-ionisation features taken from \citet{Bruch2023}.}
\label{fig:FF_comp}
\end{figure*}

\begin{figure}
  \centering
      \includegraphics[width=0.5 \textwidth]{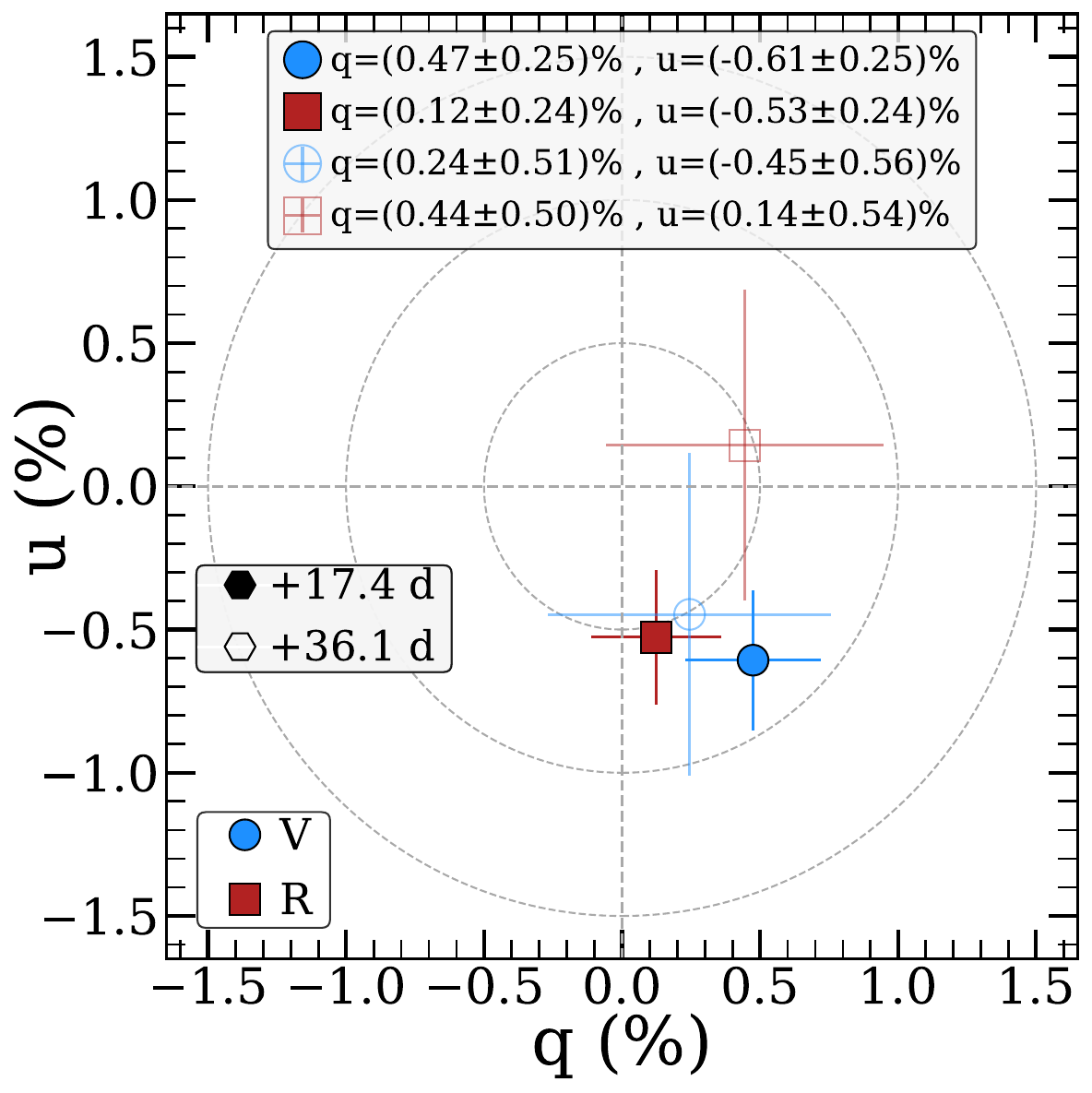}
  \caption{Intensity-normalised Stokes $q$ and $u$ parameters, from ALFOSC imaging polarimetry, in the $V$ (blue circles) and $R$ (red squares) bands, at phases $+17.4$\,d (filled markers) and $+36.1$\,d (open markers). The values are corrected for the ISP contribution. The dashed circles mark the 0.5\%, 1.0\% and 1.5\% polarisation values ($p$). The first epoch (with a good S/N ratio $\sim$ 300) shows that SN~2022lxg is intrinsically polarised to a $p\sim(0.5-1.0) \%$ level.}
  \label{fig:impol_corr}
\end{figure}

\begin{figure}[h]
\centering
\includegraphics[width=1 \textwidth]{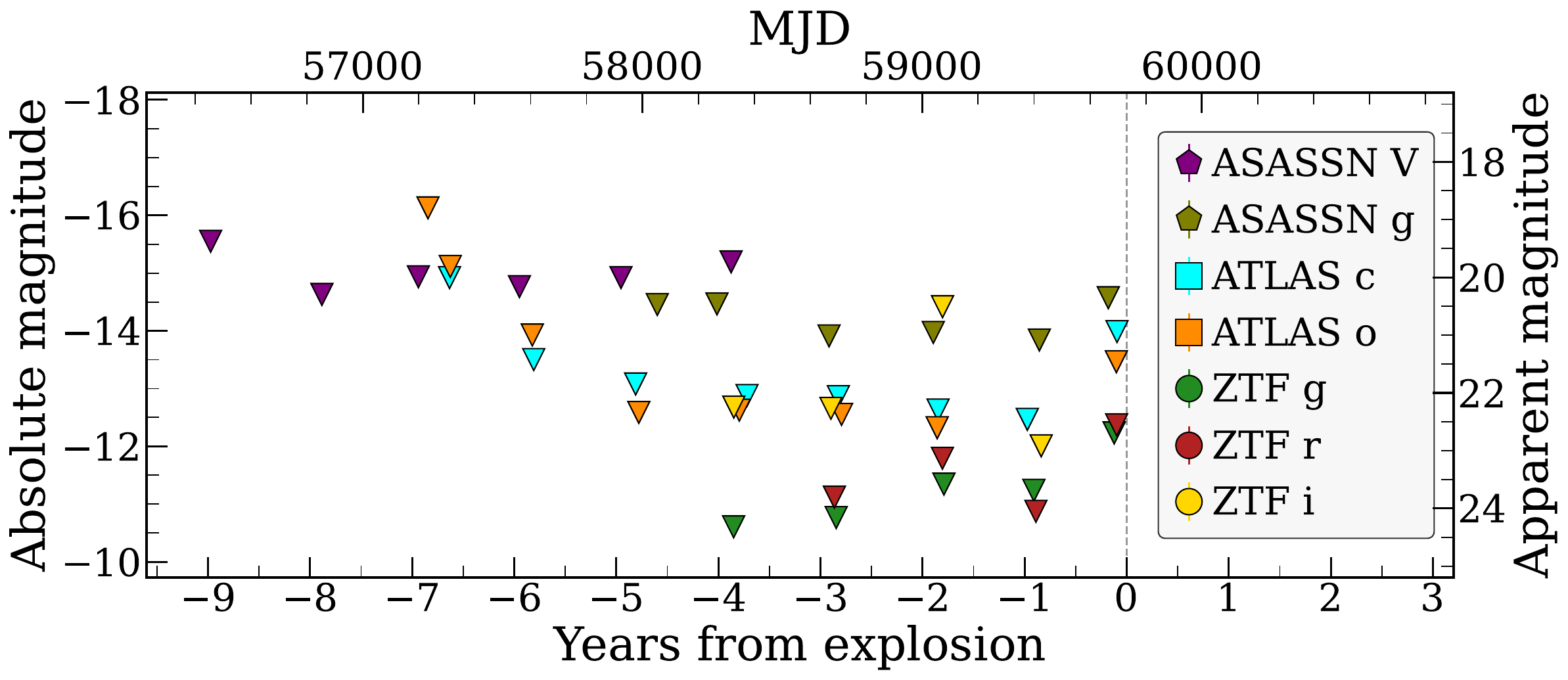}
\caption{Pre-explosion forced photometry at the location of SN~2022lxg from ZTF, ATLAS and ASAS-SN. Individual epochs have been binned per season. Non-detections are shown as downward-facing triangles. The peak time is shown as a dashed vertical line. No significant pre-explosion emission is detected for SN~2022lxg. The forced photometry rules out long-lasting precursors with absolute magnitude $M\lesssim-11$~mag that might have happened within four years prior to explosion, but before that, they are not deep enough to rule out outbursts similar to those of other SNe.}
\label{fig:historic}
\end{figure}

\clearpage

\end{document}